\documentclass{aa}

\usepackage{graphicx,natbib,txfonts, float,footmisc}
\usepackage[english]{babel}
\usepackage[dvipsnames]{xcolor}
\usepackage[colorlinks=true,linkcolor=Blue,citecolor=Blue]{hyperref}

\newcommand{\Phantom}{\textsc{Phantom}}

\newcommand{\iw}{\texttt{res}} \newcommand{\wsh}{\texttt{space}}
\newcommand{\iwlong}{\texttt{i\_wind\_resolution}} \newcommand{\wshlong}{\texttt{wind\_shell\_spacing}}

\newcommand{\Msol}{\rm{M_{\odot}}}
\newcommand{\Mjup}{\rm{M_{Jup}}}

\newcommand{\Rsol}{\rm{R_{\odot}}}
\newcommand{\kms}{\rm{km\,s^{-1}}}
\newcommand{\Msolyr}{\Msol\,yr^{-1}}
\newcommand{\AU}{{\rm au}}

\newcommand{\vwind}{v_{\rm{wind}}}

\newcommand{\vcomp}{v_{\rm{comp}}}
\newcommand{\vagb}{v_{\rm{AGB}}}
\newcommand{\vterm}{v_{\infty}}
\newcommand{\vtermtm}{v_{\infty}^{\rm tm}}
\newcommand{\Mcomp}{M_{\rm{comp}}}
\newcommand{\MAGB}{M_{\rm{AGB}}}
\newcommand{\vini}{v_{\rm{ini}}}
\newcommand{\Mdot}{\dot{M}}

\newcommand{\Rcapt}{R_{\rm{capt}}}
\newcommand{\RHill}{R_{\rm{Hill}}}

\newcommand{\Cnp}{\textsc{P90fast}}
\newcommand{\Cfp}{\textsc{P40fast}}
\newcommand{\Ctp}{\textsc{P25fast}}
\newcommand{\Cns}{\textsc{S90fast}}
\newcommand{\Css}{\textsc{S60fast}}
\newcommand{\Cfs}{\textsc{S40fast}}
\newcommand{\Cts}{\textsc{S25fast}}
\newcommand{\Onp}{\textsc{P90slow}}
\newcommand{\Ofp}{\textsc{P40slow}}
\newcommand{\Otp}{\textsc{P25slow}}
\newcommand{\Ons}{\textsc{S90slow}}
\newcommand{\Oss}{\textsc{S60slow}}
\newcommand{\Ofs}{\textsc{S40slow}}
\newcommand{\Ots}{\textsc{S25slow}}

\begin{document}

\title{ SPH modelling of companion-perturbed AGB outflows including a new morphology classification scheme }

\author{ S. Maes \inst{1}\and W. Homan \inst{2,1} \and J. Malfait \inst{1} \and L. Siess \inst{2} \and J. Bolte  \inst{1} \and F. De Ceuster \inst{3,1}\and L. Decin \inst{1,4} }

\offprints{silke.maes@kuleuven.be}

\institute{ Instituut voor Sterrenkunde, KU Leuven, Celestijnenlaan 200D,
3001 Leuven, Belgium \and
 Institut d'Astronomie et d'Astrophysique, Universit\'e Libre de Bruxelles (ULB), CP 226, 1050 Brussels, Belgium \and
 Department of Physics and Astronomy, University College London, Gower Place, London, WC1E 6BT, United Kingdom \and
 School of Chemistry, University of Leeds, Leeds LS2 9JT, United Kingdom}
\date{Received  / Accepted}


\begin{hyphenrules}{nohyphenation}


\abstract
{Asymptotic giant branch (AGB) stars are known to lose a significant amount of mass by a stellar wind, which controls the remainder of their stellar lifetime. High angular-resolution observations show that the winds of these cool stars typically exhibit mid- to small-scale density perturbations such as spirals and arcs, believed to be caused by the gravitational interaction with a\break (sub-)stellar companion. }
{We aim to explore the effects of the wind-companion interaction on the 3D density and velocity distribution of the wind, as a function of three key parameters: wind velocity, binary separation and companion mass. For the first time, we compare the impact on the outflow of a planetary companion to that of a stellar companion. We intend to devise a morphology classification scheme based on a singular parameter. }
{We ran a small grid of high-resolution polytropic models with the smoothed particle hydrodynamics (SPH) numerical code \Phantom\ to examine the 3D density structure of the AGB outflow in the orbital and meridional plane and around the poles. By constructing a basic toy model of the gravitational acceleration due to the companion, we analysed the terminal velocity reached by the outflow in the simulations.}
{We find that models with a stellar companion, large binary separation and high wind speed obtain a wind morphology in the orbital plane consisting of a single spiral structure, of which the two edges diverge due to a velocity dispersion caused by the gravitational slingshot mechanism. In the meridional plane the spiral manifests itself as concentric arcs, reaching all latitudes. When lowering the wind velocity and/or the binary separation, the morphology becomes more complex: {in the orbital plane a double spiral arises, which is irregular for the closest systems, and the wind material gets focussed towards the orbital plane}, with the formation of an equatorial density enhancement (EDE) as a consequence. Lowering the companion mass from a stellar to a planetary mass, reduces the formation of density perturbations significantly.}
{With this grid of models we cover the prominent morphology changes in a companion-perturbed AGB outflow: slow winds with a close, massive binary companion show a more complex morphology. Additionally, we prove that massive planets are able to significantly impact the density structure of an AGB wind. We find that the interaction with a companion affects the terminal velocity of the wind, which can be explained by the gravitational slingshot mechanism. We distinguish between two types of wind focussing to the orbital plane resulting from distinct mechanisms: {global flattening of the outflow} as a result of the AGB star's orbital motion and the formation of an EDE as a consequence of the companion's gravitational pull. We investigate different morphology classification schemes and uncover that the ratio of the gravitational potential energy density of the companion to the kinetic energy density of the AGB outflow yields a robust classification parameter for the models presented in this paper.
}


\maketitle

\section{Introduction}\label{sect:intro}
Stars with an initial mass between {about 0.8 and 8 }solar masses evolve through the asymptotic giant branch (AGB) phase, situated near the end of the stellar nuclear burning cycles. Stars in this phase are characterised by a significant mass-loss rate of about $10^{-7}$ to $10^{-4}\,\Msolyr$ and terminal wind velocities in a range from $5$ to $20\,\kms$ \citep{Knapp1998,Habing2003,Ramstedt2009}, {which determines their further stellar evolution}. The wind-launching mechanism is believed to emerge from a combination of stellar surface pulsations and dust formation (\citealp{Bowen1988}; see recent review by \citealp{HofnerOlofsson}). Due to the pulsations, stellar material overshoots the surface reaching cooler regions, such that the dense material cools down to about 1600\,K. Consequently, favourable conditions are met for gaseous species to condense into dust grains. These grains absorb the infrared stellar radiation, accelerate outwards and drag along the gas \citep{Liljegren2016, Freytag2017}. Therefore, AGB stars are typically embedded in a dense and dusty circumstellar envelope (CSE). It is exactly because of this favourable combination of high densities and low temperatures that the circumstellar environment of AGB stars exhibits such a rich chemistry, with over 100 molecules and 15 dust species detected so far \citep{Habing1996,Habing2004agbs.book,Heras2005,Verhoelst2009, Water2011, Gail2013, HofnerOlofsson}. 
\\ \indent Furthermore, stars in the AGB phase experience third dredge-up events \citep{Iben1975, Sugimoto1975}. {During these events, the convective envelope penetrates the region previously occupied by the thermal pulse}, and subsequently brings freshly synthesized $^{12}$C to surface layers \citep{Schwarzschild1965}. Due to this mixing, the surface abundance of carbon changes significantly throughout the evolution on the AGB track (see e.g.\ \citealp{LattandWood2004} for a detailed overview). Hence, the initially oxygen-rich (O-rich) star will gradually become more carbon-rich (C-rich) and will eventually attain a C/O-ratio larger than one. The change from O-rich to C-rich directly affects the dust composition in the outflow of the AGB star. Dust grains formed around O-rich AGB stars typically have lower opacities than the grains in C-rich outflows. Since the dust opacity controls the acceleration profile of the stellar wind, O-rich AGB outflows experience a more progressive acceleration and hence, only reach the terminal velocity at much larger distances from the star than C-rich outflows \citep{Decin2010, Decin2020}.
\\ \indent {High-resolution observations of the circumstellar environments of AGB stars, as taken for example with the ALMA interferometer, reveal the presence of complex, asymmetric morphologies on spatial scales from about 10 to several 100\,\AU, including (combinations of) spirals, arcs, disks and bipolarity (e.g.\ \citealp{Mauron2006,Decin2012,Ramstedt2014,Kervella2016,Decin2020,Homan2021_RHya}). The structures observed in AGB outflows show resemblance with the complex asymmetrical morphologies found in post-AGB stars and planetary nebulae (PNe), believed to be the descendants of AGB stars (e.g.\ \citealp{ODell2002,Guerrero2003,Cohen2004, Ertel2019}), suggesting the formation mechanism to be the same, only captured at a different evolutionary stage \citep{Decin2020}. Therefore, by studying the morphologies of AGB stars, we also gain more insight in the shaping mechanism of post-AGB stars and PNe. Though, it has been proven to be difficult to model AGB outflows with a general approach. The only common denominator seems to be the presence of a binary companion, which is observed for some targets (e.g.\ L$_2$ Pup,  \citealp{Homan2017_L2Pup}; $\pi^1$ Gru, \citealp{Homan2020_pi1gru}).} Hence, the dominant shaping mechanism of the CSE is believed to be the gravitational interaction of the AGB wind with a (sub-)stellar binary companion orbiting around the AGB star \citep{Nordhaus2006,Decin2020}. This assumption is supported by population synthesis, since the binarity rate of AGB progenitors is found to be above 50\%, reaching even $\sim$100\% when also planetary companions are included in the binarity rate \citep{Burke2015, Moe2017}.
\\ \indent Over the last three decades, 3D hydrodynamical simulations of AGB outflows with a companion embedded in the CSE have been carried out using particle-based (smoothed particle hydrodynamics; SPH) as well as grid-based (adaptive mesh refinement; AMR) numerical codes. \cite{Theuns1993} were the first to perform such simulations and uncovered that the gravitational potential of the companion is able to shape the AGB outflow in a spiral structure close to the binary system. Depending on the equation of state of the gas, an accretion disk is formed around the companion. \citeauthor{Mastrodemos1998} (\citeyear{Mastrodemos1998, Mastrodemos1999}) included a basic formulation for some molecular cooling and came to similar results:  the formation of an accretion disk and a global spiral morphology in the AGB outflow. They found that the specific shape and complexity of the spiral morphology heavily depends on the initial binary configuration and wind velocity. Some configurations lead to bipolarity in the outflow such that the majority of the outflow is compressed towards the orbital plane -- by \cite{ElMellah2020} called `equatorial density enhancement' (EDE) -- which is observed for some AGB stars \citep{Decin2020}. The simulations by \citeauthor{Chen2017} (\citeyear{Chen2017, Chen2020}) contain more complex cooling, surface pulsations and a basic prescription of radiative transfer, which results in more complex morphologies such as circumbinary disks. Not only the morphology is studied, also mass accretion onto the companion as a result of the wind-companion interaction is already investigated in several works, for example by \cite{Liu2017} and \citeauthor{Saladino2018} (\citeyear{Saladino2018,Saladino2019}).
\\ \indent The different observational and numerical studies expose the complexity of dust-driven winds and the difficulty to model and understand all physical and chemical processes taking place. A recurring parameter of interest determining the global shape of the outflows is the velocity, more precisely the proportion of the orbital velocity of the companion and AGB star to the wind velocity (e.g.\ \citealp{Theuns1993,Saladino2018, ElMellah2020}). In this paper, we analyse the wind morphology of a limited set of 3D SPH models, in which we vary three key parameters. For the first time, we compare the effects of a planetary companion on the CSE of an AGB star with the effects of a stellar companion. 
We aim to find a consistent parameter that is able to indicate the type of morphology, {as a stepping-stone to a systematic classification method for AGB-wind morphologies and as a guide to better constrain AGB system parameters in observations}.
In particular, we will focus on the different key velocities in the simulation and the vertical extent of the wind.
\\ \indent This paper is organised as follows. In Sect.\ \ref{sect:method} we introduce the numerical set-up used and the parameter space of the simulations. In Sect.\ \ref{sect:results} we present our resulting models and discuss the changes in the wind morphology when the set-up parameters are altered. In Sect.\ \ref{sect:discussion} we examine the effects of the different combinations of the model parameters on the terminal wind velocity and vertical extent of the wind. We also investigate different wind morphology classification parameters that have been mentioned throughout the literature. In Sect.\ \ref{sect:conclusions} we review and conclude.

\section{Methodology} \label{sect:method}
\subsection{Numerical set-up}\label{sect:numerics}
In this study, we adopted the numerical technique of smoothed particle hydrodynamics (SPH; \citealp{Lucy1977,GandM1977}), which solves the equations of fluid dynamics in a mesh-free, Lagrangian way using smoothing lengths and kernels to ensure a proper representation of the fluid. More specifically, we used the code \Phantom\ to perform the modelling (\citealp{PriceFederarth2010,LodatoPrice2010,PricePhantom2012}; see extensive overview by \citealp{Price_Phantom2018}).
\\ \indent The simulations consist of (i) two sink particle with a certain mass, representing the AGB star and its companion and (ii) a distribution of SPH particles representing the AGB wind. Both sink particles are so-called `gravity-only', namely they are considered gravitational point sources of which the internal structure is not modelled. The SPH particles interact according to the laws of hydrodynamics, generally given by the following conservation laws, here given in the Lagrangian form:
\begin{eqnarray}
&&\frac{\rm{d}\rho}{\rm{d}\textit{t}}+\rho \vec{\nabla}\cdot\textit{\textbf{v}} = 0 \quad {\rm{(continuity\ equation)} } \label{eq:continuity}\\
&&\rho \frac{\rm{d}\textit{\textbf{v}}}{\rm{d}\textit{t}} = -\vec{\nabla} p + \textit{\textbf{F}} \quad {\rm{(equation\ of\ motion)}}\label{eq:momentum_cons}\\
&&\frac{\rm{d}\textit{u}}{\rm{d}\textit{t}} = (\gamma-1) \frac{u}{\rho}\frac{\rm{d}\rho}{\rm{d}\textit{t}} + \Lambda \quad { \rm{(conservation\ of\ internal\ energy)}} \label{eq:energy_cons}
\end{eqnarray}
where $\rho$ is the gas density, $\textit{\textbf{v}}$ the velocity, $p$ the gas pressure and $u$ the internal energy of the gas. In Eq.\ (\ref{eq:momentum_cons}), which represents the conservation of momentum, the term $\textit{\textbf{F}}$ contains all additional forces that may act on the fluid, for example gravity and radiation pressure. Eq.\ (\ref{eq:energy_cons}) contains information about the thermodynamics via the polytropic index $\gamma$ and the term $\Lambda$ includes additional processes which may change the internal energy, namely cooling/heating terms from diverse processes if present. 
\\ \indent The wind itself is modelled as an ideal gas of which the thermodynamics obeys the polytropic equation of state
\begin{equation}\label{eq:polytropic}
p \propto \rho^\gamma.
\end{equation}
Since the temperature profile for AGB outflows is found to follow the power-law behaviour in a spherical symmetric approximation 
\begin{equation}
T(r) \propto r^{-\xi},
\end{equation}
with $\xi\approx0.6-0.7$ \citep{Millar2004}, it can be shown that this corresponds to a polytropic index $\gamma$ of 1.2. As a consequence, the thermodynamical behaviour of the gas lies in a regime between the adiabatic ($\gamma = 7/5$ for a diatomic gas) and isothermal ($\gamma=1$) extremes.
\\ \indent In AGB outflows, adiabatic cooling regulates the temperature of the wind, but there is also a significant contribution of various heating and cooling processes, which {are included in Eq.\ (\ref{eq:energy_cons}) via the term $\Lambda$, if taken into account in the modelling}. In reality, the internal energy of the gas component of the AGB outflow is mainly determined by (i) gas-dust grain collisions, (ii) the photoelectric heating from dust grains, (iii) heat exchange between the dust and gas component and (iv) by cosmic rays. The cooling in AGB winds happens mainly by collisional excitation of rotational levels of abundant molecules, such as H$_2$O (O-rich winds), HCN (C-rich winds) and CO, and by  vibrational excitations of H$_2$ \citep{Decin2006}. Additionally, the reaction enthalpies of chemical processes such as the formation and dissociation of molecules (e.g.\ H$_2$) are able to contribute significantly to the heating and cooling of the wind \citep{Omukai2000}. However, in this paper we did not include these processes (i.e.\ $\Lambda = 0$ in Eq.\ (\ref{eq:energy_cons})), {since coupling a chemical network and/or radiative transfer approach with hydrodynamical modelling at the current resolution, necessary to predict molecular and dust species abundances and their interaction with radiation, is computationally unfeasible to date.} As a direct consequence, the dynamics of the SPH fluid and thus the morphology, is invariant to density scaling {or, equivalently, the mass-loss rate} (see Eqs.\ (\ref{eq:continuity}) -- (\ref{eq:energy_cons})).
\\ \indent These ingredients (dust-gas chemistry and radiation) are in principle also needed to simulate the launch of an AGB wind \citep{Woitke2006,Boulangier2019}. To overcome the issue of their absence, the wind acceleration mechanism due to radiation pressure on the dust is mimicked by introducing an additional effective potential in the equation of motion to reduce the effect of the gravity of the AGB star \citep{LamersCassinelli}. Thus, the radial component of the force term in Eq. (\ref{eq:momentum_cons}) becomes
\begin{equation}\label{eq:wind_acc}
F_r = -\frac{G\MAGB}{r_1^2}(1-\Gamma)-\frac{G\Mcomp}{r_2^2},
\end{equation}
where the constant $\Gamma \geq 0$. {Here, $r_1$ and $r_2$ are the distances to the AGB star and companion, respectively, and $\MAGB$ and $\Mcomp$ the masses,} $G$ is the gravitational constant. To actually launch a ballistic wind, we followed the `free wind' case introduced by \cite{Theuns1993}: we assume that the gravity of the AGB star is balanced exactly by the radiation pressure on the dust by setting $\Gamma=1$. Otherwise, the ejected gas will either fall back on the AGB star ($\Gamma < 1$) or will accelerate indefinitely ($\Gamma > 1$). {The resulting velocity profile for a single AGB star is shown in Fig.\ \ref{fig:vel_profile} in black.} This method is also adopted by \cite{Mastrodemos1998}, \cite{KimTaam2012} and \cite{Liu2017}, for example. We note that the prescription of the momentum equation, Eq.\ (\ref{eq:momentum_cons}), using the force term of Eq.\ (\ref{eq:wind_acc}) corresponds to a fast acceleration of the AGB wind, that is for a single AGB star the terminal velocity is reached within the first few stellar radii of the simulated region. Further, we did not include surface pulsations and spin of the AGB star in our simulations, although they may have an important effect on the morphology of the outflow (e.g.\ \citealp{ Mastrodemos1999,Chen2017}). We also did not include tidal distortions, which may become significant when the AGB star starts to fill a substantial part of its Roche lobe. Lastly, effects due to radiation and magnetic fields were not taken into account. 
\\ \indent The wind is injected into the simulation as a number of collisionless SPH particles with constant mass, isotropically placed on a spherical shell at the effective {radius of the AGB star, set up at $1.267\,\AU\approx 272\,\Rsol$}. In other words, this radius can be considered as the location at which the outflow has reached a significant speed in order that a wind is launched. {Such shells are successively launched from the effective radius with a time interval $\Delta t$.
	The first five shells in the simulation are stationary, so that they initialise the pressure and density gradient in the wind, whereafter the equations of hydrodynamics take over to determine the further development of the outflow. The resolution parameters set the distance between the shells and the amount of particles initialised on each shell (for details, see Appendix \ref{sect:resolution}). Hence, the resolution sets $\Delta t$ in combination with the input wind velocity $\vini$, and the mass of the individual SPH particles $M_{\rm SPH}$ in combination with the input mass-loss rate $\Mdot$}. In order to minimise artificial perturbations and artefacts, the distribution of equidistant SPH particles on the spherical shells is given a random orientation before the shell is launched. A comprehensive overview on the details of the numerical implementation of stellar winds in \Phantom\ is in preparation.  
\\ \indent The companion is able to accrete wind material. This is realised by a basic wind-accretion mechanism, so-called `prompt accretion'. When an SPH particle comes within the predefined radius of the companion, which acts as its physical surface, and satisfies a number of accretion checks, it is removed from the simulation (for details, see \citealp{Price_Phantom2018}). The state of the companion sink particle (e.g.\ mass, position, velocity) is updated so that mass, linear and angular momentum are conserved by the accretion of the SPH particle. {For the stellar companions this radius is set to $0.00456\,\AU \approx 1\,\Rsol$ and for the planetary companions to $0.000912\,\AU\approx2\,{\rm R_{Jup}}$. However, since no accurate cooling is implemented here, the wind accretion by the companion cannot be modelled accurately and only serves as a way to control the amount of particles in the simulation and optimise the resolution.} Moreover, the models are constrained by an outer boundary: if SPH particles cross this boundary, they are also removed from the simulation. Similarly as before, the main reason to set up this outer boundary is to optimise the resolution of the simulations according to the maximum amount of SPH particles and runtime. The simulations contain on average $10^6$ SPH particles and are evolved up to a state of convergence or so-called `self-similar' behaviour, that is the wind morphology no longer changes in snapshots taken at time intervals equal to the orbital period. Self-similarity is reached after about 7 orbits here, depending on the binary separation, wind velocity and outer boundary of the modelled domain. 
\\ \indent We note that, although the resolution in our models is fairly high, it is still hard or even impossible to feasibly resolve the region within a few tenths of \AU\ around the companion accurately in the case of low-mass companions, such as planets, since only a dozen of SPH particles will be present in that region. Therefore, the gravitational interaction between the wind and the companion may not be optimally modelled for such simulations. For more massive companions, such as stellar companions, this is not a problem, since the gravitational potential of the companion is strong enough to attract a significant amount of SPH particles, resolving the interaction adequately.

\subsection{Parameter set-up}
\begin{table}
	\begin{center}
	\caption{Overview of the configuration of the twelve models (input values and orbital velocities).}
	\begin{tabular}{   l   c  c  c  c  c     }
		\hline \hline \\[-2ex]
		model &$\vini$ &  $a$ & $\Mcomp$ & $\vagb$& $\vcomp$\\ 
		      & {\scriptsize$[\kms]$} & {\scriptsize$[\AU]$} & {\scriptsize$[\Msol]$} &{\scriptsize$[\kms]$} &{\scriptsize$[\kms]$} \\ \hline
		\Cns& 20.0& 9.0& 1.0  & 6.3 & 9.4 \\
		\Cfs& 20.0& 4.0& 1.0  & 9.4 & 14.1\\
		\Cts& 20.0& 2.5& 1.0  & 11.9& 17.9\\
		\Cnp& 20.0& 9.0& 0.01 & 0.1 & 12.1\\
		\Cfp& 20.0& 4.0& 0.01 & 0.1 & 18.2\\
		\Ctp& 20.0& 2.5& 0.01 & 0.2 & 23.0\\ 
		\Ons& 5.0 & 9.0& 1.0  & 6.3 & 9.4  \\
		\Ofs& 5.0 & 4.0& 1.0  & 9.4 & 14.1 \\
		\Ots& 5.0 & 2.5& 1.0  & 11.9& 17.9 \\
		\Onp& 5.0 & 9.0& 0.01 & 0.1 & 12.1 \\
		\Ofp& 5.0 & 4.0& 0.01 & 0.1 & 18.2 \\
		\Otp& 5.0 & 2.5& 0.01 & 0.2 & 23.0 \\ \hline
	\end{tabular}
	\label{tab:modelsDetails}
	\end{center}
	{\footnotesize \textbf{Notes.} Here, $\vini$ is the initial velocity (input), $a$ the binary separation (input) and $\Mcomp$ the input mass of the companion ($0.01\,\Msol\approx10\,\Mjup $). The naming of the models is abbreviated according to `$\Mcomp+10a\ +\ $windtype', where $\Mcomp$ is given by `\textsc{S}' or `\textsc{P}', standing for `stellar' and `planetary', respectively. `\textsc{fast}' refers to a wind initiated with $\vini = 20\,\kms$ and $\Mdot = 10^{-4}\,\Msolyr$, and `\textsc{slow}' to a wind with $\vini = 5\,\kms$ and $\Mdot = 2\times10^{-7}\,\Msolyr$. $\vagb$ and $\vcomp$ are the orbital velocity\footref{foot:v_orb} of the AGB star and companion, respectively. Following applies for all models: $\MAGB = 1.5\,\Msol$, $e=0$, $\gamma = 1.2$. 
	}	
\end{table}
The aim of this study is to analyse the morphology of a companion-perturbed AGB outflow in different binary configuration with circular orbits (i.e.\ eccentricity $e=0$). We fixed the mass of the AGB star for all models at 1.5\,$\Msol$ and made use of two different companion masses: a stellar companion of 1\,$\Msol$ and a planetary companion of $0.01\,\Msol\approx10\,\Mjup$. We varied the binary separation between three values: $9.0\,\AU$, $4.0\,\AU$ and $2.5\,\AU$. These values were chosen such that the modelled system correspond to a so-called detached binary system \citep{EggletonBinaryBook}. {More precisely, the effective radius of the AGB star is smaller than the distance to the first Lagrangian point from the centre of the AGB star.} According to population synthesis and statistics, the binary configurations modelled here are expected to be observed approximately 20\% of the time \citep{Moe2017, Fulton2019, Decin2020}.  
\\ \indent Lastly, we varied between two different wind types: a fast wind and a slow wind. These two set-ups mimic roughly a C-rich and O-rich outflow, respectively. More specifically, the wind velocity at the location of the companion in the simulations will be lower for the slow-wind set-up, as would be the case in an O-rich AGB outflow. This is due to the progressive velocity profile, given by the well-known $\beta$-velocity law for stellar winds \citep{LamersCassinelli}:
\begin{equation}\label{eq:betalaw}
{ v(r) \simeq v_0 + (\vterm-v_0)\left(1-\frac{R_{\rm 0}}{r}\right)^\beta,}
\end{equation}
\begin{figure}
	\centering
	\includegraphics[width=0.45\textwidth]{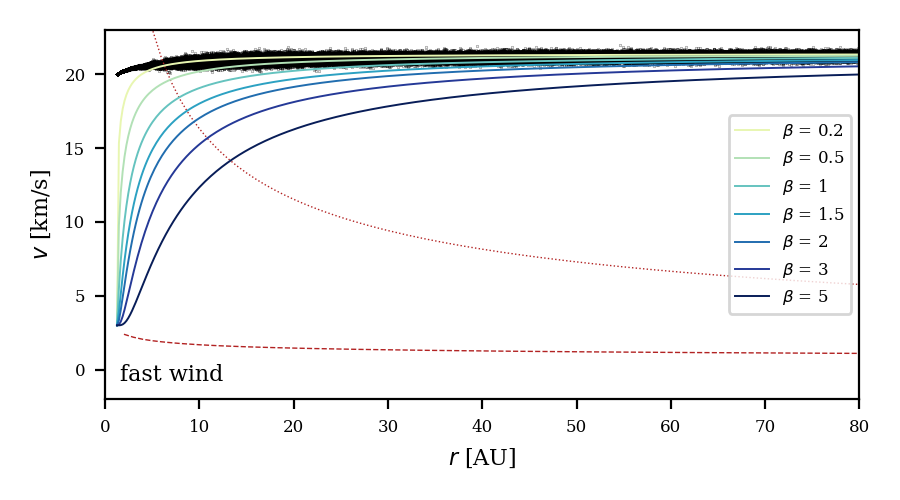}
	\\[-2.3ex]
	\includegraphics[width=0.45\textwidth]{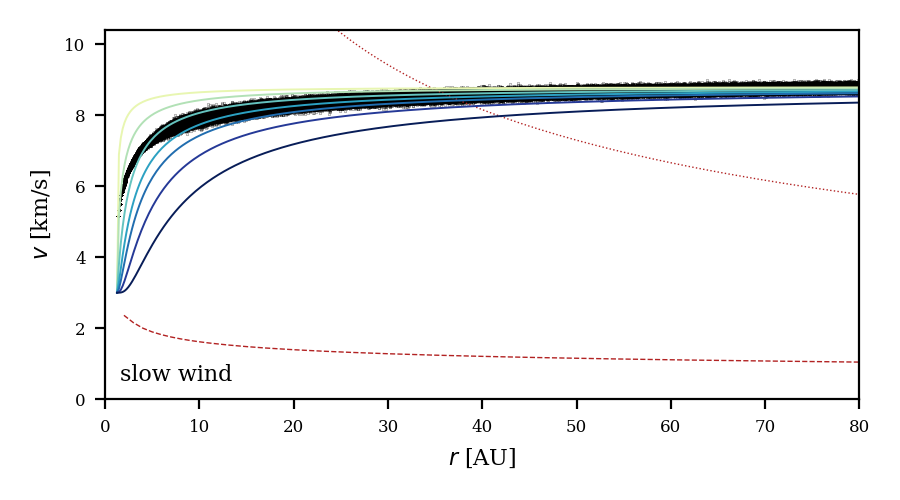}
	\\[-2.3ex]
	\caption{Radial velocity profile of a single-star AGB wind modelled with \Phantom\ in black, including $\beta$ wind velocity profiles in colour according to Eq.\ (\ref{eq:betalaw}), as indicated by the legend. \textit{Upper panel}: fast wind ($\vini = 20\,\kms$). \textit{Bottom panel}: slow wind ($\vini = 5\,\kms$). {For information the mean sound speed and the escape velocity are given in red in dashed and dotted lines, respectively}.}
	\label{fig:vel_profile}
\end{figure}
where for O-rich winds the value of $\beta$ ranges between 1 and 2, but can go up to $\sim$5 \citep{Decin2010}. For C-rich winds, the value for $\beta$ is generally found to be lower than 1/2 \citep{Decin2015}, which corresponds to a fast acceleration. In Eq.\ (\ref{eq:betalaw}) $\vterm$ is the terminal velocity and $v_0$ the velocity at radius $R_{0}$. Hence, given Eq.\ (\ref{eq:wind_acc}), the fast-wind models are set up with an initial velocity of $20\,\kms$ and the slow-wind models with $5\,\kms$. {In fig.\ \ref{fig:vel_profile} we compare the velocity profile of a single-star AGB wind with the $\beta$-velocity law from Eq.\ (\ref{eq:betalaw}) for different values of $\beta$, where we used $v_0 = 3\,\kms$ as minimum velocity following \cite{Danilovich2014} and $R_0 = 1.267\,\AU$ the effective radius of the AGB star in the simulations. For the fast and slow wind $\vterm$ was set to $21.4\,\kms$ and $8.8\,\kms$, respectively, as found in the simulations. From Fig.\ \ref{fig:vel_profile} we conclude that the wind acceleration mechanism employed here, Eq.\ (\ref{eq:wind_acc}), is reasonable in comparison with the $\beta$- velocity law. We see that the velocity profile for a fast wind corresponds to values $\beta<1$ and for a slow wind $\beta > 1$, as expected from observations.} Although in our simulations the mass-loss rate will not matter for the wind morphology (Sect.\ \ref{sect:numerics}), the fast and slow wind velocity can be associated with mass-loss rates of $10^{-4}\,\Msolyr$ and $2\times10^{-7}\,\Msolyr$, respectively, following the quasi-linear trend between those quantities \citep{Ramstedt2009}. The input parameters of the twelve models and the orbital velocity\footnote{The orbital velocity for the components in a binary system is given by $v_i=2\pi r_i/P$, where $P$ is the period according to Kepler's third law: $P=2\pi\sqrt{\frac{a^3}{G(m_1+m_2)}}$, and $r_i$ the distance from component $i$ to the centre-of-mass: $r_1 = \frac{a}{1-q^{-1}}$, $r_2 = a-r_1$, with $a$ the binary separation and $q=m_2/m_1\leq1$ the mass ratio.\label{foot:v_orb}} of both binary components are given in Table \ref{tab:modelsDetails}, including the naming of the different simulations.

\section{Results} \label{sect:results}
The different simulations showed a clear appearance of structure formation in the physical properties of the AGB outflow. In this section, we discuss the basic morphological effects of adjusting each parameter individually. {The morphology is presented in a slice through the $xy$-plane and $xz$-plane, corresponding to the orbital plane and meridional plane, respectively, at snapshots where self-similarity is reached.}

\subsection{General description of the wind morphology}\label{sect:morphology_indetail}
To facilitate the comparison between the models, we opted to select one reference model, with which all other models will be compared. Accounting for the known effects that shape the post-interaction AGB wind, we selected \Cns\ as reference model. This model shows that the dominant shaping mechanism is the well-understood funnelling of wind material into a spiral tail. 
\begin{figure*}
	\centering
	\includegraphics[width=0.31\textwidth]{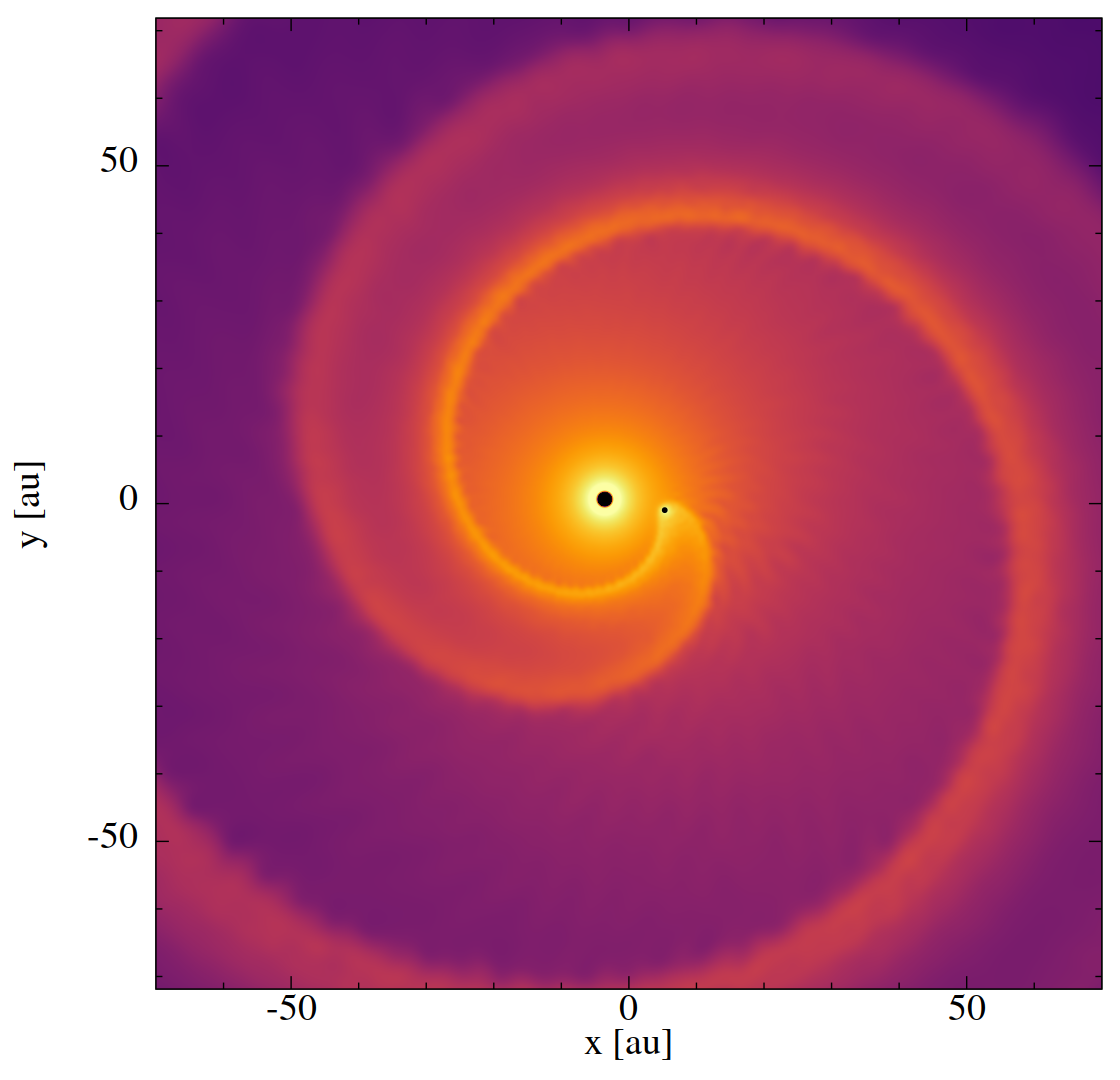}
	\includegraphics[width=0.31\textwidth]{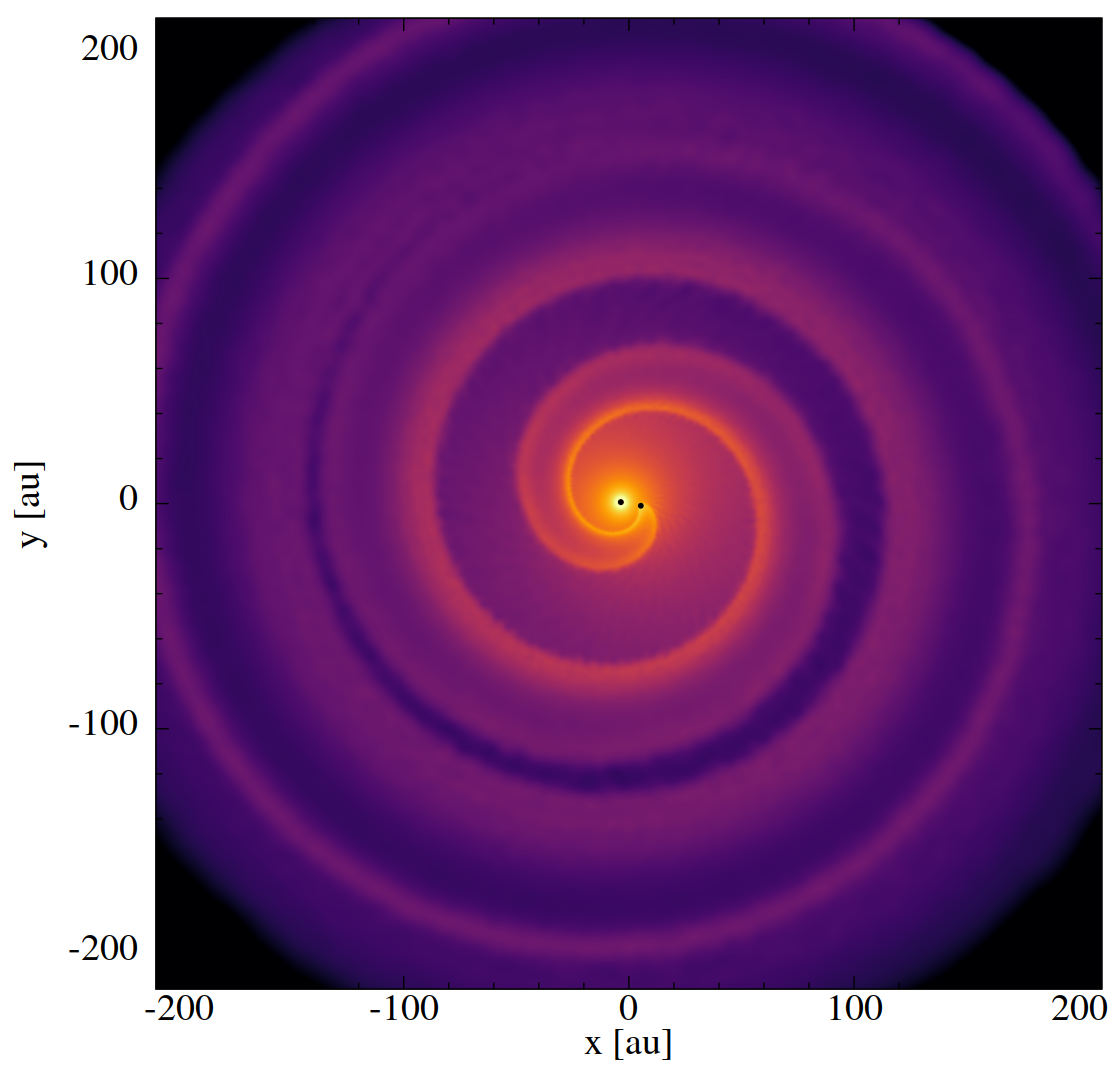}
	\includegraphics[width=0.355\textwidth]{{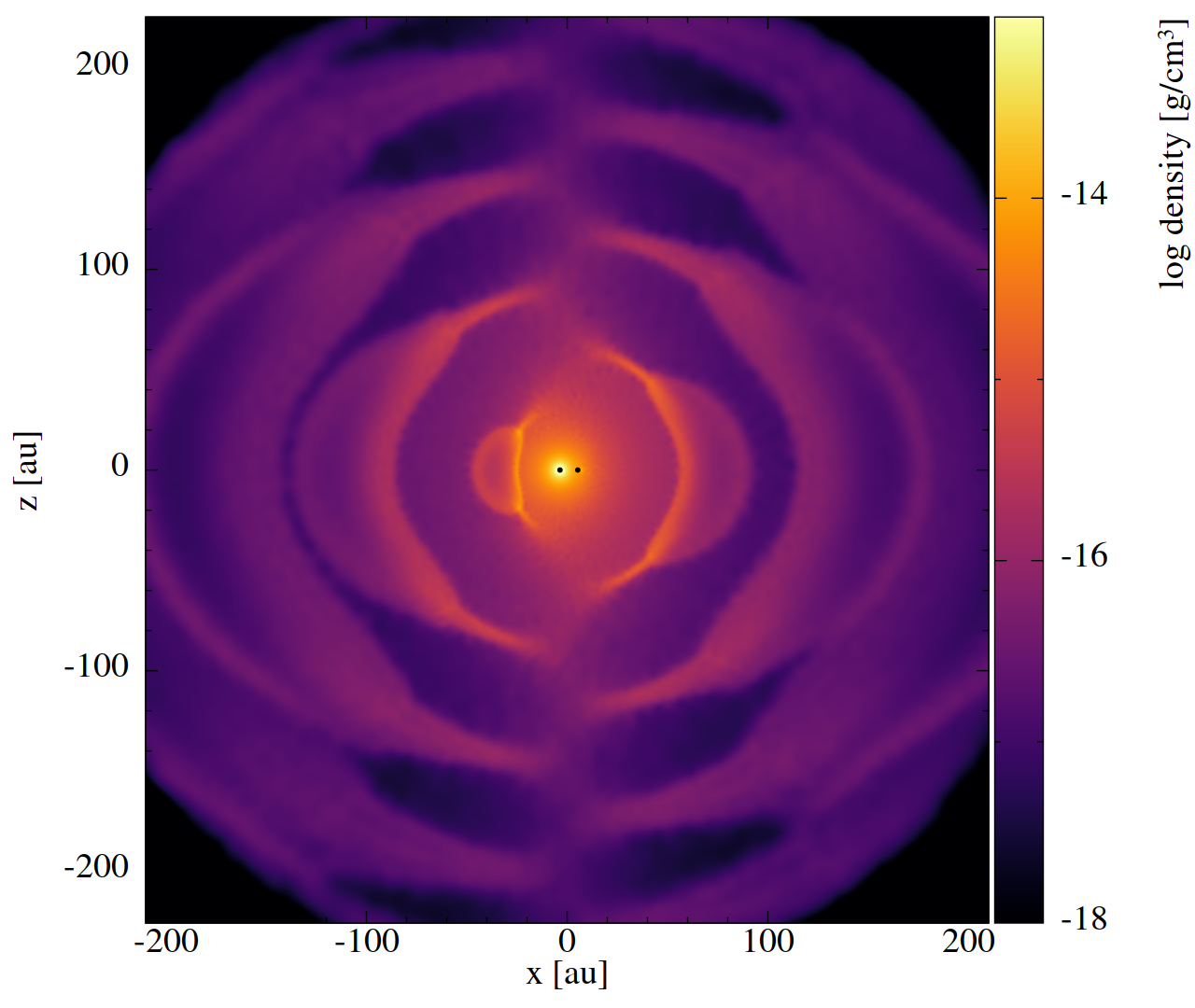}}
	\includegraphics[width=0.31\textwidth]{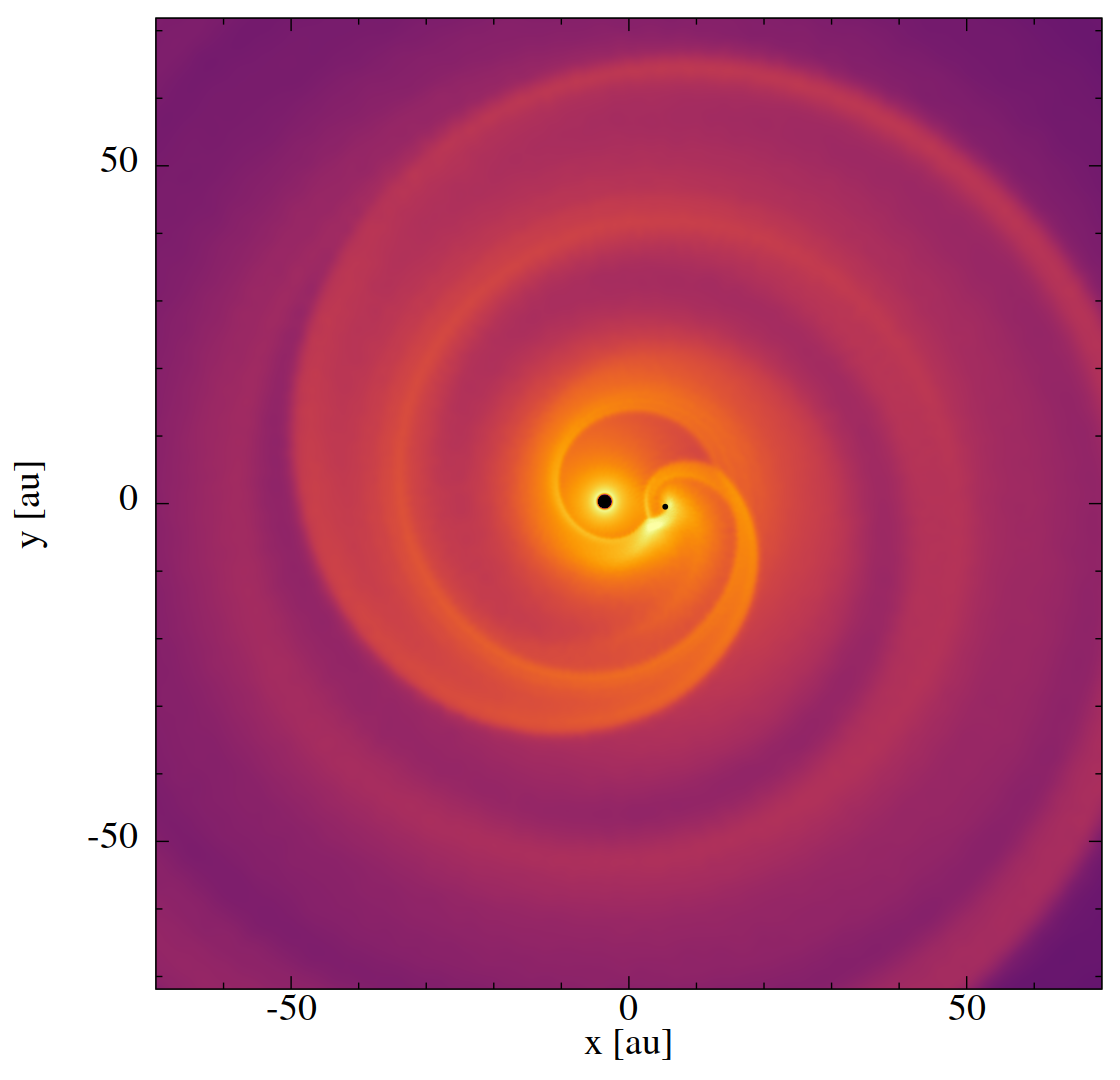}
	\includegraphics[width=0.31\textwidth]{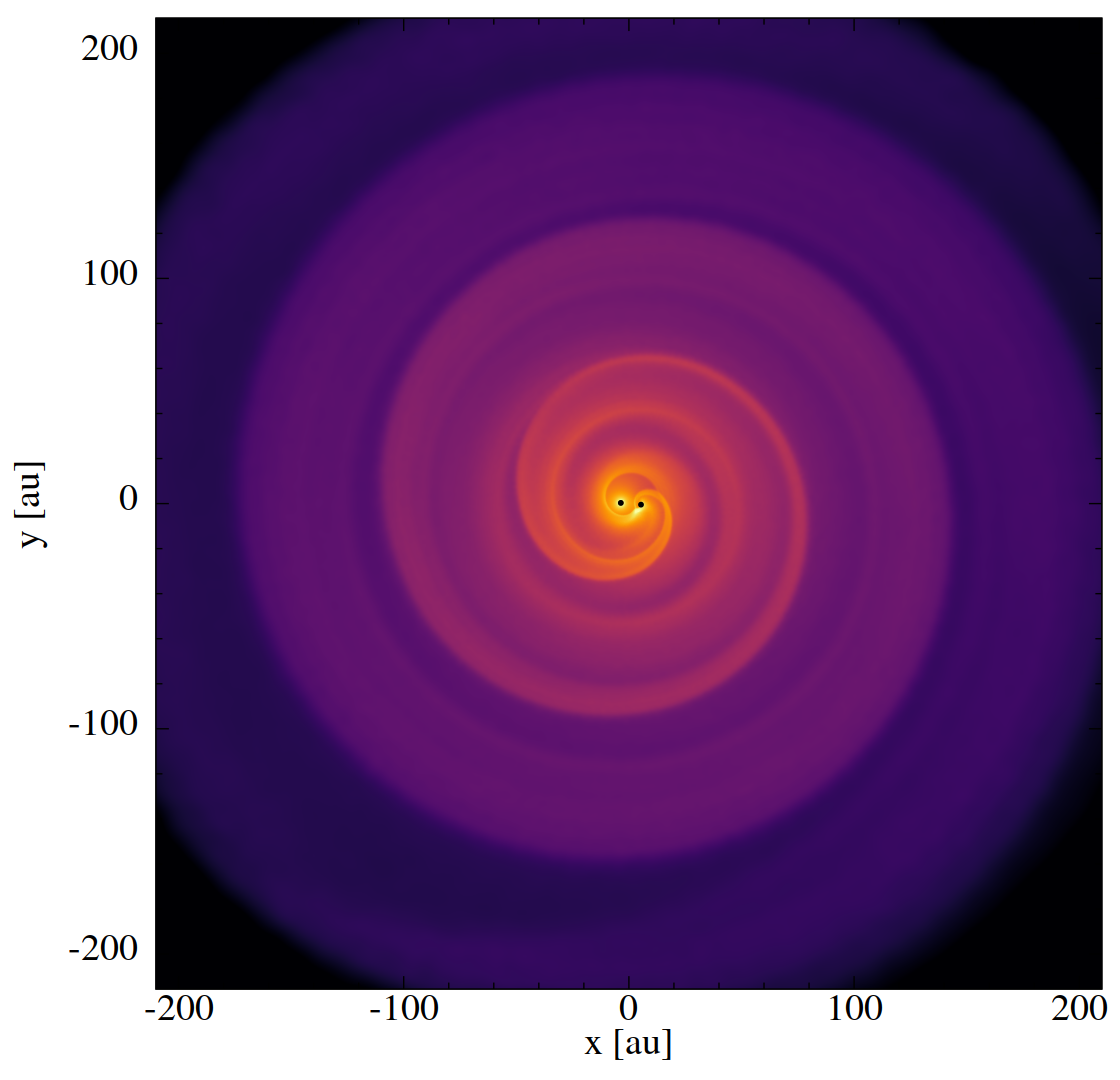}
	\includegraphics[width=0.355\textwidth]{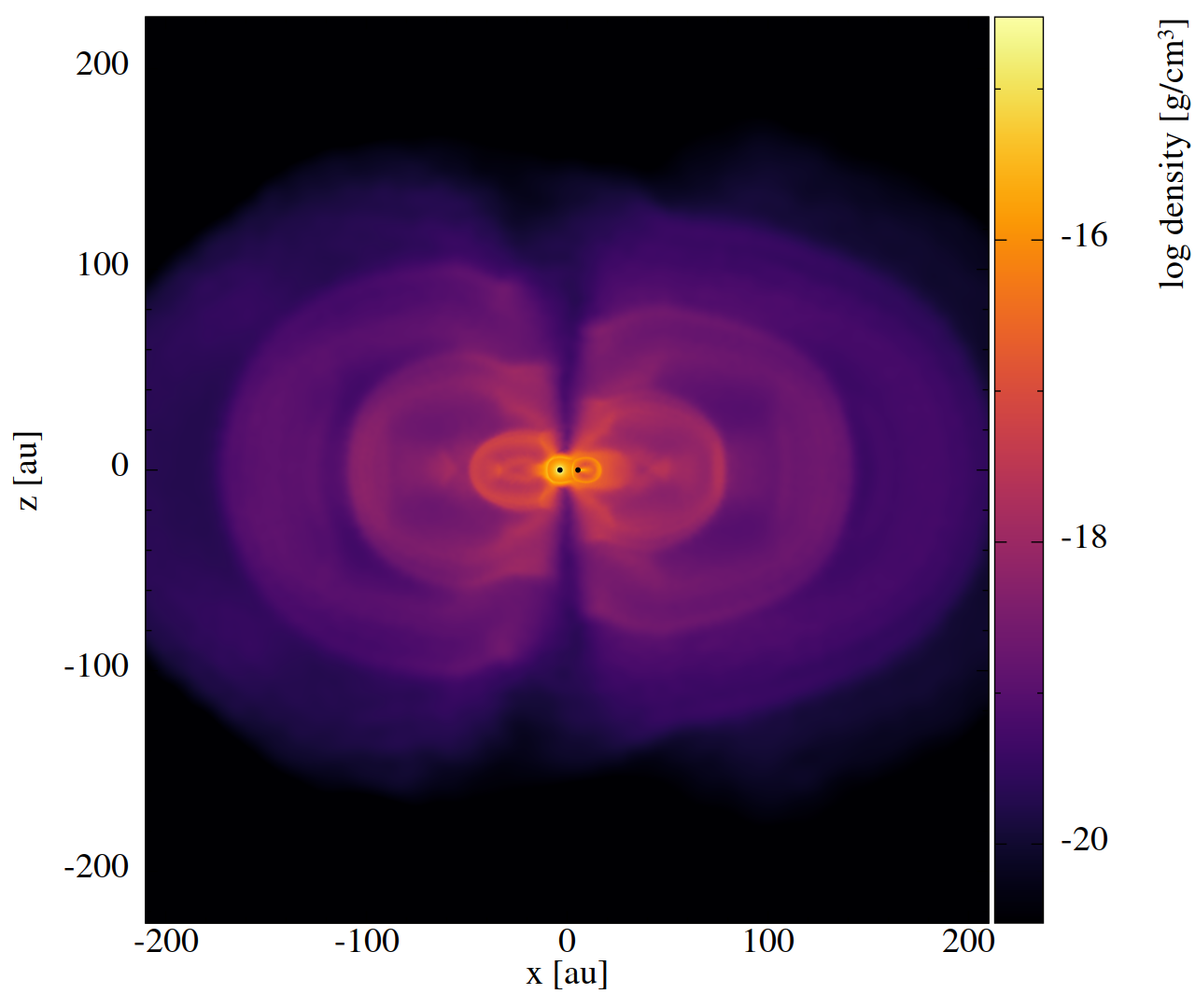}
	\caption{Density distribution of model \Cns\ (\textit{top row}) and model \Ons\ (\textit{bottom row}) of the orbital plane {(\textit{left panels} zoom-in of \textit{middle panels})} and meridional plane (\textit{right panels}).  Left and right black dots represent the AGB star and companion, respectively, not to scale. {Snapshot taken after 5 orbits, at which self-similarity is reached.}}
	\label{fig:90s_density}
\end{figure*}
\\ \indent The gravitational interaction between the AGB wind and the companion results in a spiral feature in the orbital plane, as can be seen in the upper left panel of Fig.\ \ref{fig:90s_density}. The two spiral edges or fronts delimit the gravity wake produced by the orbiting companion and propagate radially outwards at different speeds. Such a gravity wake was first described by Bondi, Hoyle and Lyttleton (BHL; \citealp{Hoyle1939,BondiHoyle1944}) for a point mass moving at constant speed on a straight path through a gas cloud free from self-gravity, and forms the basis for the mechanism here. The outer spiral front originates from part of the material that is accelerated as a result of a gravitational sling around the companion, also called slingshot. In Sect.\ \ref{sect:term_vel} we elaborate on the slingshot and establish that it is a valid model. When this material meets with the wind, which is relatively slower, a shock is generated. Therefore, we will henceforth call this outer front the `front shock'. The inner edge of the spiral wake, which is the one wrapped closest around the AGB star, is slower. This is due to part of the wind material that is decelerated by the gravitational sling and thus propagates more slowly than the unobstructed wind coming from the AGB star in the centre-of-mass restframe. Hence, a bow shock is formed when the faster wind collides with this slow material, which we name the `back edge'. Since the velocity of the unobstructed wind is affected by the motion of the AGB star (reflex motion), induced by the presence of the companion, the AGB star's orbital velocity influences the location of the back edge. As a result of the velocity dispersion caused by the slingshot, the gravity wake will widen. Consequently, for the reference model we see that after a few orbital revolutions the front shock catches up with the back edge of the previous orbit, at a radius of about 120\,\AU\ in the upper middle panel of Fig.\ \ref{fig:90s_density}. Both fronts will merge and combine into a global spiral structure, which differs from the spiral wake structure closer to the binary system.
\\ \indent In the meridional plane (plane perpendicular to the orbital plane through both sink particles) of model \Cns\ the morphology shows arcs that bend towards the polar axis, extending over all latitudes (Fig.\ \ref{fig:90s_density}, upper right panel). This feature is already carefully discussed by \cite{Mastrodemos1999}: the arcs in the vertical direction are the cross sections of the surface of the spiral in the orbital plane. The edges of the different arcs on the side closest to the binary system, correspond to the back edge of the spiral in the orbital plane and the outer edge of the arcs to the front shock. The thickening in the arcs is caused by the funnelling of the wind by the companion's gravitational potential (gravity wake), and is suppressed within a limited height from the orbital plane, as discussed by \cite{KimTaam2012b}. This funnelling will become more effective when the  gravitational interaction becomes stronger and hence the vertical extent of the gravity wake will be more compressed towards the orbital plane. Depending on the initial conditions of the binary system and AGB wind, this may cause an equatorial density enhancement (EDE, Sect.\ \ref{sect:EDE}), where a large fraction of the wind material is confined close to orbital plane. The narrow ends of the arcs near the poles are the unperturbed signatures of the spiral that is caused by the reflex motion of the AGB star, which is not visible in the orbital plane. \cite{KimTaam2012b} investigated the sole effect of this reflex motion of the AGB star with respect to its orbital and wind velocity. They found that the vertical arc structure becomes more oblate or `flattened' when the orbital-to-wind velocity ratio of the AGB star increases, due to the action of the centrifugal force. Since it is not clear if this flattening and the presence of an EDE are connected, we investigate this in Sect.\ \ref{sect:EDE}.

\subsection{Effect of the wind velocity on the morphology}\label{sect:windvel_morphology}
We discuss model \Ons, which is the slow-wind counterpart of the reference model, in order to map the morphological changes w.r.t.\ the wind velocity.
\\ \indent The density structure in the orbital plane close to the binary systems shows a double spiral structure (Fig.\ \ref{fig:90s_density}, bottom left panel). The first spiral is wrapped closely around the AGB star and the second emerges from a bow shock in front of the companion, dominating the global spiral morphology. This inner structure is described as a `vortex structure' by \cite{Liu2017} and can no longer be described as a simple BHL gravity wake. Accordingly, it differs significantly from the single spiral structure found in the fast-wind reference model, also on a global scale (Fig.\ \ref{fig:90s_density}, lower middle panel). This difference is due to the longer time that is associated with the gravitational interaction between the AGB wind and the companion when the wind propagates more slowly. Therefore, the main contribution to the morphology shifts from the reflex motion of the AGB star to the gravitational pull of the companion, based on the comparison of the kinetic energy density in the wind with the gravitational energy density of the companion (more in Sect.\ \ref{sect:morphological_param}). 
\\ \indent In the meridional plane, arcs are present in the morphology, displayed in the bottom right panel of Fig.\ \ref{fig:90s_density}. The wind material is confined within a range of $\sim$$150\,\AU$ in the $z$-direction, while in the $x$- and $y$-direction the wind expands up to $\sim$$250\,\AU$. Hence, we see that the outflow is {focussed towards} the orbital plane, which is in contrast with the morphology of the reference model. This is due to the lower wind velocity, which results in the stronger gravitational interaction between the wind and the companion. In this case, we suspect that a {global flattening of the CSE} as well as an EDE is present in the model, on which we will elaborate in Sect.\ \ref{sect:EDE}. 
\\ \indent Hence, we find that the wind velocity is an important parameter for determining the morphology in the outflow. This result was already stated by \cite{Mastrodemos1998}. More specifically, by lowering the wind velocity we find a transition in morphology from a single BHL spiral wake to a double spiral structure, which is in agreement with previous studies (e.g.\ \citealp{Saladino2018}).

\subsection{Effect of the binary separation on the morphology}
\begin{figure*}
	\centering
	\includegraphics[width=0.31\textwidth]{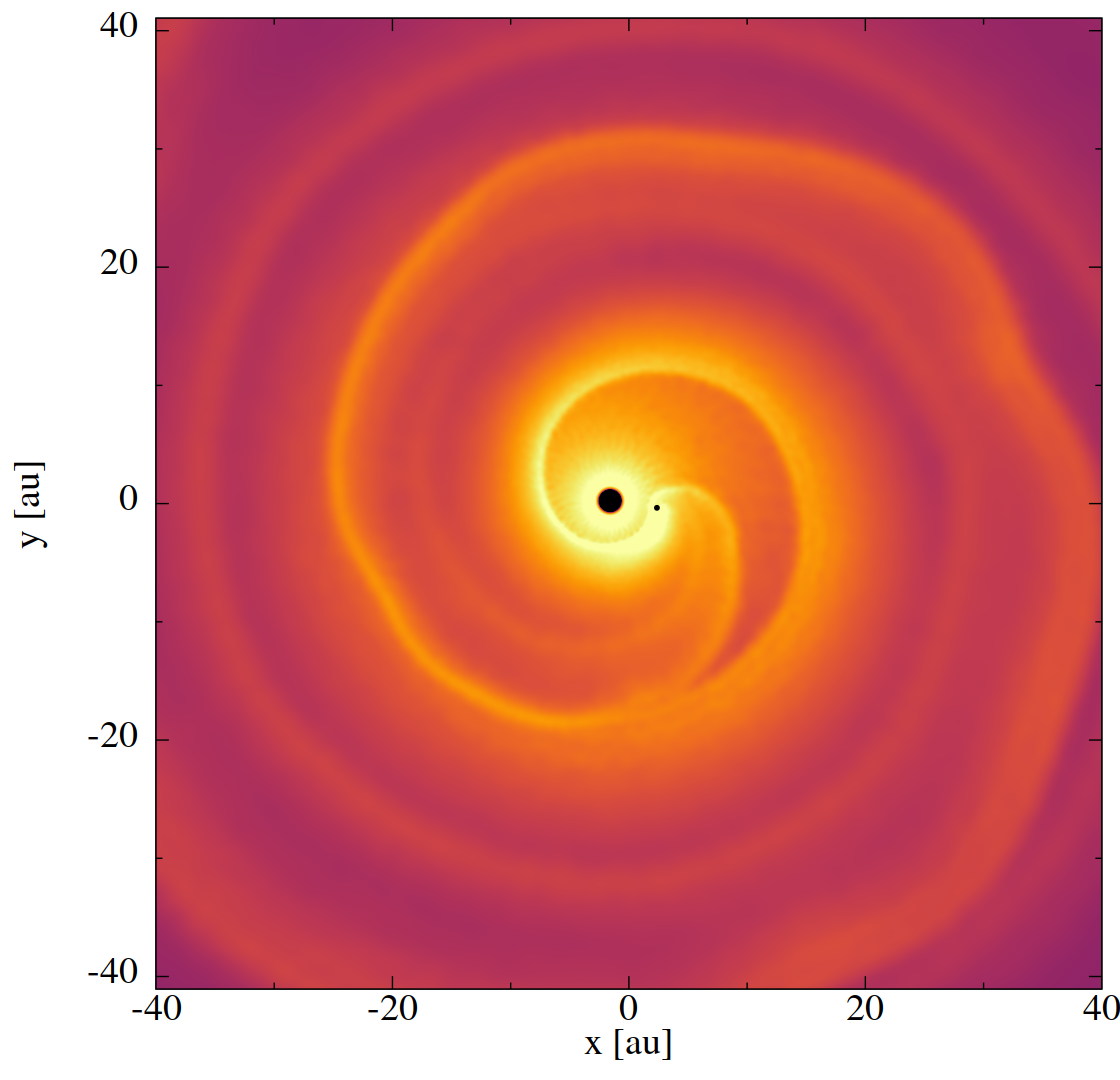}
	\includegraphics[width=0.31\textwidth]{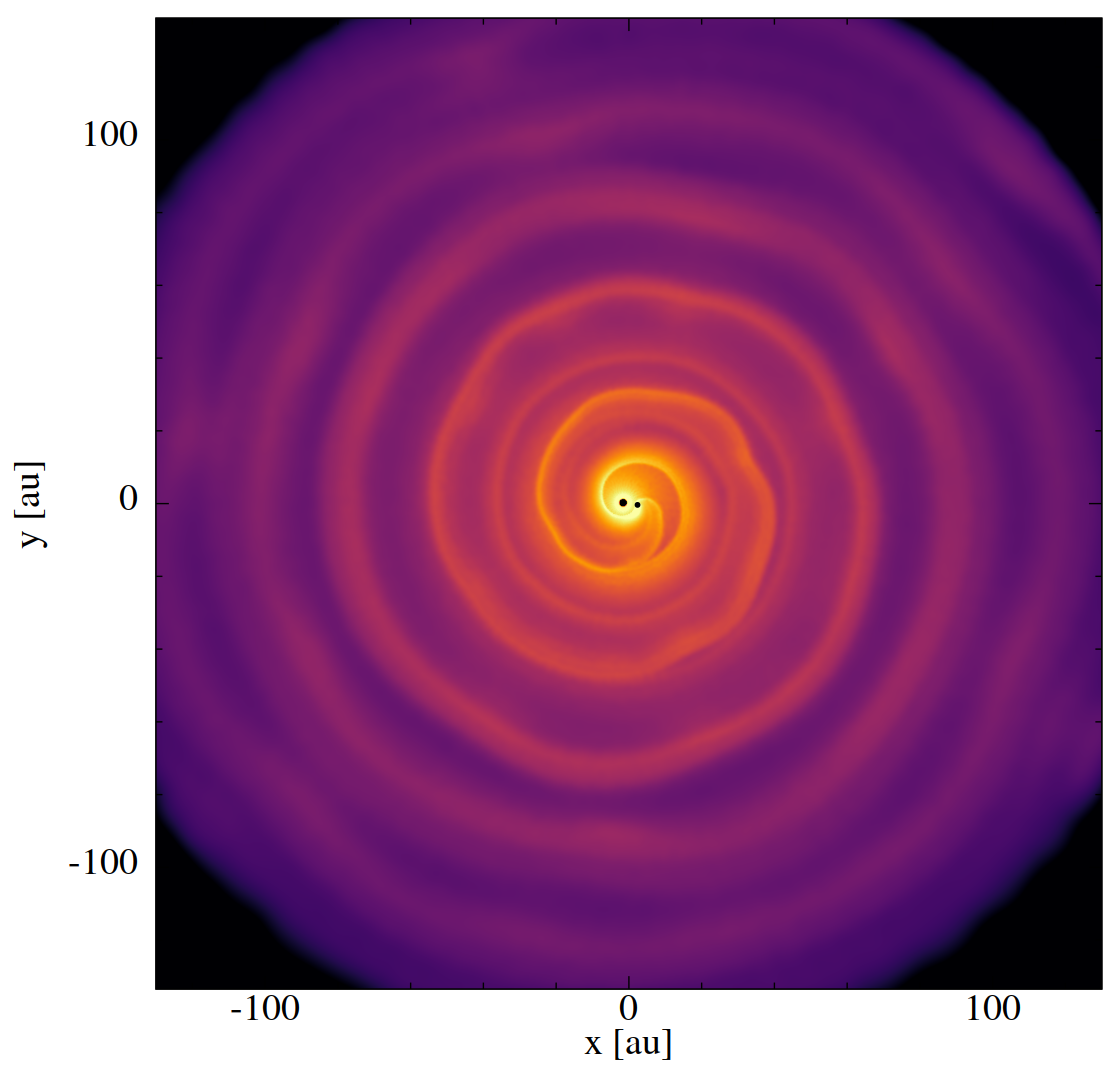}
	\includegraphics[width=0.355\textwidth]{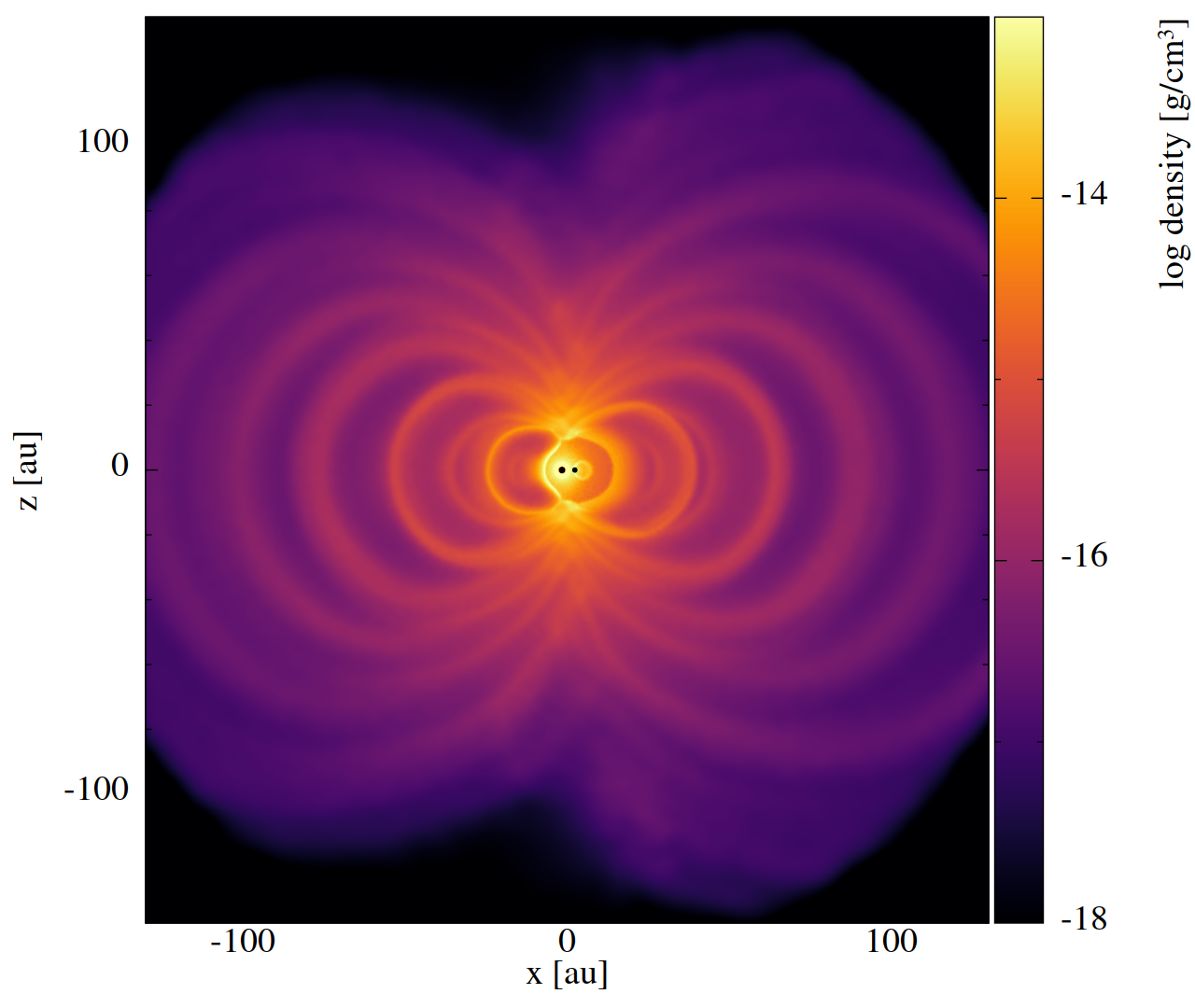}
	\includegraphics[width=0.31\textwidth]{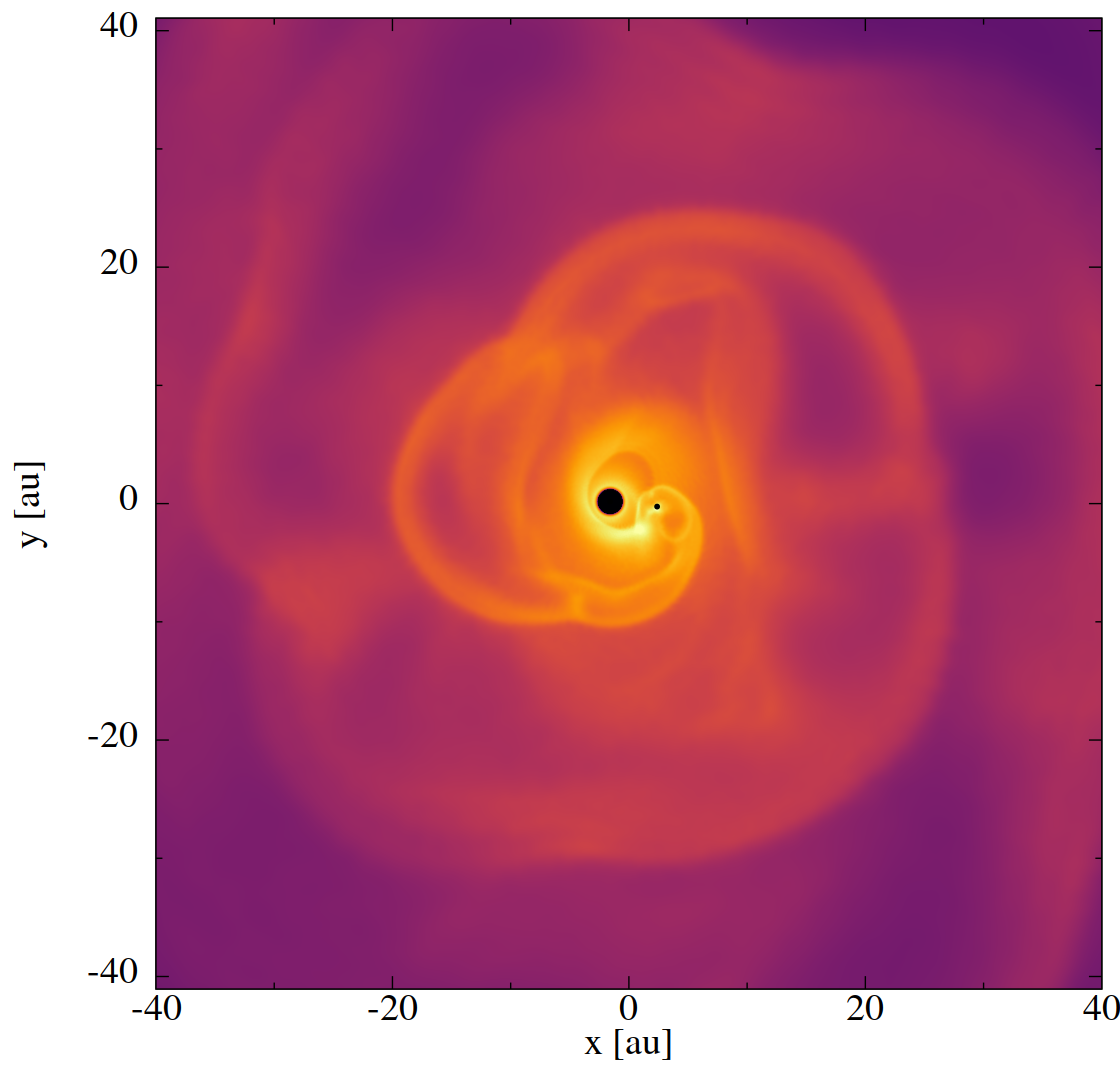}
	\includegraphics[width=0.31\textwidth]{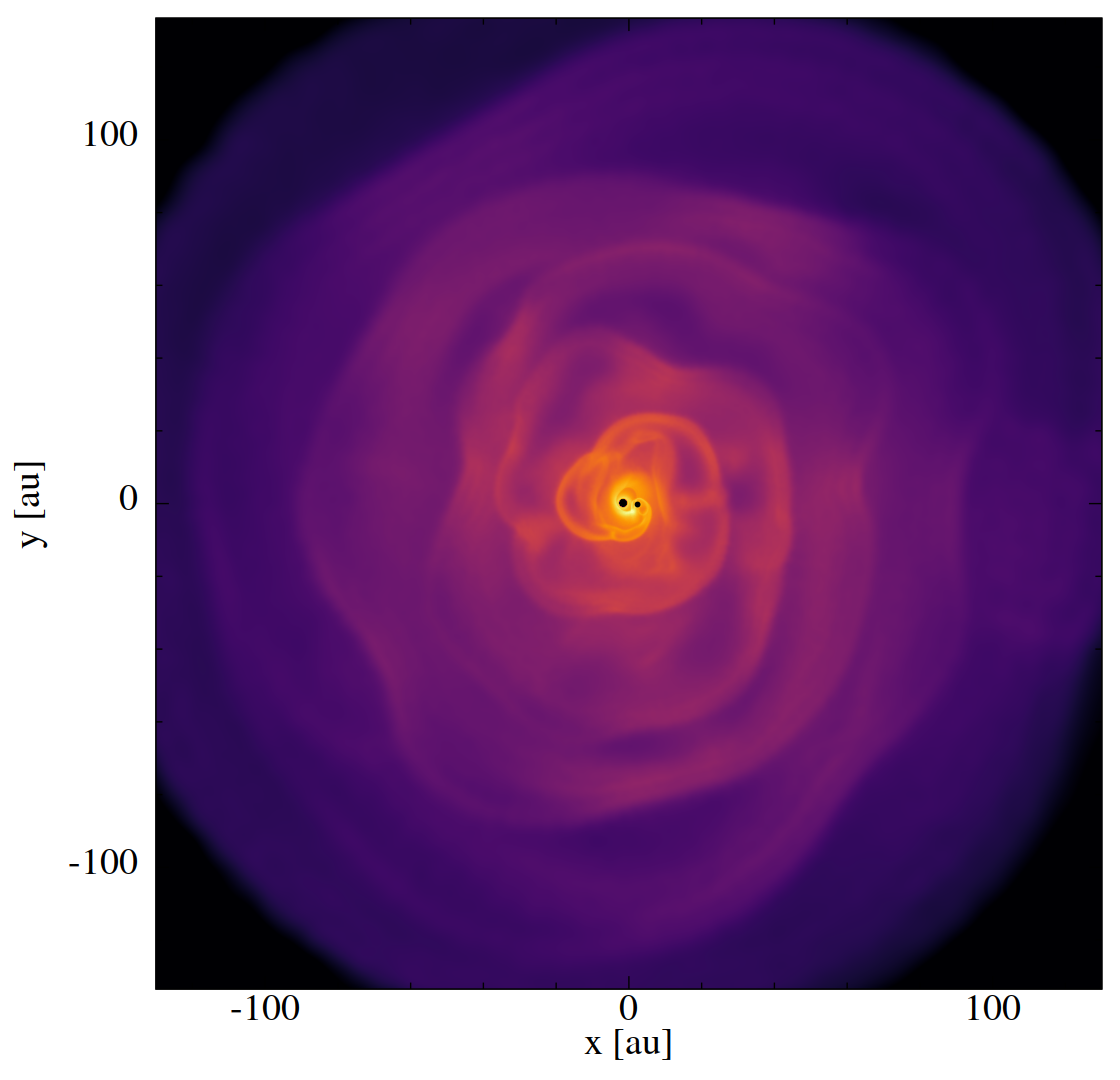}
	\includegraphics[width=0.355\textwidth]{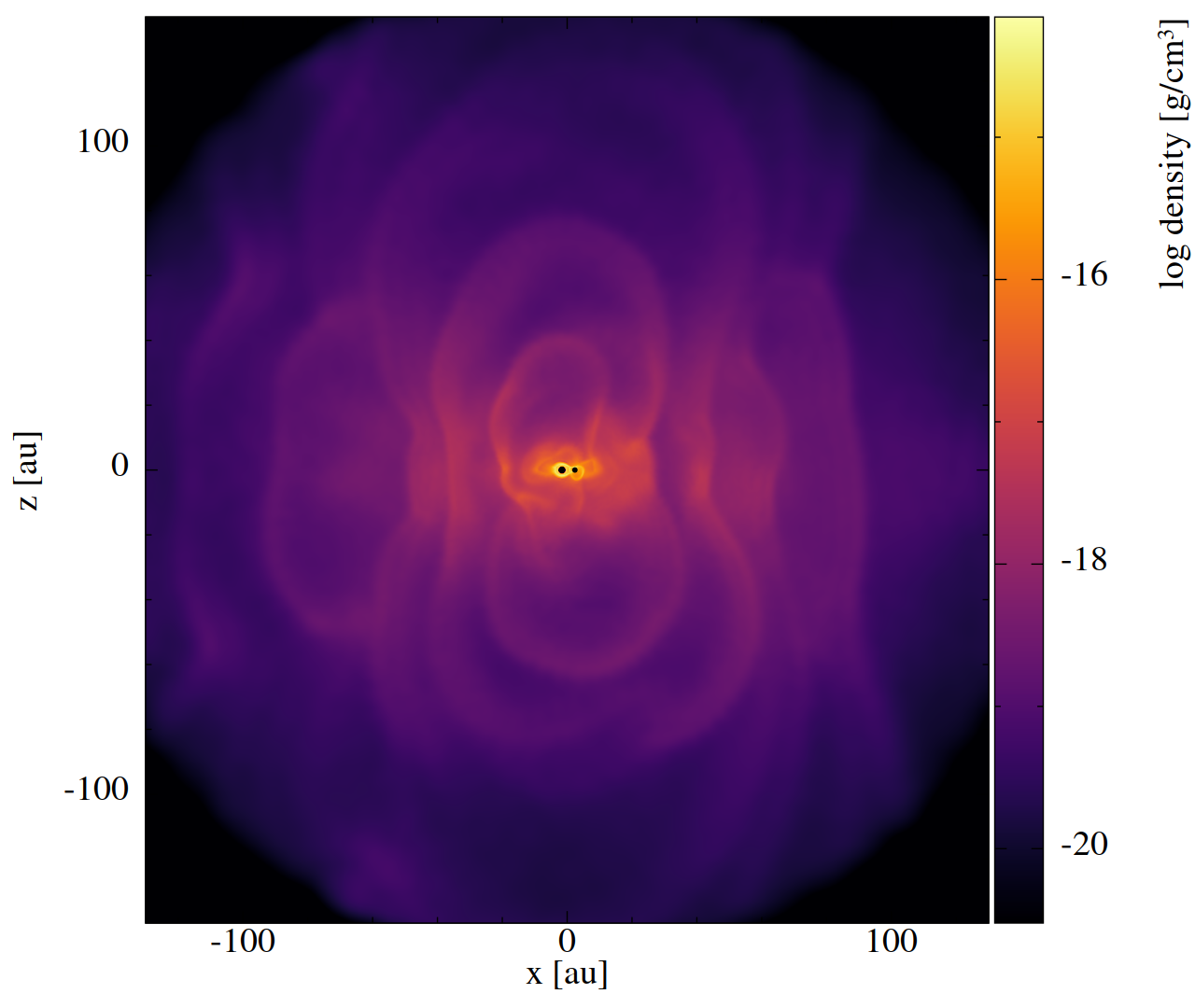}
	\includegraphics[width=0.31\textwidth]{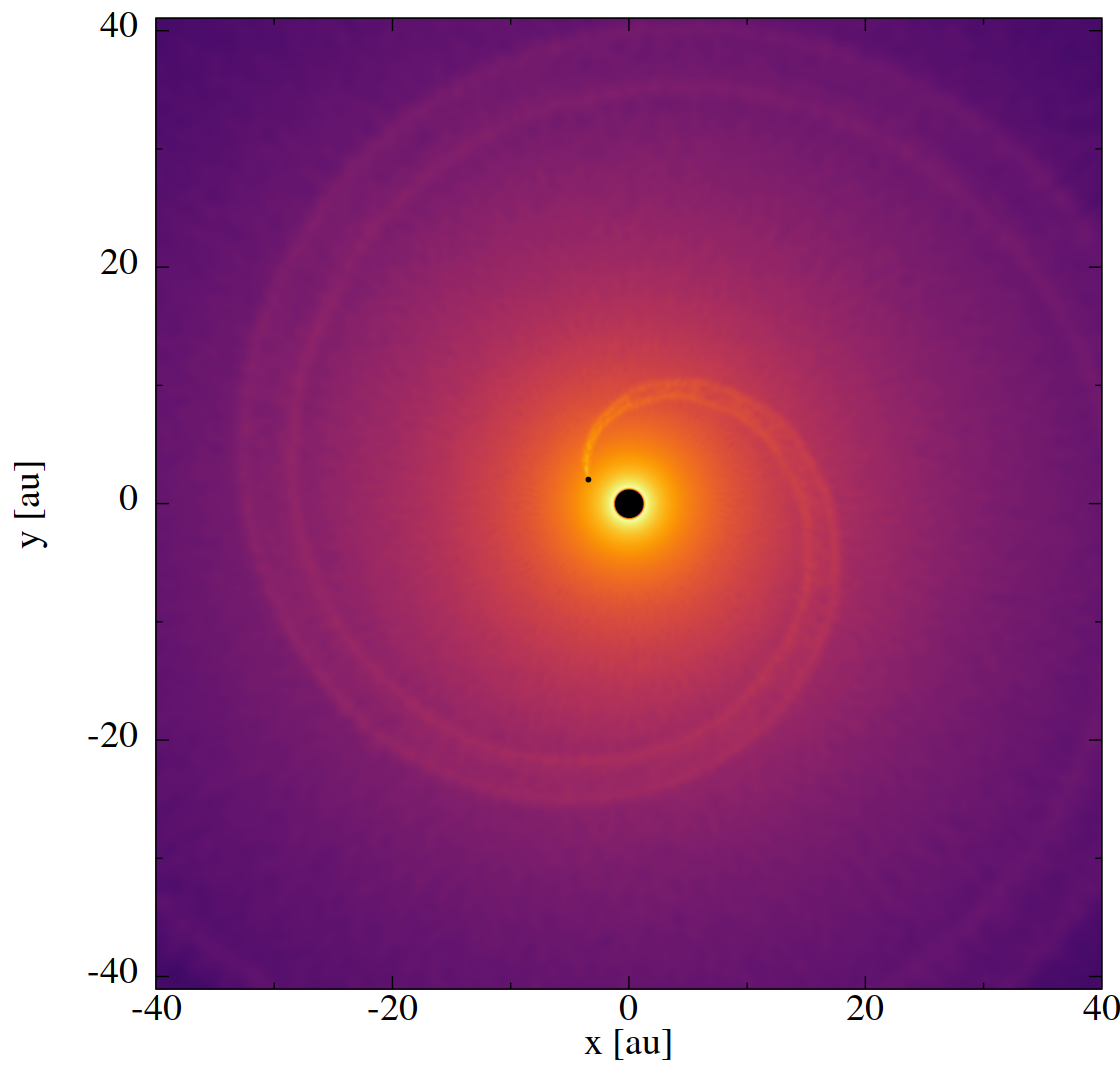}
	\includegraphics[width=0.31\textwidth]{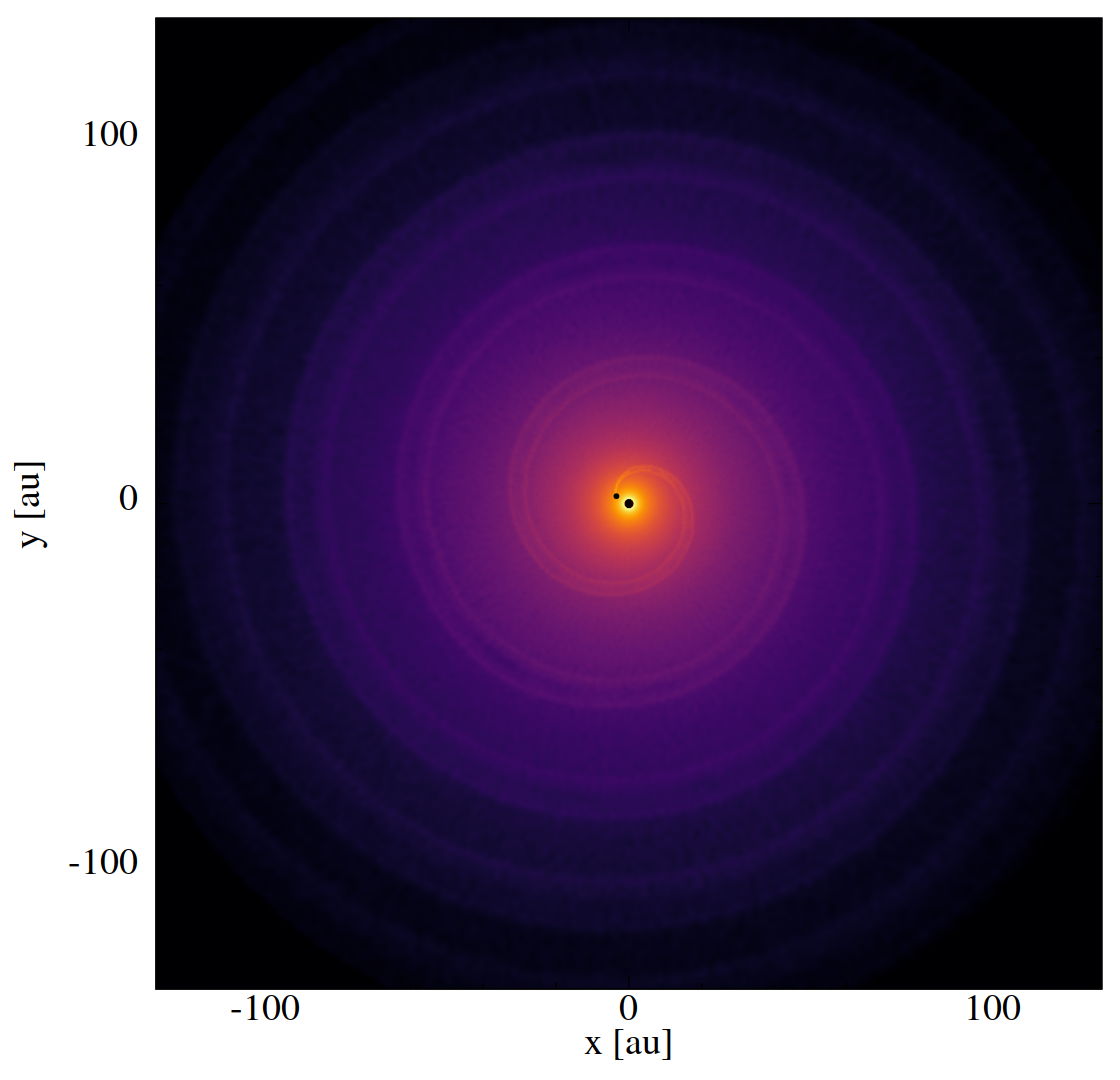}
	\includegraphics[width=0.355\textwidth]{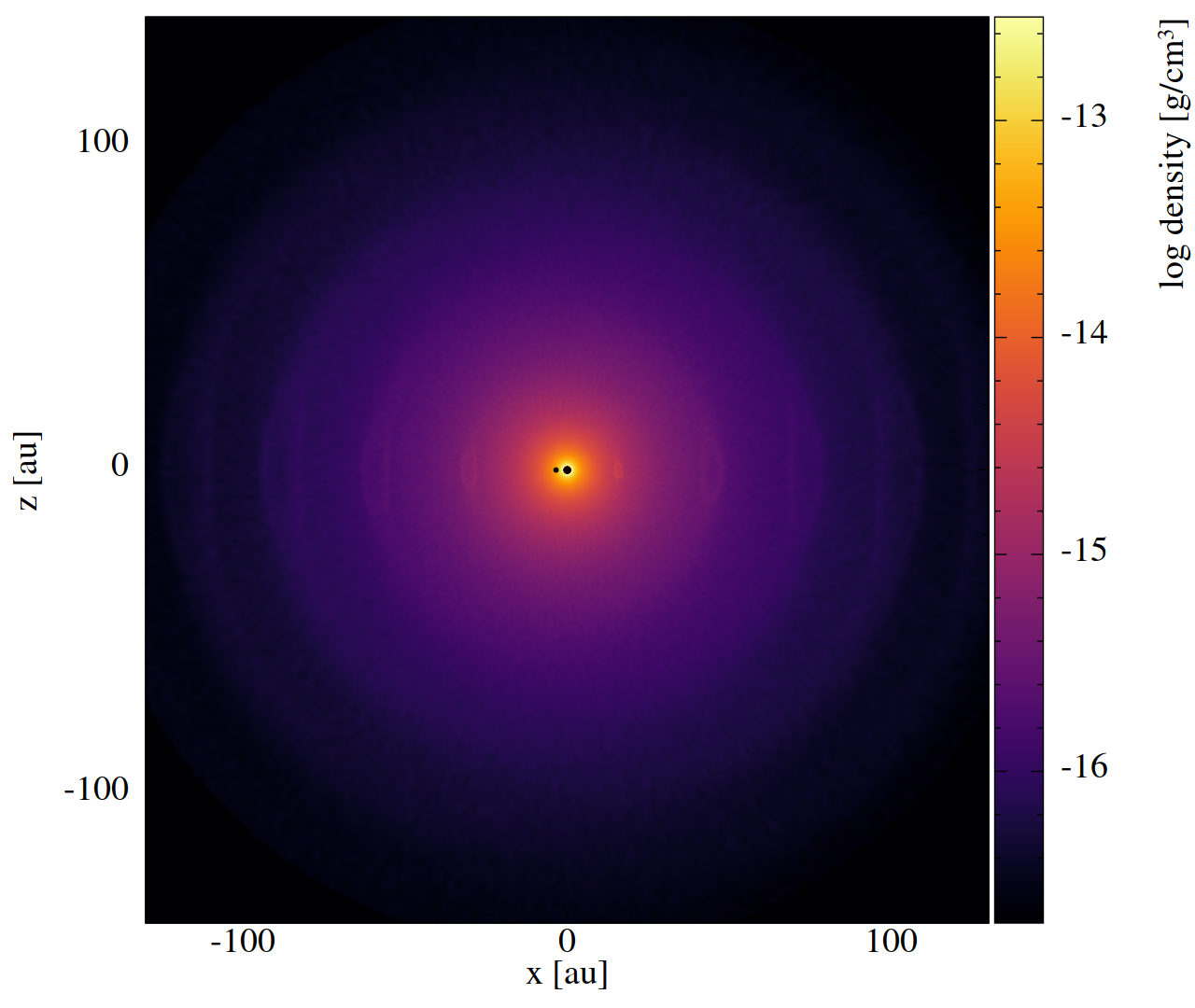}
	\caption{Density distribution of model \Cfs\ (\textit{top row}), model \Ofs\ (\textit{middle row}) and model \Cfp\ (\textit{bottom row}) of the orbital plane {(\textit{left panels} zoom-in of \textit{middle panels})} and meridional plane (\textit{right panels}).  Left and right black dots represent the AGB star and companion, respectively, not to scale. {Snapshot taken after 7 orbits, at which self-similarity is reached.}}
	\label{fig:4au_density}
\end{figure*}
Here, we discuss the models where the companion is located closer to the AGB star, namely at 4\,\AU. As a result, the orbital movement of both components becomes faster, the gravitational interaction stronger and {a larger portion of the wind interacts with the companion, since the density is higher closer to the AGB star.} We find that the effect on the morphology is threefold.
\\ \indent First of all, the morphology in the orbital plane of model \Cfs\ displays again a global spiral structure, resulting from a double spiral close to the binary system (upper left panel of Fig.\ \ref{fig:4au_density}). This is different from the reference model \Cns, but agrees with the vortex structure found in the slow-wind model \Ons. Therefore, we find that decreasing the binary separation has a comparable effect on the wind morphology as lowering the wind velocity, as was established by \cite{Mastrodemos1999}. Secondly, the density in the meridional plane (upper right panel of Fig.\ \ref{fig:4au_density}) reveals arcs which are globally flattened: the horizontal extent is significantly larger than its vertical counterpart. When the binary separation is further decreased to 2.5$\,\AU$, this flattening becomes stronger, see upper right panel of Fig.\ \ref{fig:gallery1} in Appendix \ref{sect:gallery}. This is mainly the effect of the larger orbital velocity of the AGB star relative to the wind velocity, which results in a larger centrifugal force on the wind, as investigated by \cite{KimTaam2012b}. Lastly, a striking transition from a smooth and regular spiral morphology to a perturbed spiral structure appears in the orbital plane (Fig.\ \ref{fig:4au_density}, upper middle panel). When the interaction between the companion and the wind is strong, as is the case here, a bow shock appears in front of the companion, which arches as an umbrella stagnation flow, as is discussed in detail by \citeauthor{Malfait2021} (\textit{submitted}). This umbrella stagnation flow feeds a high-density region that trails behind the companion. \citeauthor{Malfait2021} (\textit{submitted}) found that for certain AGB-wind binary set-ups, the umbrella stagnation flow is unstable and periodically brings wind particles to this high density region behind the companion, creating the irregularity in the outflow. However, the exact cause of the instability remains yet unknown.
\\ \indent For the slow-wind counterpart, we expect a large flattening and potentially a strong EDE to be present in the meridional density distribution, in analogy to the findings of Sect.\ \ref{sect:windvel_morphology} and taking into account that the gravitational interaction will be enhanced by decreasing the distance to the companion. However, for model \Ofs\ (Fig.\ \ref{fig:4au_density}, middle right panel) the fundamental arc structure can no longer be recognised, as loops of higher density material seem to extend to high latitudes. 
In the orbital plane (middle left panels of Fig.\ \ref{fig:4au_density}) a heavily disturbed spiral structure is present, originating from the double spiral structure in the centre. We attribute this broken and irregular structure in both planes (orbital and meridional) to the same phenomenon described before (\citeauthor{Malfait2021} \textit{submitted}), which is enhanced in this case by the {longer interaction between the companion and the wind due to the lower wind velocity}, in comparison with model \Cfs. A similar morphology is found by \cite{ElMellah2020}, labelled as a `concentric petals pattern', for such a model with a low wind velocity.  

\subsection{Effect of the companion mass on the morphology}
We investigate the effect of a 10\,$\Mjup$\ companion on the wind morphology in a fast wind, displayed in the bottom panels of Fig.\ \ref{fig:4au_density}. The amount of wind material that will experience the gravitational attraction of the companion will be limited in this case, hence the global interaction with the AGB wind is expected to be much weaker.
\\ \indent In the orbital plane a narrow, two-edged single spiral structure emerges, which can be described by the simple BHL gravity wake. This is similar to the reference model, albeit less distinct and with a smaller relative density contrast between the spiral arms and the region in between the arms in the case of a planetary companion. This is because only a limited amount of wind particles experiences the gravitational sling around the planetary companion. The shock, created by the small fraction of the total mass of the outflow, will quickly lose its energy and thus the perturbation in the wind is limited. In the meridional plane, the fundamental arc structure is again retrieved. The vertical extent is limited in this case, {since the induced orbital motion of the AGB star by the low-mass companion is negligible.} This is in agreement with the results of \cite{KimTaam2012b}. In the study by \cite{Mastrodemos1999}, they established that decreasing the companion mass has a similar effect on the shape of the morphology of the outflow as increasing the wind velocity or also as increasing the binary separation, which we confirm. 
\\ \indent These models (see also Figs.\ \ref{fig:gallery1} and \ref{fig:p_slow} in Appendix \ref{sect:gallery}) reveal that even massive planets are able to alter the morphology of an AGB outflow. If the structure formation is detectable, observing AGB winds may lead to an indirect way of finding exoplanets.

\section{Discussion}\label{sect:discussion}
In this section we elaborate on three different aspects of the models: (i) the terminal velocity reached in the models, (ii) the vertical extent and distribution of the outflow and (iii) different proposed parameters that may be able to indicate the type of morphology present in the outflow. To improve the quality of the discussion, we employed two models from \cite{Malfait2021} , namely v20e00 and v05e00, which have the same numerical set-up, but the stellar companion is located at 6\,\AU\ for both the fast- and slow-wind model. We renamed them \Css\ and \Oss, respectively, in accordance with the naming of our models.

\subsection{Terminal velocity}\label{sect:term_vel}
Up until today the effect of the gravitational interaction between a companion and the AGB wind on the terminal velocity $\vterm$ is not studied. However, we expect an altered terminal velocity compared to a single AGB outflow: due to the interaction between the companion and the outflow, wind particles are accelerated and decelerated by the gravitational potential of the companion and by the so-called gravitational sling(shot) or gravity assist when the wind passes close to the companion. Studying the effect on the terminal velocity is crucial since it is an important parameter concerning stellar wind observations.
\subsubsection{Toy model}
In order to verify the terminal velocity found in the simulations, we constructed a toy model of the gravitational interaction of the companion on the AGB wind. {We here neglected the gravitational pull of the AGB star, since in the simulations it is assumed that this is balanced out by the wind acceleration in the `free wind' case adopted, see Eq.\ \ref{eq:wind_acc}. In doing so, we implicitly took into account the wind acceleration mechanism of the simulations in the toy model.} We modelled the evolution of the velocity of a particle, given an initial velocity, as it passes close ($d=0.01$\,\AU) to the companion. Following a straight path $\ell$, illustrated in Fig. \ref{fig:geometry_tm}, the gravitational acceleration due to the companion's potential is given by
\begin{eqnarray}
	g_{\ell}(r) &=& ||\textit{\textbf{g}}(r)||\cos\delta \nonumber \\
	&=& \frac{G\Mcomp}{r^2}\cos\delta = \frac{G\Mcomp}{a^2+\ell^2-2a\ell\cos\theta}\cos\delta, \label{eq:grav_pot}
\end{eqnarray}
where $r$ is the distance from the wind particle to the companion, $a$ the binary separation, $\theta$ the angle between the path $\ell$ and the binary axis, given by $\sin\theta = d/a$ and $\cos\delta$ the projection of $\textit{\textbf{g}}(r)$ on $\ell$; see Fig.\ \ref{fig:geometry_tm} for a visual representation of the geometry used in the toy model. We note that in reality a particle will not stay on this straight path $\ell$ after the interaction, but for simplicity this is not taken into account in the toy model, since it will only have a minor impact on the velocity evolution. To predict the final velocity $v_n$ of this toy model, namely the velocity at some chosen distance from the AGB star that is taken as a proxy for the terminal velocity, we used an iterative approach. Based on simple mechanics, the evolution of the wind velocity in function of time is given by
\begin{eqnarray}
	v_j = v_{j-1} + g_{\ell,j-1}t_j, \quad {\rm with}\quad t_j &=& \frac{|\ell_j-\ell_{j-1}|}{v_{j-1}}, \label{eq:vel_evo}\\
	\quad {\rm for} \quad j &=& 1,2,3,...,n. \nonumber
\end{eqnarray}
\begin{figure}
	\centering
	\includegraphics[width=0.49\textwidth]{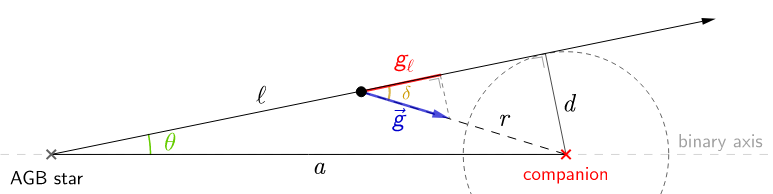}
	\caption{Schematic representation of the geometry of the toy model of a wind particle (black dot) at timestep $t_j$.}
	\label{fig:geometry_tm}
\end{figure}
Thus, the terminal velocity of the toy model $\vtermtm = v_n$. In Eq.\ (\ref{eq:vel_evo}) $t_j$ is the timestep and $v_0$ equals the terminal velocity of the simulation of a single AGB outflow. The number of steps $n$ in the toy models depends on the outer boundary of the corresponding hydro simulation, which varies between 30 and 230 \AU, in order that the stepwidth is $10^{-5}$\,\AU\ for all toy models.
\\ \indent In our toy model, we included a gravitational slingshot or also called gravity assist. In general, when a low-mass object $m$ (e.g.\ satellite) passes closely by a massive body $M$ (e.g.\ a planet), {its velocity and path are altered due to the movement of the massive body, as a result of conservation of momentum and energy. Thus, using these conservation laws in 1D,} the final velocity $v_{\rm f}$ of the object is calculated as
\begin{equation}
v_{\rm f}  = v_{\rm in} +2u,
\end{equation}
where $v_{\rm in}$ is the initial velocity of object $m$ and $u$ is the velocity of $M$, assuming $m \ll M$. This can be used to calculate the final velocity in 2D using a vector sum, thus taking into account that the motion of the massive body is not along the same axis as the low-mass object:
\begin{eqnarray}
\textit{\textbf{v}}_{\rm f} &=& \textit{\textbf{v}}_{\rm in} + 2\textit{\textbf{u}} \nonumber\\
&=& (-v_{\rm in}\cos\theta+2u\cos\zeta)\,\hat{\textit{\textbf{x}}}+(v_{\rm in}\sin\theta+2u\sin\zeta)\,\hat{\textit{\textbf{y}}} \label{eq:slingshot_2D}
\end{eqnarray}
where $\theta$ is the angle of $\textit{\textbf{v}}_{\rm in}$ with respect to the $x$-axis and $\zeta$ the angle of $\textit{\textbf{u}}$ with respect to the $x$-axis; $\hat{\textit{\textbf{x}}}$ and $\hat{\textit{\textbf{y}}}$ are the 2D Cartesian unit vectors. Hence, the magnitude of the final velocity becomes 
\begin{equation}\label{eq:slingshot_vel}
v_{\rm f} = (v_{\rm in} +2u) \sqrt{1-\frac{4v_{\rm in}u\,\big(1-\cos(\theta-\zeta)\big)}{(v_{\rm in} +2u)^2}}.
\end{equation}
In our case, a wind particle along the path $\ell$ will experience a gravitational slingshot due to its passage close to the companion. Therefore, in Eq.\ (\ref{eq:slingshot_vel}) $\theta$ is the same as defined in Eq.\ (\ref{eq:grav_pot}) and $u$ corresponds to the orbital velocity of the companion $\vcomp$ so that $\zeta = 90^\circ$. 
\\ \indent From this toy model, we derived two values for the terminal velocity $\vtermtm$: (i) we calculated the final velocity without taking into account the slingshot and (ii) we included the slingshot instantaneously at the moment of closest approach. Thus for the latter, $v_{\rm f}$ from Eq.\ (\ref{eq:slingshot_vel}) was added to $v_j$ from Eq.\ (\ref{eq:vel_evo}) at timestep $j$ when the distance to the companion $r_j$ is minimal, thus when $r_j = d$. The results will be discussed in the following section.

\subsubsection{Comparison of the simulations to the toy model}
From the simulations, we calculated the terminal velocity in the orbital plane, in order to accurately compare the results to the 2D toy model. The terminal velocity was obtained by averaging the velocity over the last $20\%$ of the modelled region per radius {(Fig.\ \ref{fig:v_C90s})}. Since structure is present in the outflow, a range of velocities are found at a certain radius, resulting in {a range in the terminal velocity, which is represented by a minimum, a mean and a maximum value.} This methodology is clarified in Fig.\ \ref{fig:v_C90s}, giving the radial velocity profile of the reference model \Cns\ in the upper panel, where the different obtained terminal velocities are indicated.
\begin{figure}
	\centering
	\includegraphics[width=0.45\textwidth]{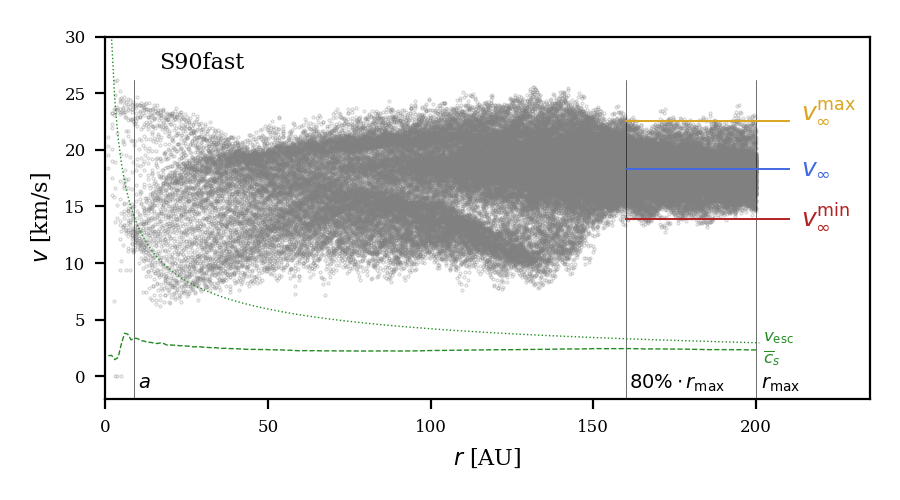}
	\\[-2.3ex]
	\includegraphics[width=0.45\textwidth]{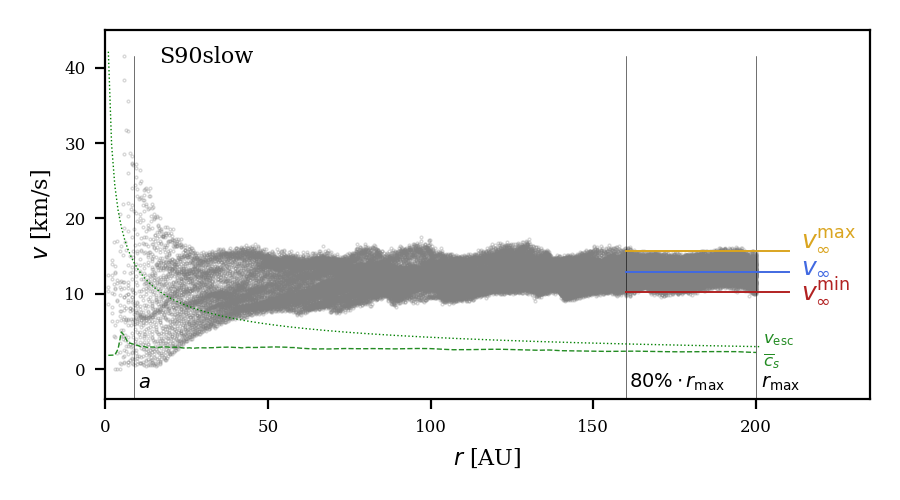}
	\\[-2.3ex]
	\caption{Radial velocity in the orbital plane of model \Cns\ (\textit{upper panel}) and \Ons\ (\textit{bottom panel}), illustrating the methodology for the determination of the terminal velocity $\vterm$. {For information the mean sound speed $\bar{c}_s$ and the escape velocity $v_{\rm esc}$ are given in green in dashed and dotted lines, respectively}.}
	\label{fig:v_C90s}
\end{figure}
\\ \indent Fig. \ref{fig:v_term} shows the results for the terminal velocity for the different simulations in black empty circles, {where its range is illustrated by the errorbars}. The results of the toy model with and without the slingshot are given by the purple triangles and green squares, respectively. 
\\ \indent For the planetary models, the terminal velocity of the simulation coincides almost exactly with the toy model without the slingshot and with the terminal velocity of the single models. This demonstrates that the amount of particles that experience a slingshot in the simulation is negligible, due to the limited interaction region around a planetary companion. Hence, the acceleration due to the slingshot is not able to alter the global velocity field in the outflow. The interaction region is approximately given by the capture radius $\Rcapt$ 
and is defined as the largest distance to a central object (in this case the companion) at which particles with speed $\vwind$ can still be affected by the gravitational potential \citep{Illarionov1975}, based on the BHL-accretion principle. It is derived from equating the kinetic with the gravitational energy of such a particle, thus $\Rcapt$ is given by
\begin{equation}\label{eq:capt_radius}
	 \Rcapt=\frac{2G\Mcomp}{v_{\rm{wind}}^2}.
\end{equation}
\begin{figure*}
	\centering
	\includegraphics[width=0.91\textwidth]{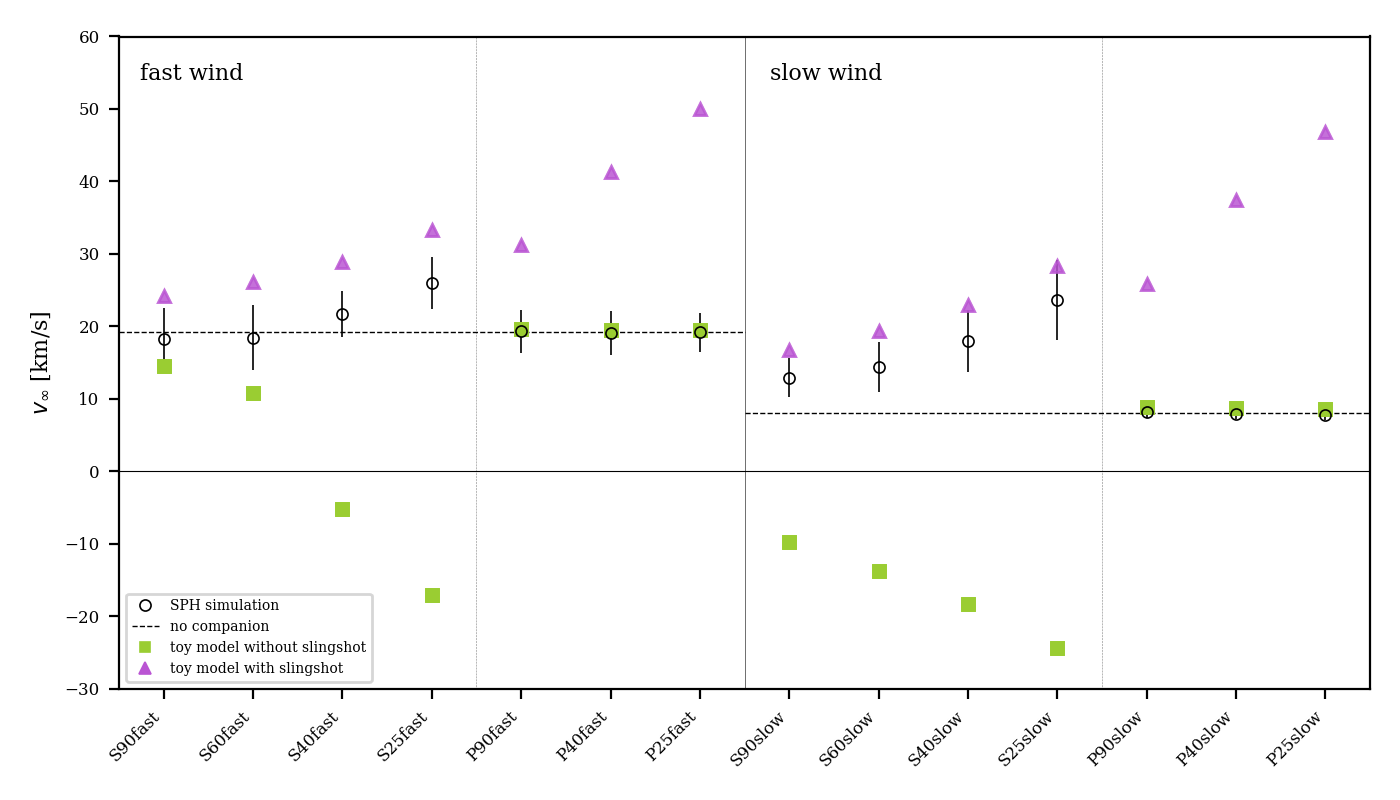}
	\caption{Terminal velocity for the fast models (\textit{left}) and the slow models (\textit{right}) as calculated from the toy model with and without a slingshot in purple triangles and green squares, respectively, and from the simulation in empty circles. The terminal velocity of the corresponding single star model is given by the black dashed lines.}
	\label{fig:v_term}
\end{figure*}
For the planetary models $\Rcapt$ is only a fraction of an \AU; about a few hundredths up to a few tenths of an \AU, which explains why only a few wind particles experience an acceleration due to the slingshot. Further, we find a slight decreasing trend in terminal velocity with binary separation, which can be explained as follows. When the binary separation is smaller, the wind has less time to accelerate up to the companion's location, and thus will reach a lower maximal velocity and accordingly a lower terminal velocity. For the slow-wind models, the effect is more distinct due to the longer interaction time, as was explained earlier.
\\ \indent For the stellar models, we see that for the toy model without the slingshot the terminal velocity becomes negative for some simulations, which is unrealistic. This is because the maximum velocity reached by a particle due to the gravitational potential of the companion does not exceed the escape velocity of the companion, given by $v_{\rm esc} = \sqrt{2G\Mcomp/r}$, at closest approach $r=d$ (see Fig.\ \ref{fig:geometry_tm}). Therefore, the extra acceleration due to the slingshot is needed in order to explain the terminal velocity of the simulations. Note here that the escape velocity of the AGB star should not be taken into account, since the SPH particles are not subject to the gravitational potential of the AGB star {in the `free wind' case (see Sect.\ \ref{sect:method})}. The importance of the slingshot in these cases also becomes clear from the value of the capture radius: since $\Rcapt$ is about 70\,\AU\ for a slow wind and about 5\,\AU\ for a fast wind, many wind particles interact with the companion and hence experience a slingshot. However, we note that this value of the capture radius may not be used as the true size of the region of influence, especially for the slow-wind models, since it is only applicable to systems of which the morphologies can be described by the BHL formulation \citep{Decin2020}. Thus, here $\Rcapt$ serves as an indicator that the interaction region is large compared to the binary separation for the stellar models, in contrast with the planetary models. In Fig.\ \ref{fig:v_term} we see that still part of the wind dodges the slingshot, considering that the terminal velocity in the stellar simulations is smaller than the one calculated for the toy model with a slingshot. Contrary to the planetary models, we find an increasing trend in the terminal velocity for decreasing binary separation. This is because the final velocity due to the slingshot depends on the orbital velocity of the companion, which will be large when the binary separation is small. Lastly, we identify that the relative increase of the terminal velocity compared to the single models is larger for the slow-wind models, which is again due to the longer interaction time in combination with the relatively larger impact of the orbital velocity of the AGB star.

\subsection{Vertical wind extent \& distribution}
\label{sect:EDE}
From previous studies it is known that binary interaction between an AGB star and a companion can cause the AGB wind material to be focussed towards the orbital plane (e.g.\ \citealp{Mastrodemos1999, Liu2017, ElMellah2020}). By including basic prescriptions of cooling, dust opacity and radiative transfer in their simulations, \cite{Chen2017, Chen2020} even obtained a circumbinary disk in certain models. This is important since circumbinary disks are often observed around post-AGB binary systems \citep{VanWinckel2003} and recently also around {AGB stars in a binary} (e.g. \citealp{Homan2017_L2Pup}). However, the formation process of these disks is still largely unknown. Hence, studying the wind focussing towards the orbital plane is a crucial step to gain insight in the formation process.
\\ \indent Figs.\ \ref{fig:tubes_solar} and \ref{fig:tubes_planet} display the radial structure of the mean density in the orbital plane $\langle \rho_{\rm orb} \rangle$ (dashed black lines) and along the polar axis $\langle \rho_{\rm polar} \rangle$ (cyan full lines) for the different models. To come to these profiles, the average density was calculated in a disk-like region of height twice the binary separation $a$ along the orbital plane and in a cylinder of diameter $2a$ along the polar axis, respectively. The bumps on these two profiles are remnants of the spiral morphology in the density that is not fully averaged out. The grey lines give the ratio of the mean densities $\langle \rho_{\rm orb} \rangle/\langle \rho_{\rm polar} \rangle$ (right-hand side $y$-axis). {From inspecting Fig.\ \ref{fig:tubes_solar}, we noticed that the wind-focussing towards the orbital plane exhibits different signatures on the polar density profiles. Hence, we distinguished between two types of focussing: global flattening of the CSE and equatorial density enhancement (EDE), which originate from different physical mechanisms. In the following two sections, flattening and EDE are studied in detail. }
\\ \indent {We note that the following analysis is performed on models that do not include cooling {other than} quasi-adiabatic cooling ($\gamma = 1.2$). When modelling the outflow with another thermodynamical prescription or by including other cooling and heating sources, these findings may change. Following the results from \cite{Chen2017, Chen2020} we suspect that the presence of a global flattening of the wind and the formation of an EDE will be strengthened and thus that more material will be present in a broad disk-like region around the orbital plane. }

{\subsubsection{Global flattening of the circumstellar envelope}}
We defined a `flattening' of the CSE due to the centrifugal effect of the orbital motion of the AGB star on its outflow, induced by the presence of the companion \citep{KimTaam2012b}. In the meridional plane of the outflow of the AGB star, the flattening is expressed as a smaller vertical than horizontal extent of the wind, {especially near the poles where a conical void-like region is visible (e.g.\ upper right plot Fig.\ \ref{fig:4au_density})}. It depends on the relative magnitude of the orbital velocity of the AGB star and its wind velocity. Therefore, we defined a flattening ratio $\Phi$ in two different ways here: (i) from the input values of the model, more specifically from the initial wind velocity and the orbital velocity via the binary separation, since it is the orbital-to-wind velocity ratio that controls the flattening  and (ii) from the {global} spatial information of the morphology of the simulation, based on the mean density around the polar axis relative to the mean density in the orbital plane.
\\ \indent When we looked at the AGB system edge-on, the orbital velocity of the AGB star is only affecting the wind velocity in the horizontal direction and not in the vertical, since the orbital motion occurs in the face-on plane. Therefore, the flattening ratio based on velocity $\Phi_v$ was defined as 
\begin{equation}\label{eq:flat_vel}
\Phi_v = 1-\frac{\vini}{\vini+\vagb},
\end{equation}
where $\vini$ is the initial wind velocity of the model and $\vagb$ the orbital velocity of the AGB star. Thus Eq.\ (\ref{eq:flat_vel}) describes the ratio of the velocity in the meridional plane to the velocity in the orbital plane. Consequently, the flattening is expected to be largest for low initial wind velocities and small binary separations (large orbital velocity). The results of $\Phi_v$ for the different models can be found in Table \ref{tab:flattening_perc}. 
\begin{table}
	\begin{center}
		\caption{Flattening ratios. }
		\begin{tabular}{ l  c  c  }
			\hline \hline \\[-2ex]
			& $\Phi_v$ & $\Phi_r$ \\ \hline
			\Cns & 0.24 & 0.01	 \\
			\Css & 0.28 & 0.30   \\
			\Cfs & 0.32 & 0.36   \\
			\Cts & 0.37 & 0.42	 \\ 
			\Cnp & 0.00 & 0.02	 \\
			\Cfp & 0.01 & 0.01	 \\
			\Ctp & 0.01 & 0.01	 \\
			\Ons & 0.56 & 0.46 	 \\
			\Oss & 0.61 & 0.21   \\
			\Ofs & 0.65 & 0.00   \\
			\Ots & 0.70 & 0.00	 \\
			\Onp & 0.02 & 0.01	 \\
			\Ofp & 0.02 & 0.02	 \\ 
			\Otp & 0.03 & 0.03	 \\ \hline 
		\end{tabular}
		\label{tab:flattening_perc}
	\end{center}
	{\footnotesize \textbf{Notes.} Here, $\Phi_v$ and $\Phi_r$ are given for the different models, according to Eqs.\ (\ref{eq:flat_vel}) and (\ref{eq:flat_spatial}), respectively. The larger the value, the more flattened the outflow is expected to be (in case of $\Phi_v$)/is found to be (in case of $\Phi_r$).
	}	
\end{table}
\begin{figure*}
	\centering
	\includegraphics[width=0.91\textwidth]{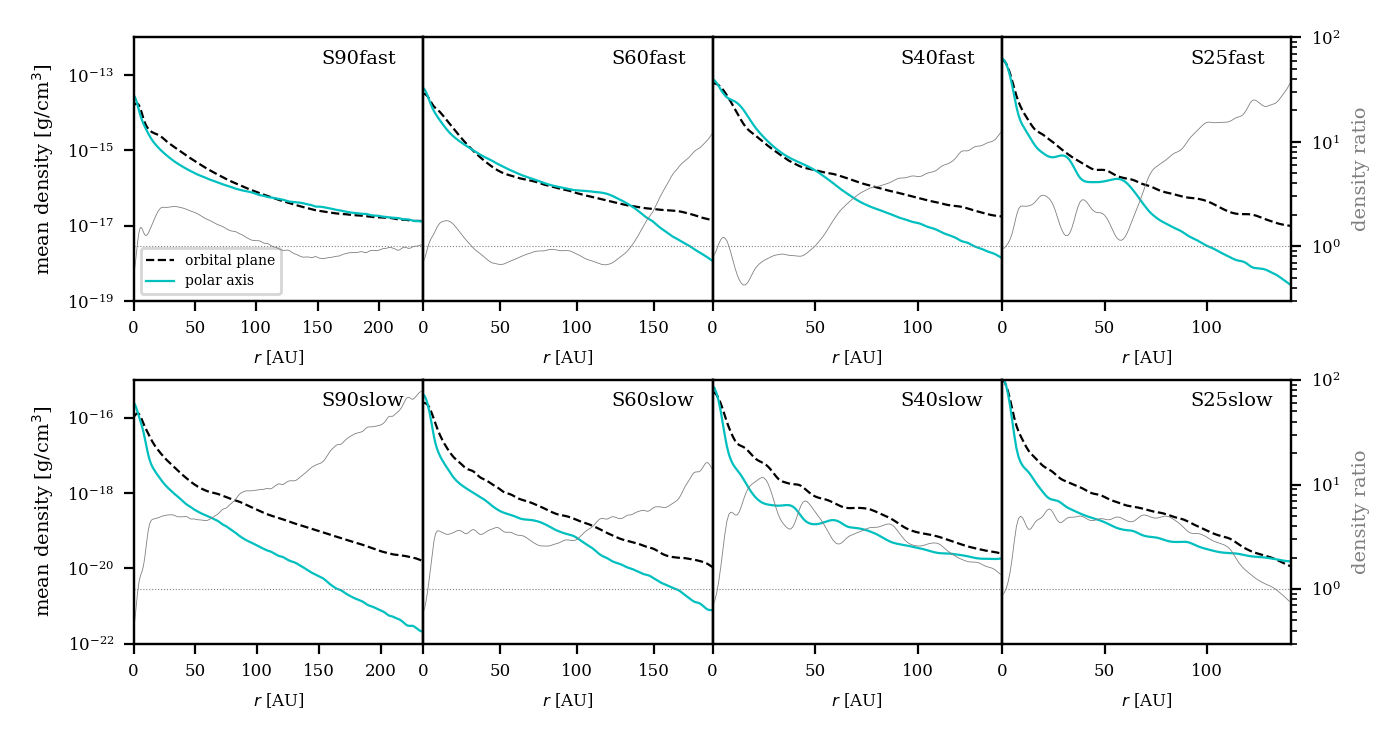}
	\caption{Radial structure profiles of the mean density of the stellar models around the polar axis $\langle \rho_{\rm polar} \rangle$ ({cyan, full lines}) and in the orbital plane $\langle \rho_{\rm orb} \rangle$ ({black, dashed lines}). The grey lines give ratio of the mean densities $\langle \rho_{\rm orb} \rangle/\langle \rho_{\rm polar} \rangle$, corresponding to the $y$-axis on the right side of the plots. \textit{Top}: fast wind models, \textit{bottom:} slow wind models. From \textit{left} to \textit{right} according to decreasing binary separation.}
	\label{fig:tubes_solar}
\end{figure*} 
\begin{figure*}
	\centering
	\includegraphics[width=0.71\textwidth]{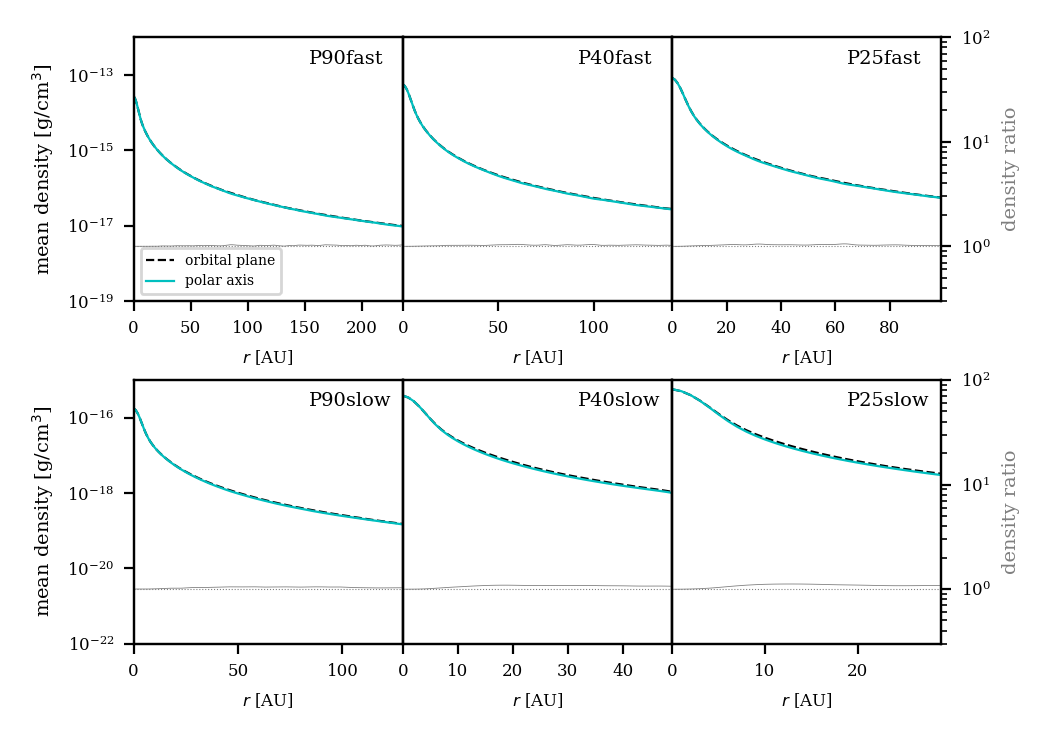}
	\caption{Radial structure profiles of the mean density of the planetary models around the polar axis $\langle \rho_{\rm polar} \rangle$ ({cyan, full lines}) and in the orbital plane $\langle \rho_{\rm orb} \rangle$ ({black, dashed lines}). The grey lines give ratio of the mean densities $\langle \rho_{\rm orb} \rangle/\langle \rho_{\rm polar} \rangle$, corresponding to the $y$-axis on the right side of the plots. \textit{Top}: fast wind models, \textit{bottom:} slow wind models. From \textit{left} to \textit{right} according to decreasing binary separation.}
	\label{fig:tubes_planet}
\end{figure*} 
\\ \indent From the definition of flattening, we identified flattening in the models when the density ratio in Figs.\ \ref{fig:tubes_solar} and \ref{fig:tubes_planet} (grey profiles) is generally increasing in function of radius from a certain point onwards, {indicating the conical void near the poles}. From these figures, we calculated the spatial flattening ratio $\Phi_r$ as follows. We took the ratio of the maximum radius in the model $r_{\rm max}$\footnote{{This maximum radius is however smaller than the predefined outer boundary of the model, to avoid physically incorrect fluid properties due to the adapted boundary conditions. Hence, in Figs.\ \ref{fig:tubes_solar} and \ref{fig:tubes_planet} we made sure to only show the physically correct region.}} to the radius where the mean density on the polar axis is the same as the final mean density in the orbital plane $r_{\rm polar}$ in logarithmic space, given by
\begin{equation}\label{eq:r_polar}
r_{\rm polar} = \max \{r \in [r_{\rm min}, r_{\rm max}] : \log\rho_{\rm polar}(r) =  \log\rho_{\rm orb}(r_{\rm max}) \}.
\end{equation}
Thus, the spatial flattening ratio $\Phi_r$ is given by
\begin{equation}\label{eq:flat_spatial}
\Phi_r = 1- \frac{r_{\rm polar}}{r_{\rm max}}.
\end{equation}
The results are also given in Table \ref{tab:flattening_perc} and are expected to be in agreement with the values for $\Phi_v$. 
\\ \indent From the mean density profiles in Fig.\ \ref{fig:tubes_solar} we see that a flattening is present for all stellar models except \Cns, \Ofs\ and \Ots. We notice that the expected flattening, according to the predictions of $\Phi_v$ in Table \ref{tab:flattening_perc}, is not present for the slow-wind models of smallest binary separation \Ofs\ and \Ots\ and smaller for the remainder slow-wind models \Ons\ and \Oss\ (see lower panels of Fig.\ \ref{fig:tubes_solar} and right column of Table \ref{tab:flattening_perc}). {Since the wind is slower in these models, the interaction will take longer and the amount of wind particles, participating in the interaction, is larger. Therefore, the contribution to the morphology of the gravitational pull of the companion is large compared to the contribution of the orbiting AGB star. Thus, this implies that the gravitational pull can alter the morphology more severely in slow winds, which results in diminishing or even cancelling out the flattening caused by the orbital motion of the AGB star itself (Table \ref{tab:flattening_perc}).} For the fast-wind models, the flattening ratios $\Phi$ of both methods seem to agree well with each other. Only for model \Cns\ the flattening ratio resulting from spatial dimensions $\Phi_r$ diverges from the one predicted from velocities $\Phi_v$. Since here the wind is fast, it can easily reach an extensive distance from the AGB star and since the binary separation is larger, the density contrast due to the interaction with the companion only sets in further out. If we compare it to the remainder fast-wind models, we notice that the flattening sets in at a smaller radius for decreasing binary separation. Therefore, we suspect some flattening is present in model \Cns, but only on a larger spatial scale than is modelled here (i.e.\ at a radius greater than 250\,\AU). 
\\ \indent For the planetary models the wind is mostly spherical, as discussed before, and which can be seen from Fig.\ \ref{fig:tubes_planet}, since the mean density in the orbital plane almost matches exactly the mean density around the polar axis. Therefore, no distinct flattening is present for these models as given by both $\Phi_v$ and $\Phi_r$ in Table \ref{tab:flattening_perc}. 
 
\subsubsection{Equatorial density enhancement (EDE)}
The gravitational pull of the companion on the AGB wind confines the wind material in the orbital plane, resulting in a so-called equatorial density enhancement or EDE in short \citep{Theuns1993, ElMellah2020}. {Different to the flattening, an EDE is expressed in the outflow as a more compact, disk-like density enhancement along the orbital plane. }Hence, we identified an EDE to be present when the mean density around the polar axis is smaller than the mean density in the orbital plane over the whole extent of the wind, indicating this disk-like structure. This was the case when in Figs.\ \ref{fig:tubes_solar} and \ref{fig:tubes_planet} the mean density ratio is larger than 1 over the whole radial range, not taking into account the first few radii\footnote{The first few radii represent the boundary of the modelled region, and therefore the physical quantities at those radii cannot be fully trusted since they are influenced by the adapted boundary conditions.}.
\\ \indent From Fig.\ \ref{fig:tubes_solar} we identify the presence of an EDE for the four slow-wind models and for the fast-wind model at smallest binary separation \Cts. This reveals that when the gravitational interaction between the companion and AGB wind is strong, more specifically when the wind is slow or the companion is located close-by and the spiral structure can no longer be described by the simple BHL wake, the presence of a companion is able to confine the majority of the wind material in a small range around the orbital plane of about 50\,\AU. For the planetary models  (Fig.\ \ref{fig:tubes_planet}) no distinct EDE is identified since the outflow is mostly spherical, as expected. However, the slow-wind models display a slightly lower mean density around the polar axis compared to the orbital plane and this becomes more plain for smaller binary separation. This confirms that even planetary sized companions are able to alter the density distribution in the wind.

\subsection{Morphology classification}\label{sect:morphological_param}
{The analysis of the previous section hints that there should exist a combination of system parameters that, when properly combined, result in one general parameter that is able to capture the nature of the companion-induce perturbation in the AGB wind. Ideally, such a `wind morphology classification parameter' (henceforward referred to as `classification parameter') should at least subdivide the wind into three major classes: (i) barely perturbed winds, which resemble smooth outflows, (ii) intermediately perturbed winds, which contain regular spirals, (iii) and the highly perturbed winds, which contain complex instabilities. Once the classification parameter is found, its classification scheme can be used to provide tighter constraints on the possible system parameters, since high-resolution observations provide a notion of the global morphology of the AGB wind. Hence, using the known system parameters and the classification parameter estimate, the ratio of the unknown remaining system parameters with respect to each other can be deduced, limiting the range of modelling options for the system.
	\\ \indent In the literature, attempts have been made to construct such a {classification parameter} from the known system properties. For example,} \cite{Theuns1993} found that an unusual morphology was more likely to arise when the wind velocity is of the same order of magnitude as the orbital velocity of the system. Further, \cite{Mastrodemos1998} pointed out the importance of the wind velocity at the location of the companion as having a crucial effect on the resulting morphology of the wind and accretion processes. Two decades later, \citeauthor{Saladino2018} (\citeyear{Saladino2018, Saladino2019}) and \cite{ElMellah2020} described their simulations in function of the ratio of the terminal velocity to the orbital velocity of the companion and could thereby classify the morphology of their models based on several input values. 
\\ \indent {Further in this section,} we investigate the validity and consistency of different candidate classification parameters adopted in the literature and introduce another candidate, using the twelve simulations in this paper as well as the two models from \cite{Malfait2021} used in the previous section. Since this set of simulations shows crucial changes in morphology (from regular to broken spiral, flattening, EDE, etc.), it is a good starting point for the analysis of the different classification parameters, despite the limited input parameter space. {In Sect.\ \ref{sect:descript_params} we introduce the four different parameters; $\eta$ being the ratio of velocities, $Q^p$ being the ratio of momenta, $\alpha$ the ratio of radii and $\varepsilon$ the ratio of energies. In Sect.\ \ref{sect:appl_params} we continue with the discussion of each parameter applied to the models, see also Fig.\ \ref{fig:morph_param_sep}.}

\subsubsection{Description of the parameters}\label{sect:descript_params}
\citeauthor{Saladino2018} (\citeyear{Saladino2018, Saladino2019}) and \cite{ElMellah2020} constructed their models in terms of scale-invariant quantities. The speed of the models is for example expressed in terms of $\vterm/\vcomp$, which is named $\eta$ by the latter. They retrieve a complex and irregular morphology when $\eta \lesssim 1$. However, considering that there exist a multitude of manners in which to measure $\vterm$ {from a modelling point-of-view}, some ambiguity is found in this definition. For example, in the case of \citeauthor{Saladino2018} (\citeyear{Saladino2018, Saladino2019}) $\vterm$ is the terminal velocity of the corresponding single star model for each binary star model. Yet, it can be shown that $\vterm$ is not always a good representation of the wind speed when it interacts with the companion, see for example Fig.\ \ref{fig:v_C90s}. A better approximation for the wind velocity $\vwind$ near the companion is the velocity of the corresponding single AGB model at the distance where the companion in the binary model is located (thus at the binary separation $a$), including the orbital motion of the AGB star:
\begin{equation}\label{eq:vwind}
	\vwind = \sqrt{v_{\rm{single}}^2(r = a)+\vagb^2}.
\end{equation}
We henceforth define the velocity-based classification parameter $\eta$ as a function of this more refined measure of the wind velocity:
\begin{equation}\label{eq:eta}
	\eta \equiv \frac{\vwind}{\vcomp}.
\end{equation}
\hspace{\parindent} \cite{Decin2020} introduced the dimensionless parameter $Q^p$, with the aim to predict the morphology of the outflow of the AGB star as a result of the interaction with the companion. $Q^p$ is defined as  the ratio of tangential momentum of the companion to the wind's radial momentum encountered by the companion in one orbit:
\begin{equation}\label{eq:Qp}
	Q^p = \frac{p_{\rm{comp}}}{p_{\rm{wind}}} = \frac{\Mcomp\vcomp}{M_{\rm{wind}}\vwind},
\end{equation}
where $\Mcomp$ is the companion's mass, $\vcomp$ the orbital velocity of the companion, $M_{\rm{wind}}$ the mass of the wind material in a torus of width twice the Hill radius\footnote{{In general}, the Hill radius of a body $m$ orbiting another body $M$ is approximately given by $R_{\rm{Hill}} = a(1-e)\left(\frac{m}{3M}\right)^{1/3}$, with $m/M<1$, $a$ the semi-major axis and $e$ the eccentricity of the orbit, defining the Hill sphere of body $m$  \citep{Hamilton1992}.} of the companion at the companion's orbit and $\vwind$ the velocity of the wind defined in Eq.\ (\ref{eq:vwind}). The morphology is expected to have little departure from a radial outflow for small values of $Q^p$, models with intermediate values are assumed to display an EDE and for large values of $Q^p$ the morphology is expected to deviate strongly from a radial outflow.
\\ \indent \cite{Mastrodemos1999} found that their models were sensitive to the ratio of the capture radius $\Rcapt$, introduced in Eq.\ (\ref{eq:capt_radius}), to the binary separation $a$, which we will call $\alpha$ henceforth. Thus, $\alpha$ is the fraction of the binary separation that is filled by $\Rcapt$:
\begin{equation}\label{eq:alpha}
	\alpha \equiv \frac{\Rcapt}{a} = \frac{2G\Mcomp}{v_{\rm{wind}}^2}\frac{1}{a}.
\end{equation}
It is expected that regular and quasi-spherical morphologies result in small $\alpha$-values and more irregular and chaotic outflows give a large value for $\alpha$.
\\ \indent Lastly, we propose a new parameter $\varepsilon$, defined as the ratio of the gravitational energy density in the Hill sphere of the companion to the kinetic energy of the wind:
\begin{equation}\label{eq:epsilon}
	\varepsilon \equiv \frac{e_{\rm{grav}}}{e_{\rm{kin}}} = \frac{\frac{G\Mcomp\rho}{\RHill}}{\frac{1}{2}\rho v_{\rm{wind}}^2} = \frac{(24G^3\Mcomp^2\MAGB)^{1/3}}{\vwind^2 a(1-e)}.
\end{equation}
\cite{ElMellah2020} already hinted to this parameter, by stating that the outflow structure depends on the amount of specific kinetic energy deposited in the wind compared to the Roche potential. This $\varepsilon$-quantity is exactly equal to the ratio of the capture radius $\Rcapt$ to the Hill radius $\RHill$, since these two radii are derived from energy conservation. The former focusses on the contribution from the kinetic energy of an incoming particle and the latter on a gravitational energy of the two components of a binary system. Also here, small values for $\varepsilon$ are expected for models where the wind kinetics dominates, thus models that show a regular or quasi-spherical morphology and larger values are predicted when the morphology is irregular or strongly flattened, because then the gravitational pull of the companion dominates over the wind kinetics.

\subsubsection{Application of the parameters}\label{sect:appl_params}
\begin{figure}
	\centering
	\includegraphics[width=0.45\textwidth]{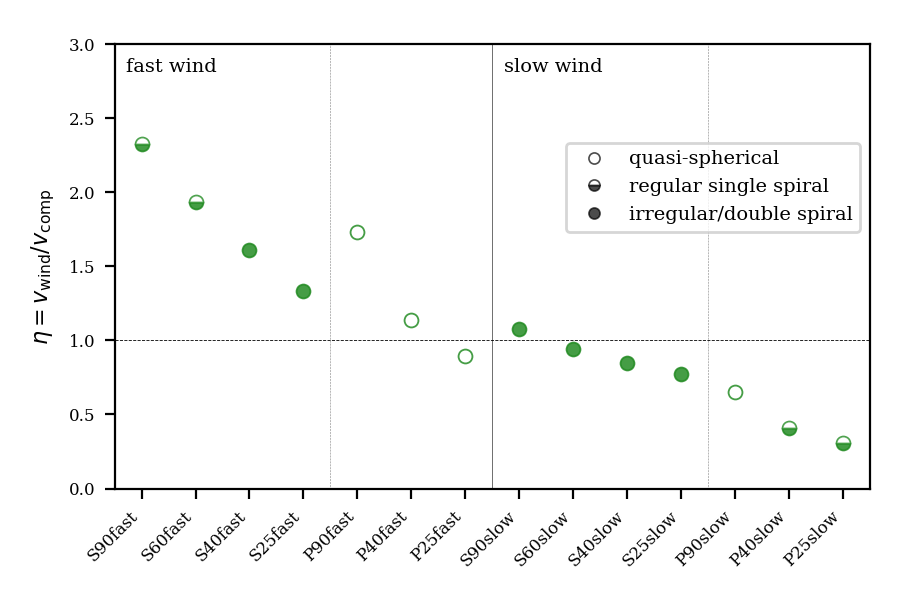}
	\\[-2.0ex]
	\includegraphics[width=0.45\textwidth]{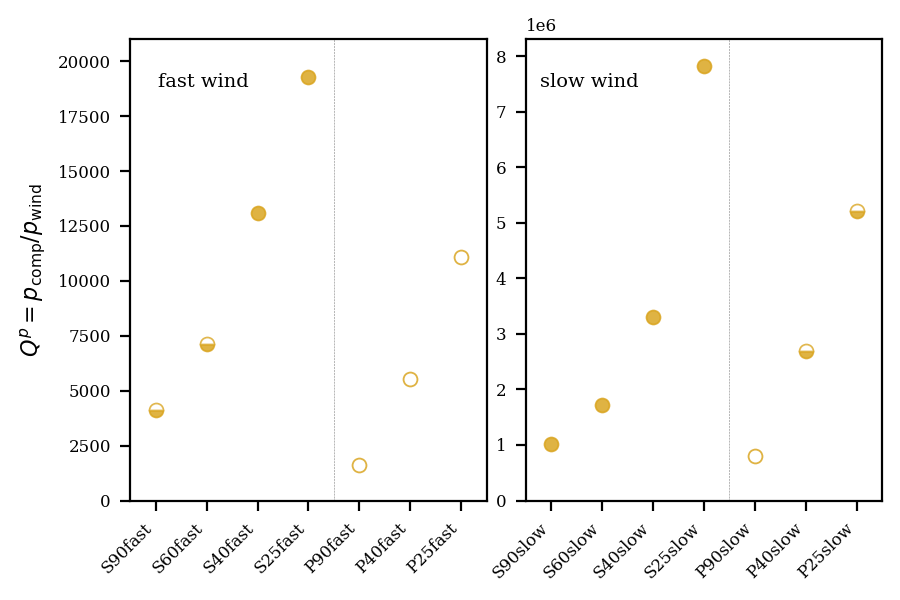}
	\\[-1.2ex]
	\includegraphics[width=0.45\textwidth]{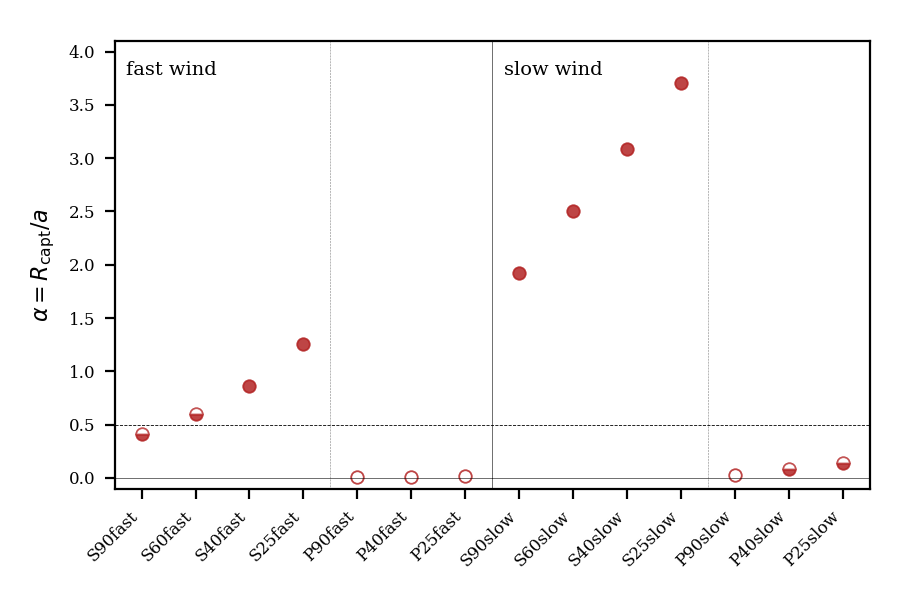}
	\\[-2.3ex]
	\includegraphics[width=0.45\textwidth]{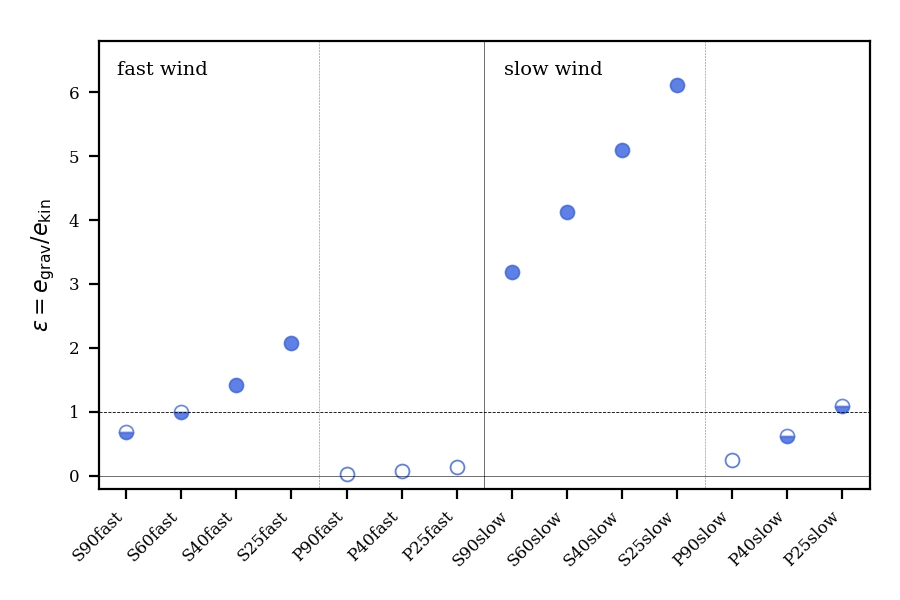}
	\\[-2.3ex]
	\caption{{Resulting values of the different classification parameters. The different models are given on the horizontal axis in each panel: fast models (\textit{left})  and  slow models (\textit{right}). The fillstyle of the symbols indicate the morphology found in the outflow of the model, as given by the legend in the top figure. For all but $\eta$, we expect more complex morphologies to correspond to higher values for the different classification parameters w.r.t.\ quasi-spherical morphologies, for $\eta$ we expect lower values.
	}}
	\label{fig:morph_param_sep}
\end{figure}
Fig.\ \ref{fig:morph_param_sep} presents the resulting values of the four classification parameters for the different models. In Sect.\ \ref{sect:results} we found different characteristics in the morphology of the simulations when changing certain input parameters. (i) The morphology gains more complexity when the binary separation is decreased, due to the stronger gravitational interaction between the wind and the companion. (ii) A slow-wind set-up results in a larger contribution of the companion's gravity to the interaction, with again a more complex outflow structure as a result, due to the longer interaction time. (iii) The planetary models show a quasi-spherical outflow with a low-density contrast spiral wake, which is different from the stellar models. Hence, if a proposed classification parameter is valid, we expect to retrieve these three features from its numerical value. We will further discuss the results of the parameters one by one {and systematically discard the ones that are not found to be valid and consistent when applied to the models in this paper}.
\\ \indent The results for $\eta$ are given in green in the upper panel of Fig.\ \ref{fig:morph_param_sep}. Feature (i) is well captured by $\eta$: models with a smaller binary separation result in a smaller value. Also, feature (ii) is visible: for the fast-wind models $\eta$ is slightly higher than for the slow-wind models. However, the distinction in value is not as clear as the visualisation of the morphology indicates, and $\eta$ does not show the clear cut-off value around 1 for more irregular morphologies, as suggested by \cite{ElMellah2020}. Lastly, feature (iii) is not at all present in the numerical values, and thus $\eta$ does not distinguish between companion type. Therefore, our simulations show more diversity than captured in the  classification proposed by \cite{ElMellah2020} and reveal that a ratio of characteristic velocities alone does not give a full quantitative description of the morphology in the AGB outflow. 
\\ \indent {Next, we discuss the results for the momentum ratio $Q^p$ (Eq.\ (\ref{eq:Qp})) given in yellow in Fig.\ \ref{fig:morph_param_sep}. Note that the results for the fast- and slow-wind models differ by two orders of magnitude. The reason for this is the different input mass-loss rate for both wind types. Hence, this is a algebraic effect, since the input mass-loss rate does not influence the morphology for polytropic winds, but only the mass of the individual SPH particles (see Sect.\ \ref{sect:numerics}), in that way affecting $M_{\rm wind}$ in Eq.\ (\ref{eq:Qp}). Therefore, only the relative differences between the results will be taken into account here, because we can rescale $Q^p$ with any chosen value.} $Q^p$ increases with binary separation, thus feature (i) is decently captured by this parameter. However, there is no clear distinction in $Q^p$-value between the models of different wind type {(taking into account what was stated before)} and of different companion type, thus $Q^p$ is lacking the ability to represent features (ii) and (iii) found in the morphology. Therefore, $Q^p$ is not able to give a decisive classification of the morphologies of our models. More information about the binary and AGB wind set-up is needed in order to describe the morphology in a quantitative way.
\\ \indent The third panel of Fig.\ \ref{fig:morph_param_sep} displays the results for the ratio of radii $\alpha$ (Eq.\ (\ref{eq:alpha})) in red. The values for $\alpha$ increase in function of binary separation, displaying feature (i). Furthermore, a clear distinction between the stellar and planetary models is present and the difference in wind velocity comes to appearance. Thus features (ii) and (iii) are well captured by $\alpha$, since $\alpha$ does not only deal with the velocity of the system, but also with gravitational interaction, contrary to $\eta$ and $Q^p$. Based on these results, we can make the following classification scheme for $\alpha$. For $\alpha \ll 0.5$ the morphology is quasi-spherical, with a small density-contrast spiral present. If $\alpha \sim 0.5$, the outflow shows a clear, regular spiral in the orbital plane and arcs in the meridional plane. Systems with $\alpha \gg 0.5$ have a more irregular or even chaotic (spiral) morphology, often flattened in the meridional plane. Hence, $\alpha$ gives a consistent representation of the different morphologies that we find in our models and is therefore a suitable candidate as classification parameter.
\\ \indent Lastly, we investigate the quantity $\varepsilon$ (Eq.\ (\ref{eq:epsilon})) given in blue in the bottom panel of Fig.\ \ref{fig:morph_param_sep} for the different models. The outcome looks similar as the results for $\alpha$: an increasing trend in function of decreasing binary separation is present and a clear distinction between the stellar and planetary models appears as well as between the fast- and slow-wind models, thus $\varepsilon$ is also able to capture well the three morphological features found in the simulations. The following classification scheme arises: models with $\varepsilon \ll 1$ show a quasi-spherical outflow, when $\varepsilon \sim 1$ a single, regular spiral is retrieved and for $\varepsilon \gg 1$ complex, irregular morphologies are found, such as double spirals and chaotic behaviour, often flattened in the meridional plane.
\\ \indent Therefore, the $\alpha$ and $\varepsilon$ are both found to be suited as classifications parameters. Because $\varepsilon$ is a ratio of energy densities, we accept three clear regimes: $\varepsilon < 1$, $\varepsilon \sim 1$ and $\varepsilon > 1$, as is found. This is not the case for $\alpha$, since the classification value, here found to be $\sim$0.5, fluctuates when varying $a$. We note here that from observations it is known that the mass-loss rate $\Mdot$ is also an important quantity for the morphology classification \citep{Decin2020}, because it impacts the chemistry and dust formation. Therefore, one would expect that $\Mdot$ should enter into some classification parameter if these processes could be taken into account in the modelling.

\section{Summary \& conclusions} \label{sect:conclusions}
In this paper, we performed twelve hydrodynamical simulations of companion-perturbed AGB outflows using the SPH code \Phantom, varying three key parameters: wind velocity, binary separations and companion mass. We investigated (i) the prominent global morphological changes in the density structure when varying the parameters, (ii) the effect of the gravitational interaction between the outflow and the companion on the terminal velocity of the wind, (iii) the vertical compression of wind material to the orbital plane due to the interaction and (iv) we carried out a morphology classification of our models using four different candidate classification parameters.
\\ \indent Regarding the morphology of the AGB outflow, we find that for a fast wind and large binary separation set-up, a single spiral structure forms in de orbital plane, where the widening of the spiral is caused by the velocity dispersion due to the gravitational slingshot or gravity assist of the companion on the wind particles. In the meridional plane the spiral manifests itself as arcs, reaching the poles of the outflow, showing the inherent 3D structure of the wind morphology. We recover that this fundamental spiral structure becomes more complex when decreasing the wind velocity or by decreasing the binary separation. A double spiral or so-called vortex structure emerges in the orbital plane close to the binary system and the arcs in the meridional plane are compressed towards the orbital plane. For smaller binary separations, the spiral becomes irregular and broken, due to periodicity in a high density region forming behind the companion. When the companion mass is lowered to the mass of a massive planet, the spiral becomes again similar to the fundamental single spiral structure. We proved that even massive planets are able to affect the wind morphology.
\\ \indent By studying the effect of a companion on the terminal wind velocity, we find that, for stellar companions, the increase in terminal velocity w.r.t.\ a single AGB wind is explained by a gravity assist on the wind particles due to the companion's motion. For planetary companions, too few wind particles experience this assist, so that the slight decrease in terminal velocity w.r.t.\ to a single AGB wind is explained by the gravitational attraction of the companion only. 
\\ \indent We analysed the vertical extent and distribution of the wind and distinguish between two signatures: {a global flattening of the CSE} as a result of the orbital motion of the AGB star and an equatorial density enhancement (EDE) caused by the gravitational pull of the companion on the wind. We find that flattening in the outflow of the models corresponds well with the predictions we made based on velocity, and that the effect is enhanced in the case of a slow AGB wind. However, for models \Ofs\ and \Ots, the stronger wind-companion interaction present diminishes or even cancels out the global flattening of wind material. An EDE is found for all slow-wind models and fast-wind model \Cts, since the structure formation in the outflow is dominated by the gravitational attraction of the companion and no longer by the orbital movement of the AGB star. 
\\ \indent Multiple morphology classification parameters from the literature, and one that we introduced ourselves, were examined to verify if they are able to probe the type of morphology present in our models. {The aim was to derive a singular parameter that can be used to retrieve and constrain some information from observations about AGB system, without the need of detailed 3D modelling}. We conclude that $\alpha$ (ratio of the capture radius to the binary separation, Eq.\ (\ref{eq:alpha})) and $\varepsilon$ (ratio of the companion's gravitational energy density to the wind's kinetic energy density Eq.\ (\ref{eq:epsilon})) both properly capture the variety of morphologies found in our simulations and result in a distinct morphology classification scheme.

\begin{acknowledgements}
S.M., W.H., J.M., J.B., F.D.C.\ and L.D. acknowledge support from the ERC consolidator grant 646758 AEROSOL. Also, W.H.\ acknowledges support from the Fonds de la Recherche Scientifique (FNRS) through grant 40000307, L.S.\ is a senior FNRS researcher, J.B.\ acknowledges support from the KU Leuven C1 grant MAESTRO C16/17/007 and F.D.C.\ is supported by the EPSRC iCASE studentship programme, Intel Corporation and Cray Inc. 
\end{acknowledgements}

\end{hyphenrules}

\bibliographystyle{aa}
{\small \bibliography{references}}

\begin{thebibliography}{67}
\expandafter\ifx\csname natexlab\endcsname\relax\def\natexlab#1{#1}\fi

\bibitem[{{Bondi} \& {Hoyle}(1944)}]{BondiHoyle1944}
{Bondi}, H. \& {Hoyle}, F. 1944, \mnras, 104, 273

\bibitem[{{Boulangier} {et~al.}(2019){Boulangier}, {Clementel}, {van Marle},
  {Decin}, \& {de Koter}}]{Boulangier2019}
{Boulangier}, J., {Clementel}, N., {van Marle}, A.~J., {Decin}, L., \& {de
  Koter}, A. 2019, \mnras, 482, 5052

\bibitem[{{Bowen}(1988)}]{Bowen1988}
{Bowen}, G.~H. 1988, \apj, 329, 299

\bibitem[{{Burke}(2015)}]{Burke2015}
{Burke}, C.~J. 2015, in AAS/Division for Extreme Solar Systems Abstracts, ed.
  501.01, Vol.~47, 501.01

\bibitem[{{Chen} {et~al.}(2017){Chen}, {Frank}, {Blackman}, {Nordhaus}, \&
  {Carroll-Nellenback}}]{Chen2017}
{Chen}, Z., {Frank}, A., {Blackman}, E.~G., {Nordhaus}, J., \&
  {Carroll-Nellenback}, J. 2017, in IAU Symposium, Vol. 323, Planetary Nebulae:
  Multi-Wavelength Probes of Stellar and Galactic Evolution, ed. X.~{Liu},
  L.~{Stanghellini}, \& A.~{Karakas}, 367--368

\bibitem[{{Chen} {et~al.}(2020){Chen}, {Ivanova}, \&
  {Carroll-Nellenback}}]{Chen2020}
{Chen}, Z., {Ivanova}, N., \& {Carroll-Nellenback}, J. 2020, \apj, 892, 110

\bibitem[{{Cohen} {et~al.}(2004){Cohen}, {Van Winckel}, {Bond}, \&
  {Gull}}]{Cohen2004}
{Cohen}, M., {Van Winckel}, H., {Bond}, H.~E., \& {Gull}, T.~R. 2004, \aj, 127,
  2362

\bibitem[{{Danilovich} {et~al.}(2014){Danilovich}, {Bergman}, {Justtanont},
  {Lombaert}, {Maercker}, {Olofsson}, {Ramstedt}, \& {Royer}}]{Danilovich2014}
{Danilovich}, T., {Bergman}, P., {Justtanont}, K., {et~al.} 2014, \aap, 569,
  A76

\bibitem[{{Decin} {et~al.}(2012){Decin}, {Cox}, {Royer}, {Van Marle},
  {Vandenbussche}, {Ladjal}, {Kerschbaum}, {Ottensamer}, {Barlow}, {Blommaert},
  {Gomez}, {Groenewegen}, {Lim}, {Swinyard}, {Waelkens}, \&
  {Tielens}}]{Decin2012}
{Decin}, L., {Cox}, N.~L.~J., {Royer}, P., {et~al.} 2012, \aap, 548, A113

\bibitem[{{Decin} {et~al.}(2010){Decin}, {De Beck}, {Br{\"u}nken},
  {M{\"u}ller}, {Menten}, {Kim}, {Willacy}, {de Koter}, \&
  {Wyrowski}}]{Decin2010}
{Decin}, L., {De Beck}, E., {Br{\"u}nken}, S., {et~al.} 2010, \aap, 516, A69

\bibitem[{{Decin} {et~al.}(2006){Decin}, {Hony}, {de Koter}, {Justtanont},
  {Tielens}, \& {Waters}}]{Decin2006}
{Decin}, L., {Hony}, S., {de Koter}, A., {et~al.} 2006, \aap, 456, 549

\bibitem[{{Decin} {et~al.}(2020){Decin}, {Montarg{\`e}s}, {Richards},
  {Gottlieb}, {Homan}, {McDonald}, {El Mellah}, {Danilovich}, {Wallstr{\"o}m},
  {Zijlstra}, {Baudry}, {Bolte}, {Cannon}, {De Beck}, {De Ceuster}, {de Koter},
  {De Ridder}, {Etoka}, {Gobrecht}, {Gray}, {Herpin}, {Jeste}, {Lagadec},
  {Kervella}, {Khouri}, {Menten}, {Millar}, {M{\"u}ller}, {Plane}, {Sahai},
  {Sana}, {Van de Sande}, {Waters}, {Wong}, \& {Yates}}]{Decin2020}
{Decin}, L., {Montarg{\`e}s}, M., {Richards}, A.~M.~S., {et~al.} 2020, Science,
  369, 1497

\bibitem[{{Decin} {et~al.}(2015){Decin}, {Richards}, {Neufeld}, {Steffen},
  {Melnick}, \& {Lombaert}}]{Decin2015}
{Decin}, L., {Richards}, A.~M.~S., {Neufeld}, D., {et~al.} 2015, \aap, 574, A5

\bibitem[{{Eggleton}(2006)}]{EggletonBinaryBook}
{Eggleton}, P. 2006, {Evolutionary Processes in Binary and Multiple Stars}

\bibitem[{{El Mellah} {et~al.}(2020){El Mellah}, {Bolte}, {Decin}, {Homan}, \&
  {Keppens}}]{ElMellah2020}
{El Mellah}, I., {Bolte}, J., {Decin}, L., {Homan}, W., \& {Keppens}, R. 2020,
  \aap, 637, A91

\bibitem[{{Ertel} {et~al.}(2019){Ertel}, {Kamath}, {Hillen}, {van Winckel},
  {Okumura}, {Manick}, {Boffin}, {Milli}, {Bertrang}, {Guzman-Ramirez},
  {Horner}, {Marshall}, {Scicluna}, {Vaz}, {Villaver}, {Wesson}, \&
  {Xu}}]{Ertel2019}
{Ertel}, S., {Kamath}, D., {Hillen}, M., {et~al.} 2019, \aj, 157, 110

\bibitem[{{Freytag} {et~al.}(2017){Freytag}, {Liljegren}, \&
  {H{\"o}fner}}]{Freytag2017}
{Freytag}, B., {Liljegren}, S., \& {H{\"o}fner}, S. 2017, \aap, 600, A137

\bibitem[{{Fulton} {et~al.}(2019){Fulton}, {Rosenthal}, {Howard}, {Hirsch}, \&
  {Isaacson}}]{Fulton2019}
{Fulton}, B., {Rosenthal}, L., {Howard}, A., {Hirsch}, L., \& {Isaacson}, H.
  2019, in AAS/Division for Extreme Solar Systems Abstracts, Vol.~51,
  AAS/Division for Extreme Solar Systems Abstracts, 401.01

\bibitem[{{Gail} \& {Sedlmayr}(2013)}]{Gail2013}
{Gail}, H.-P. \& {Sedlmayr}, E. 2013, {Physics and Chemistry of Circumstellar
  Dust Shells}

\bibitem[{{Gingold} \& {Monaghan}(1977)}]{GandM1977}
{Gingold}, R.~A. \& {Monaghan}, J.~J. 1977, \mnras, 181, 375

\bibitem[{{Guerrero} {et~al.}(2003){Guerrero}, {Chu}, {Manchado}, \&
  {Kwitter}}]{Guerrero2003}
{Guerrero}, M.~A., {Chu}, Y.-H., {Manchado}, A., \& {Kwitter}, K.~B. 2003, \aj,
  125, 3213

\bibitem[{{Habing}(1996)}]{Habing1996}
{Habing}, H.~J. 1996, \aapr, 7, 97

\bibitem[{{Habing} \& {Olofsson}(2003)}]{Habing2003}
{Habing}, H.~J. \& {Olofsson}, H., eds. 2003, {Asymptotic giant branch stars},
  ed. H.~J. {Habing} \& H.~{Olofsson}

\bibitem[{{Habing} \& {Olofsson}(2004)}]{Habing2004agbs.book}
{Habing}, H.~J. \& {Olofsson}, H. 2004, {Asymptotic Giant Branch Stars}

\bibitem[{{Hamilton} \& {Burns}(1992)}]{Hamilton1992}
{Hamilton}, D.~P. \& {Burns}, J.~A. 1992, \icarus, 96, 43

\bibitem[{{Heras} \& {Hony}(2005)}]{Heras2005}
{Heras}, A.~M. \& {Hony}, S. 2005, \aap, 439, 171

\bibitem[{{H{\"o}fner} \& {Olofsson}(2018)}]{HofnerOlofsson}
{H{\"o}fner}, S. \& {Olofsson}, H. 2018, \aapr, 26, 1

\bibitem[{{Homan} {et~al.}(2020){Homan}, {Montarg{\`e}s}, {Pimpanuwat},
  {Richards}, {Wallstr{\"o}m}, {Kervella}, {Decin}, {Zijlstra}, {Danilovich},
  {de Koter}, {Menten}, {Sahai}, {Plane}, {Lee}, {Waters}, {Baudry}, {Tat
  Wong}, {Millar}, {Van de Sande}, {Lagadec}, {Gobrecht}, {Yates}, {Price},
  {Cannon}, {Bolte}, {De Ceuster}, {Herpin}, {Nuth}, {Philip Sindel}, {Kee},
  {Grey}, {Etoka}, {Jeste}, {Gottlieb}, {Gottlieb}, {McDonald}, {El Mellah}, \&
  {M{\"u}ller}}]{Homan2020_pi1gru}
{Homan}, W., {Montarg{\`e}s}, M., {Pimpanuwat}, B., {et~al.} 2020, \aap, 644,
  A61

\bibitem[{{Homan} {et~al.}(2021){Homan}, {Pimpanuwat}, {Herpin}, {Danilovich},
  {McDonald}, {Wallstr{\"o}m}, {Richards}, {Baudry}, {Sahai}, {Millar}, {de
  Koter}, {Gottlieb}, {Kervella}, {Montarg{\`e}s}, {Van de Sande}, {Decin},
  {Zijlstra}, {Etoka}, {Jeste}, {M{\"u}ller}, {Maes}, {Malfait}, {Menten},
  {Plane}, {Lee}, {Waters}, {Tat Wong}, {Lagadec}, {Gobrecht}, {Yates},
  {Price}, {Cannon}, {Bolte}, {De Ceuster}, {Nuth}, {Sindel}, {Kee}, {Gray}, \&
  {El Mellah}}]{Homan2021_RHya}
{Homan}, W., {Pimpanuwat}, B., {Herpin}, F., {et~al.} 2021, arXiv e-prints,
  arXiv:2104.07297

\bibitem[{{Homan} {et~al.}(2017){Homan}, {Richards}, {Decin}, {Kervella}, {de
  Koter}, {McDonald}, \& {Ohnaka}}]{Homan2017_L2Pup}
{Homan}, W., {Richards}, A., {Decin}, L., {et~al.} 2017, \aap, 601, A5

\bibitem[{{Hoyle} \& {Lyttleton}(1939)}]{Hoyle1939}
{Hoyle}, F. \& {Lyttleton}, R.~A. 1939, Proc. Camb. Phil. Soc., 35, 405

\bibitem[{{Iben}(1975)}]{Iben1975}
{Iben}, I., J. 1975, \apj, 196, 525

\bibitem[{{Illarionov} \& {Sunyaev}(1975)}]{Illarionov1975}
{Illarionov}, A.~F. \& {Sunyaev}, R.~A. 1975, \aap, 39, 185

\bibitem[{{Kervella} {et~al.}(2016){Kervella}, {Homan}, {Richards}, {Decin},
  {McDonald}, {Montarg{\`e}s}, \& {Ohnaka}}]{Kervella2016}
{Kervella}, P., {Homan}, W., {Richards}, A.~M.~S., {et~al.} 2016, \aap, 596,
  A92

\bibitem[{{Kim} \& {Taam}(2012{\natexlab{a}})}]{KimTaam2012}
{Kim}, H. \& {Taam}, R.~E. 2012{\natexlab{a}}, \apj, 744, 136

\bibitem[{{Kim} \& {Taam}(2012{\natexlab{b}})}]{KimTaam2012b}
{Kim}, H. \& {Taam}, R.~E. 2012{\natexlab{b}}, \apj, 759, 59

\bibitem[{{Knapp} {et~al.}(1998){Knapp}, {Young}, {Lee}, \&
  {Jorissen}}]{Knapp1998}
{Knapp}, G.~R., {Young}, K., {Lee}, E., \& {Jorissen}, A. 1998, \apjs, 117, 209

\bibitem[{{Lamers} \& {Cassinelli}(1999)}]{LamersCassinelli}
{Lamers}, H. J.~G.~L.~M. \& {Cassinelli}, J.~P. 1999, {Introduction to Stellar
  Winds}

\bibitem[{{Lattanzio} \& {Wood}(2004)}]{LattandWood2004}
{Lattanzio}, J.~C. \& {Wood}, P.~R. 2004, {Evolution, Nucleosynthesis, and
  Pulsation of AGB Stars}, 23--104

\bibitem[{{Liljegren} {et~al.}(2016){Liljegren}, {H{\"o}fner}, {Nowotny}, \&
  {Eriksson}}]{Liljegren2016}
{Liljegren}, S., {H{\"o}fner}, S., {Nowotny}, W., \& {Eriksson}, K. 2016, \aap,
  589, A130

\bibitem[{{Liu} {et~al.}(2017){Liu}, {Stancliffe}, {Abate}, \&
  {Matrozis}}]{Liu2017}
{Liu}, Z.-W., {Stancliffe}, R.~J., {Abate}, C., \& {Matrozis}, E. 2017, \apj,
  846, 117

\bibitem[{{Lodato} \& {Price}(2010)}]{LodatoPrice2010}
{Lodato}, G. \& {Price}, D.~J. 2010, \mnras, 405, 1212

\bibitem[{{Lucy}(1977)}]{Lucy1977}
{Lucy}, L.~B. 1977, \aj, 82, 1013

\bibitem[{{Malfait} {et~al.}(2021){Malfait}, {Homan}, {Maes}, {Bolte}, {De
  Ceuster}, \& {Decin}}]{Malfait2021}
{Malfait}, J., {Homan}, W., {Maes}, S., {et~al.} 2021, submitted to A\&A

\bibitem[{{Mastrodemos} \& {Morris}(1998)}]{Mastrodemos1998}
{Mastrodemos}, N. \& {Morris}, M. 1998, \apj, 497, 303

\bibitem[{{Mastrodemos} \& {Morris}(1999)}]{Mastrodemos1999}
{Mastrodemos}, N. \& {Morris}, M. 1999, \apj, 523, 357

\bibitem[{{Mauron} \& {Huggins}(2006)}]{Mauron2006}
{Mauron}, N. \& {Huggins}, P.~J. 2006, \aap, 452, 257

\bibitem[{{Millar}(2004)}]{Millar2004}
{Millar}, T.~J. 2004, {Molecule and Dust Grain Formation}, 247--289

\bibitem[{{Moe} \& {Di Stefano}(2017)}]{Moe2017}
{Moe}, M. \& {Di Stefano}, R. 2017, \apjs, 230, 15

\bibitem[{{Nordhaus} \& {Blackman}(2006)}]{Nordhaus2006}
{Nordhaus}, J. \& {Blackman}, E.~G. 2006, \mnras, 370, 2004

\bibitem[{{O'Dell} {et~al.}(2002){O'Dell}, {Balick}, {Hajian}, {Henney}, \&
  {Burkert}}]{ODell2002}
{O'Dell}, C.~R., {Balick}, B., {Hajian}, A.~R., {Henney}, W.~J., \& {Burkert},
  A. 2002, \aj, 123, 3329

\bibitem[{{Omukai}(2000)}]{Omukai2000}
{Omukai}, K. 2000, \apj, 534, 809

\bibitem[{{Price}(2012)}]{PricePhantom2012}
{Price}, D.~J. 2012, J. Comput. Phys., 231, 759

\bibitem[{{Price} \& {Federrath}(2010)}]{PriceFederarth2010}
{Price}, D.~J. \& {Federrath}, C. 2010, \mnras, 406, 1659

\bibitem[{{Price} {et~al.}(2018){Price}, {Wurster}, {Tricco}, {Nixon},
  {Toupin}, {Pettitt}, {Chan}, {Mentiplay}, {Laibe}, {Glover}, {Dobbs},
  {Nealon}, {Liptai}, {Worpel}, {Bonnerot}, {Dipierro}, {Ballabio}, {Ragusa},
  {Federrath}, {Iaconi}, {Reichardt}, {Forgan}, {Hutchison}, {Constantino},
  {Ayliffe}, {Hirsh}, \& {Lodato}}]{Price_Phantom2018}
{Price}, D.~J., {Wurster}, J., {Tricco}, T.~S., {et~al.} 2018, \pasa, 35, e031

\bibitem[{{Ramstedt} {et~al.}(2014){Ramstedt}, {Mohamed}, {Vlemmings},
  {Maercker}, {Montez}, {Baudry}, {De Beck}, {Lindqvist}, {Olofsson},
  {Humphreys}, {Jorissen}, {Kerschbaum}, {Mayer}, {Wittkowski}, {Cox},
  {Lagadec}, {Leal-Ferreira}, {Paladini}, {P{\'e}rez-S{\'a}nchez}, \&
  {Sacuto}}]{Ramstedt2014}
{Ramstedt}, S., {Mohamed}, S., {Vlemmings}, W.~H.~T., {et~al.} 2014, \aap, 570,
  L14

\bibitem[{{Ramstedt} {et~al.}(2009){Ramstedt}, {Sch{\"o}ier}, \&
  {Olofsson}}]{Ramstedt2009}
{Ramstedt}, S., {Sch{\"o}ier}, F.~L., \& {Olofsson}, H. 2009, \aap, 499, 515

\bibitem[{{Saladino} {et~al.}(2019){Saladino}, {Pols}, \&
  {Abate}}]{Saladino2019}
{Saladino}, M.~I., {Pols}, O.~R., \& {Abate}, C. 2019, \aap, 626, A68

\bibitem[{{Saladino} {et~al.}(2018){Saladino}, {Pols}, {van der Helm},
  {Pelupessy}, \& {Portegies Zwart}}]{Saladino2018}
{Saladino}, M.~I., {Pols}, O.~R., {van der Helm}, E., {Pelupessy}, I., \&
  {Portegies Zwart}, S. 2018, \aap, 618, A50

\bibitem[{{Schwarzschild} \& {H{\"a}rm}(1965)}]{Schwarzschild1965}
{Schwarzschild}, M. \& {H{\"a}rm}, R. 1965, \apj, 142, 855

\bibitem[{{Siess} {et~al.}(2021){Siess}, {Homan}, {Toupin}, {Price}, \&
  {Jorissen}}]{Siess2021}
{Siess}, L., {Homan}, W., {Toupin}, S., {Price}, D.~J., \& {Jorissen}, A. 2021,
  \aap, in prep.

\bibitem[{{Sugimoto} \& {Nomoto}(1975)}]{Sugimoto1975}
{Sugimoto}, D. \& {Nomoto}, K. 1975, \pasj, 27, 197

\bibitem[{{Theuns} \& {Jorissen}(1993)}]{Theuns1993}
{Theuns}, T. \& {Jorissen}, A. 1993, \mnras, 265, 946

\bibitem[{{van Winckel}(2003)}]{VanWinckel2003}
{van Winckel}, H. 2003, \araa, 41, 391

\bibitem[{{Verhoelst} {et~al.}(2009){Verhoelst}, {van der Zypen}, {Hony},
  {Decin}, {Cami}, \& {Eriksson}}]{Verhoelst2009}
{Verhoelst}, T., {van der Zypen}, N., {Hony}, S., {et~al.} 2009, \aap, 498, 127

\bibitem[{{Waters}(2011)}]{Water2011}
{Waters}, L.~B.~F.~M. 2011, in Astronomical Society of the Pacific Conference
  Series, Vol. 445, Why Galaxies Care about AGB Stars II: Shining Examples and
  Common Inhabitants, ed. F.~{Kerschbaum}, T.~{Lebzelter}, \& R.~F. {Wing}, 227

\bibitem[{{Woitke}(2006)}]{Woitke2006}
{Woitke}, P. 2006, \aap, 452, 537

\end{thebibliography}

\appendix
\section{Resolution set-up}\label{sect:resolution}
{In the \Phantom\ simulations, the numerical resolution is set by two parameters, called \iwlong\ and \wshlong, we shall abbreviate them as \iw\ and \wsh\ in the formulae, respectively. Together they determine the amount of particles that are launched into the simulation as follows. 
	\\ \indent The \iwlong\ is an integer directly associated with a fixed amount of particles, $N$, which can be placed on a sphere isotropically and is calculated as
	\begin{equation}
	N = 20 \times (2\times\iw\times(\iw-1)) + 12.
	\end{equation} 
	The interparticle distance on the sphere $D_{\rm part}$ can be determined by using the radius $R$ of the sphere:
	\begin{equation}\label{eq:interparticle_distance}
	D_{\rm part} = R \times\frac{2}{(2\times \iw-1)\times \sqrt{\sqrt{5}\times \varphi}},
	\end{equation}
	with $\varphi$ the golden ratio and where $R$ is in this case the effective radius of the AGB star. 
	\\ \indent The \wshlong\ sets the distance $D$ between the successively launched shells via $D_{\rm part}$:
	\begin{equation}
	D = \wsh \times D_{\rm part}.
	\end{equation}
	\quad \quad Since in this paper we aim to model the global morphology of the outflow with about $10^6$ SPH particles in each model, we tweak the two resolution parameters per model individually to optimally resolve the morphology. In other words, due to the different input wind velocity, binary separation and companion mass adopted, we cannot use the same resolution set-up for all models. The resolution input values for the different models can be found in Table \ref{tab:res_models}.}
\begin{table}
	\begin{center}
		\caption{Resolution set-up of the simulations. }
		\begin{tabular}{ l  c  c  }
			\hline \hline \\[-2ex]
			& \iw & \wsh \\ \hline
			\Cns & 6 & 3	 \\
			\Cfs & 4 & 0.3   \\
			\Cts & 4 & 0.25	 \\ 
			\Cnp & 4 & 0.65	 \\
			\Cfp & 4 & 0.3	 \\
			\Ctp & 4 & 0.25	 \\
			\Ons & 5 & 0.4 	 \\
			\Ofs & 5 & 0.18   \\
			\Ots & 5 & 0.15	 \\
			\Onp & 4 & 0.19	 \\
			\Ofp & 5 & 0.17	 \\ 
			\Otp & 5 & 0.15	 \\ \hline 
		\end{tabular}
		\label{tab:res_models}
	\end{center}
	{\footnotesize \textbf{Notes.} Here, \iw\ is the \iwlong\ input value and \wsh\ the \wshlong\ for \Phantom. 
	}
\end{table}

\section{Gallery of models}\label{sect:gallery}

\begin{figure*}
	\centering
	\includegraphics[width=0.31\textwidth]{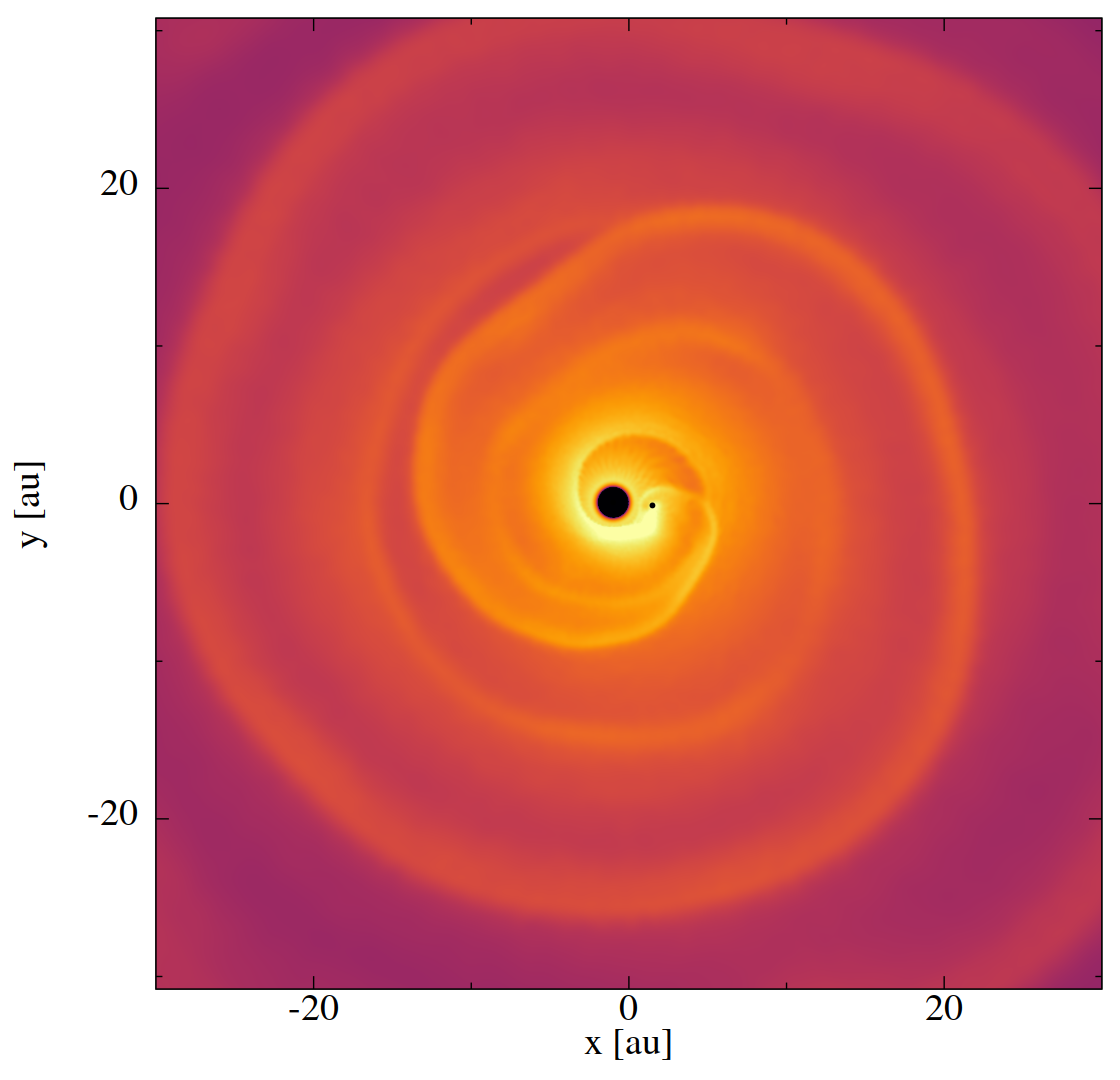}
	\includegraphics[width=0.31\textwidth]{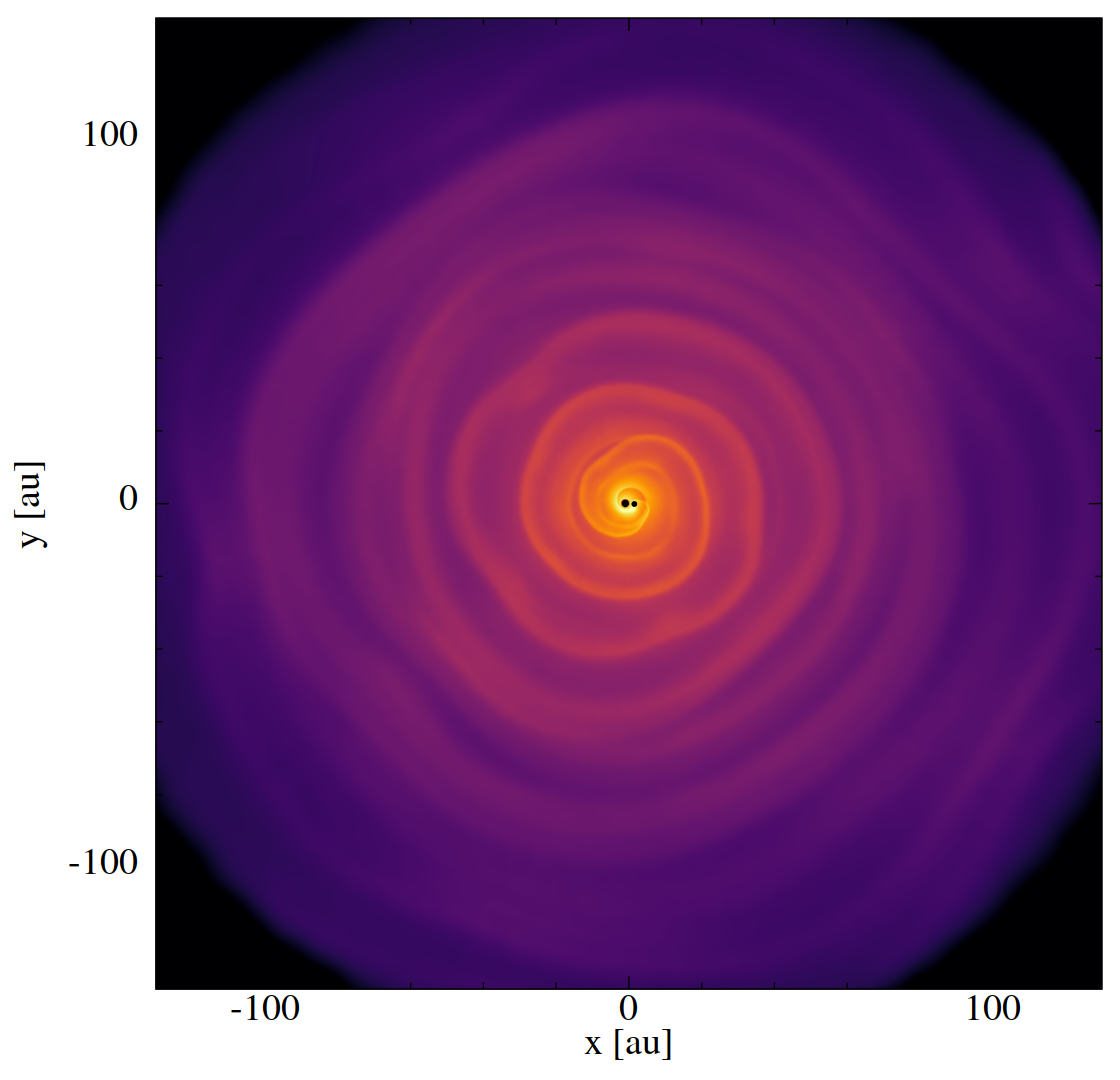}
	\includegraphics[width=0.355\textwidth]{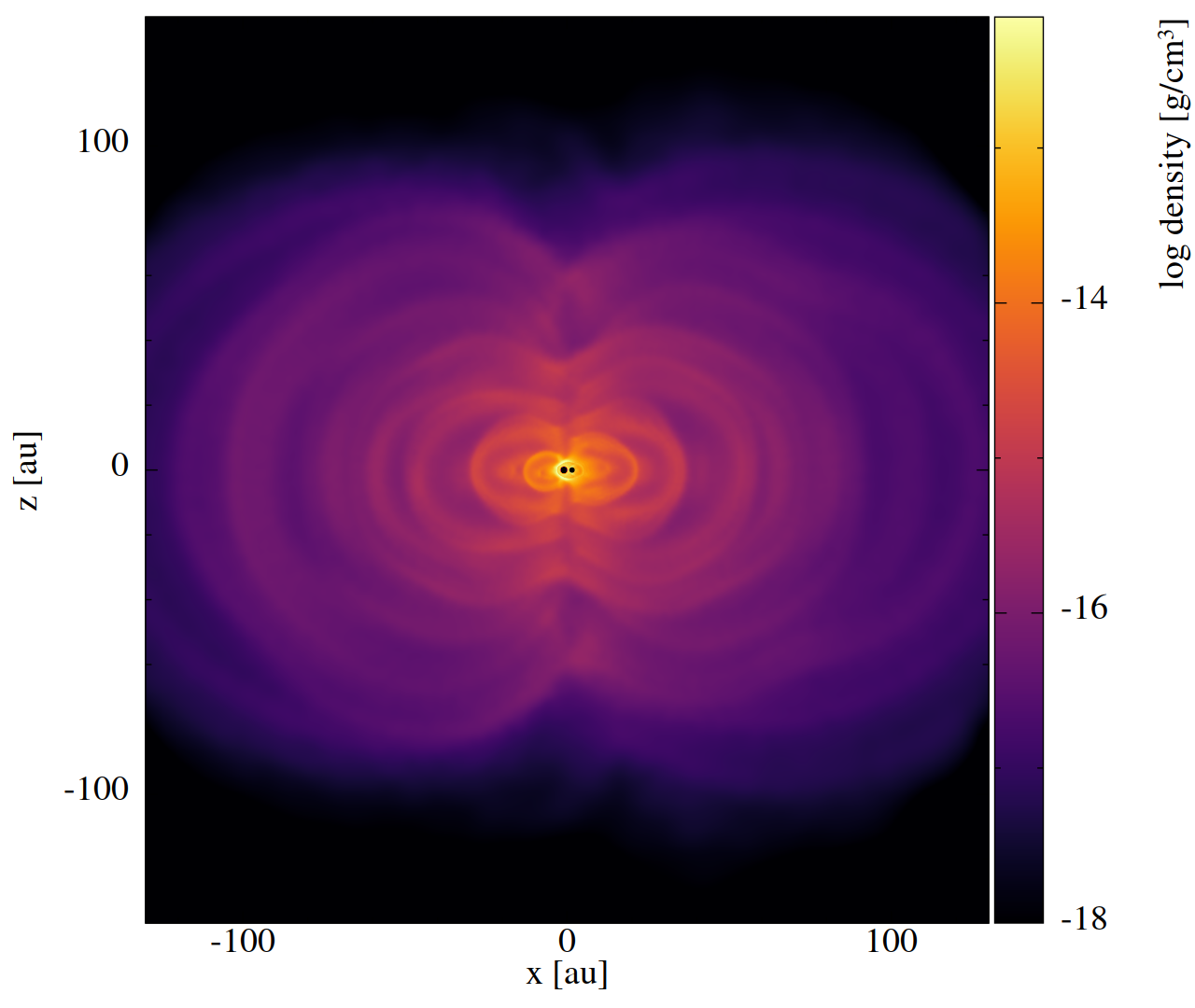}
	\includegraphics[width=0.31\textwidth]{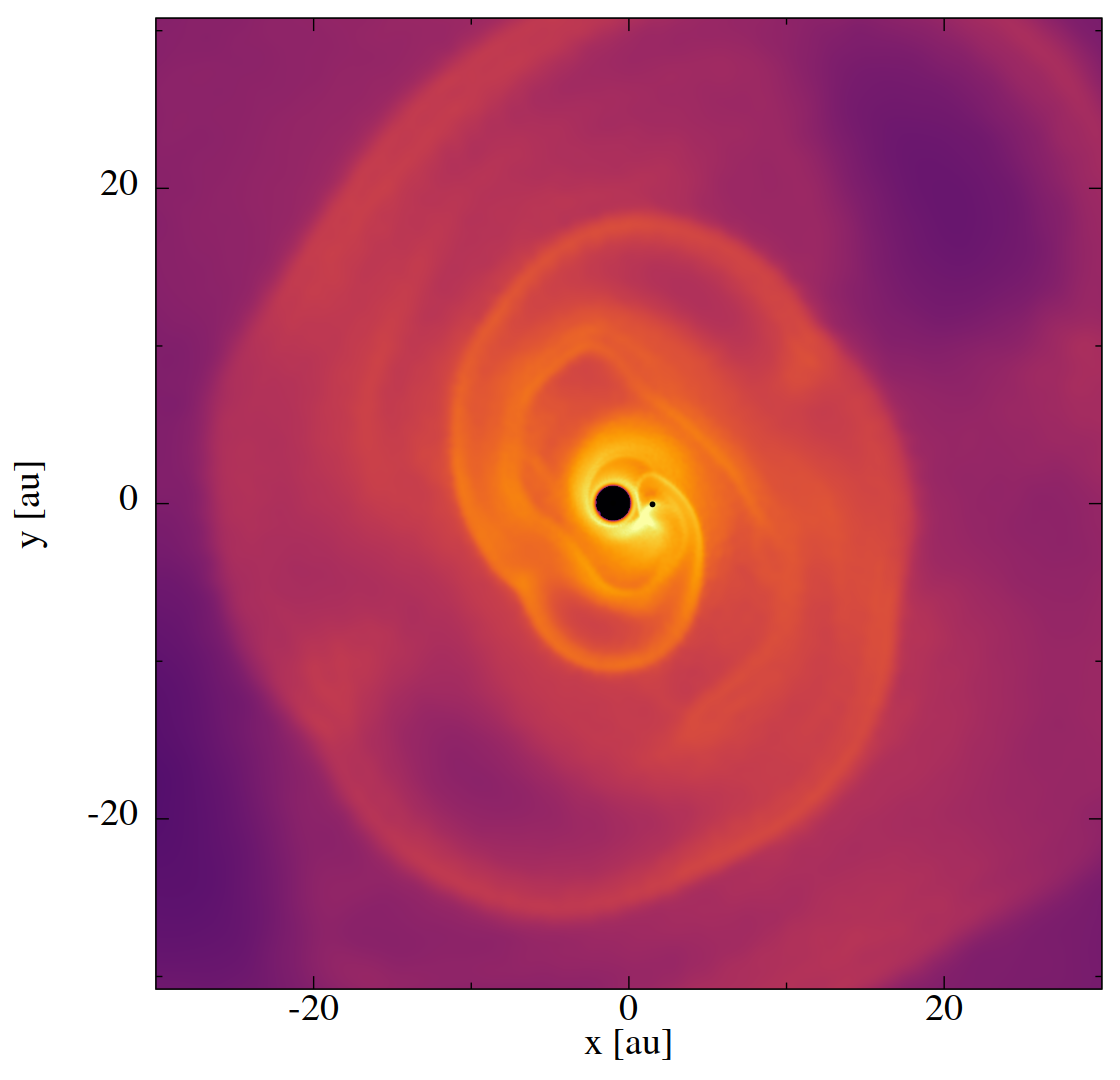}
	\includegraphics[width=0.31\textwidth]{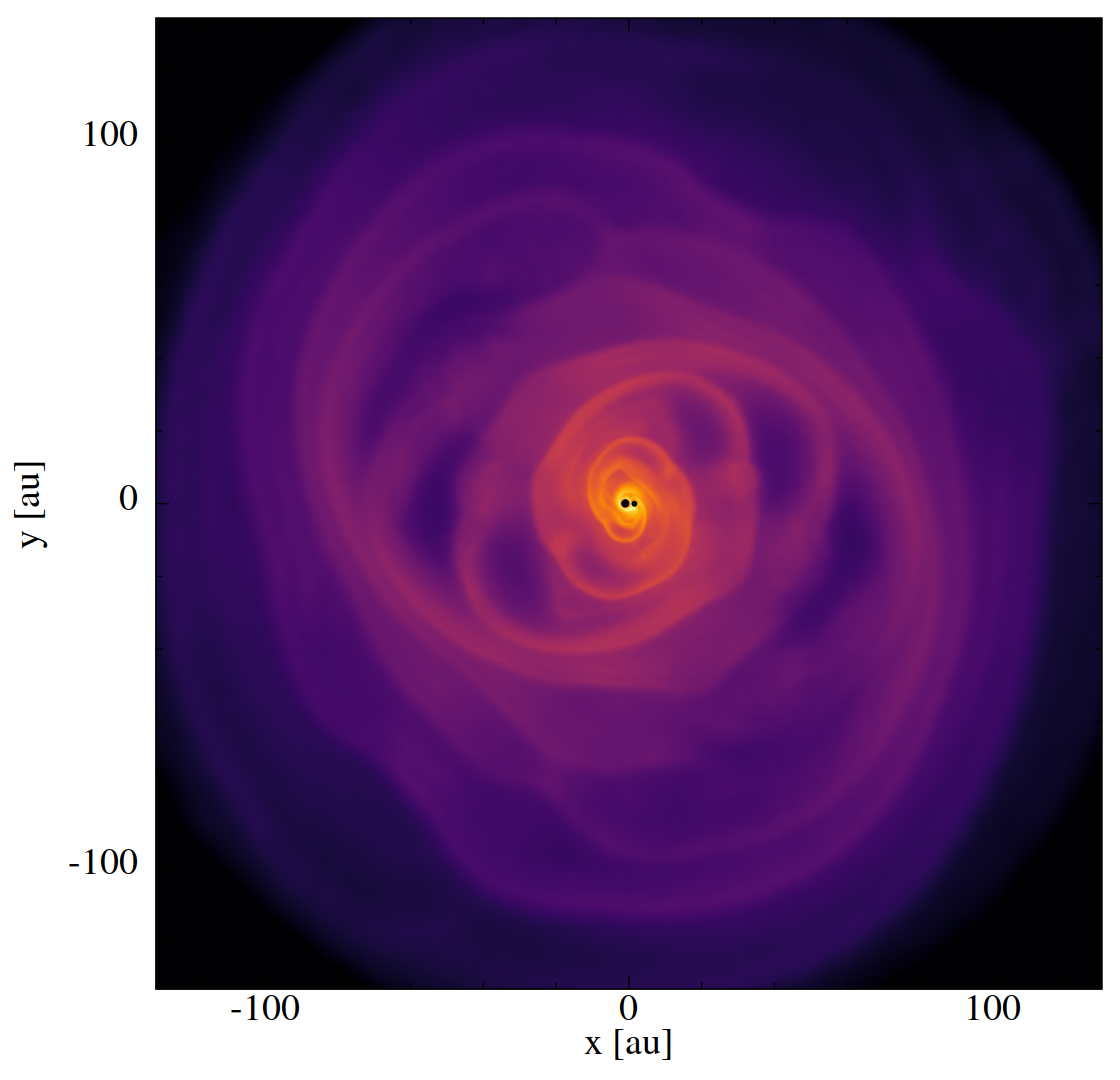}
	\includegraphics[width=0.355\textwidth]{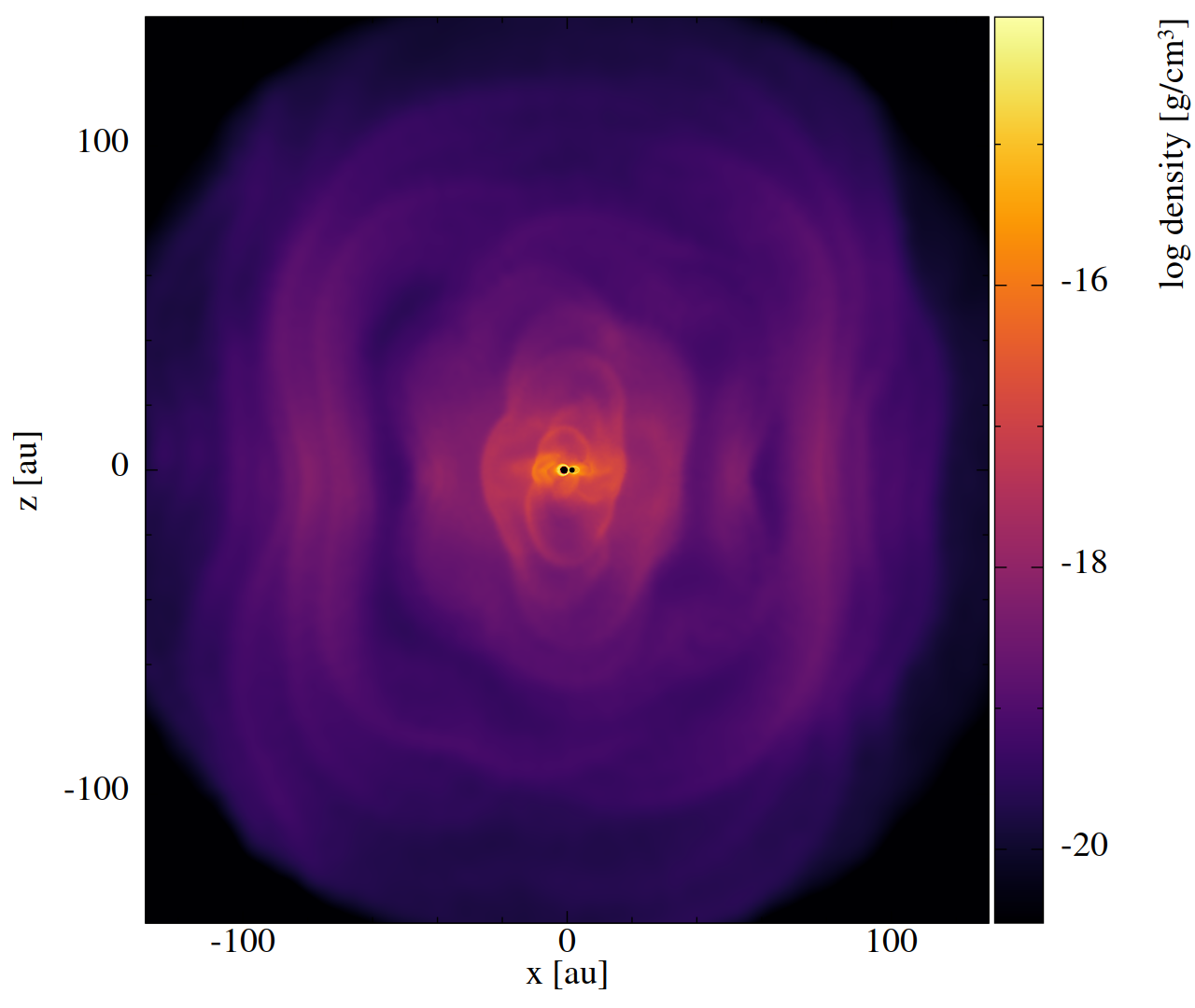}
	\includegraphics[width=0.31\textwidth]{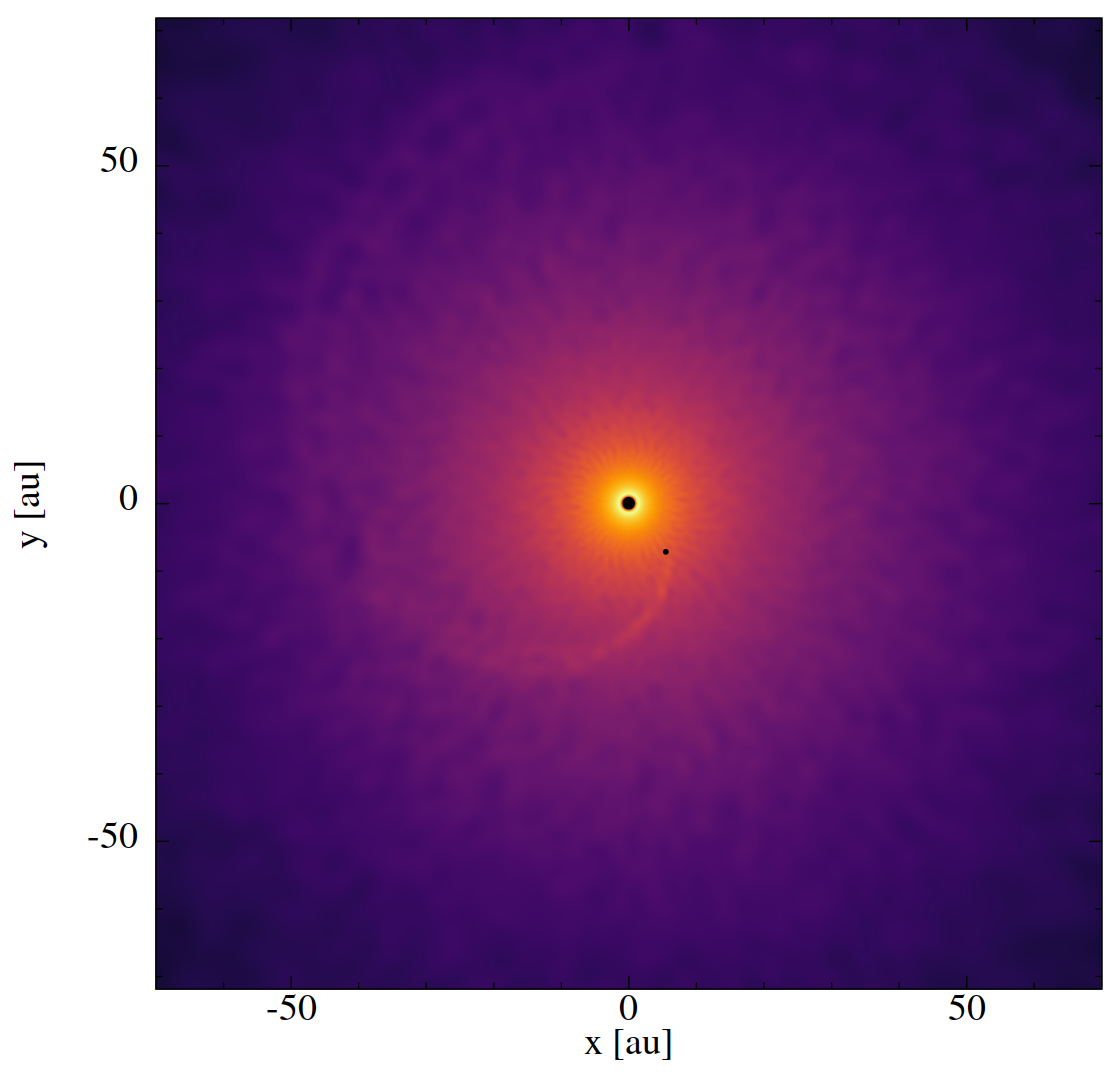}
	\includegraphics[width=0.31\textwidth]{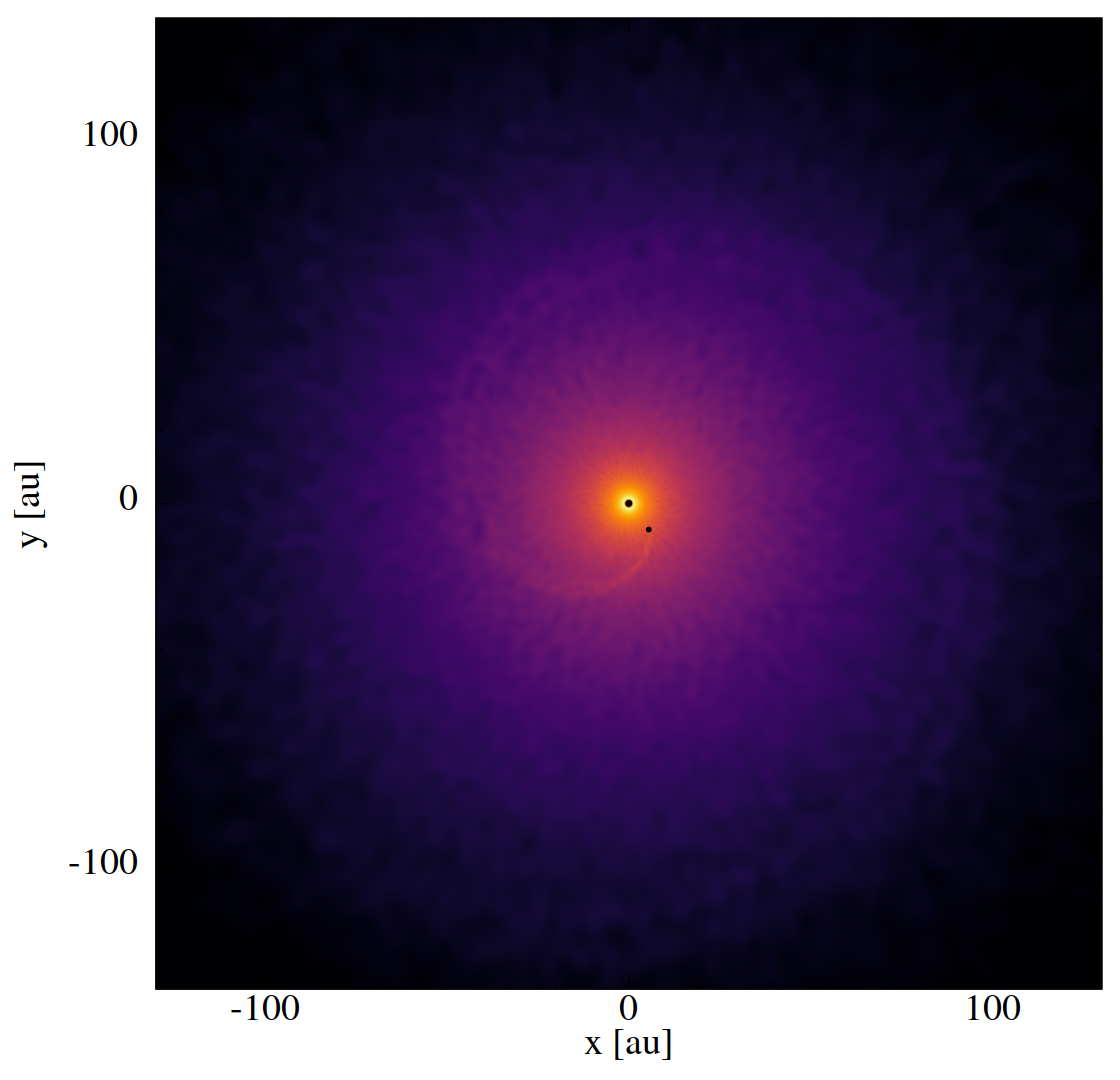}
	\includegraphics[width=0.355\textwidth]{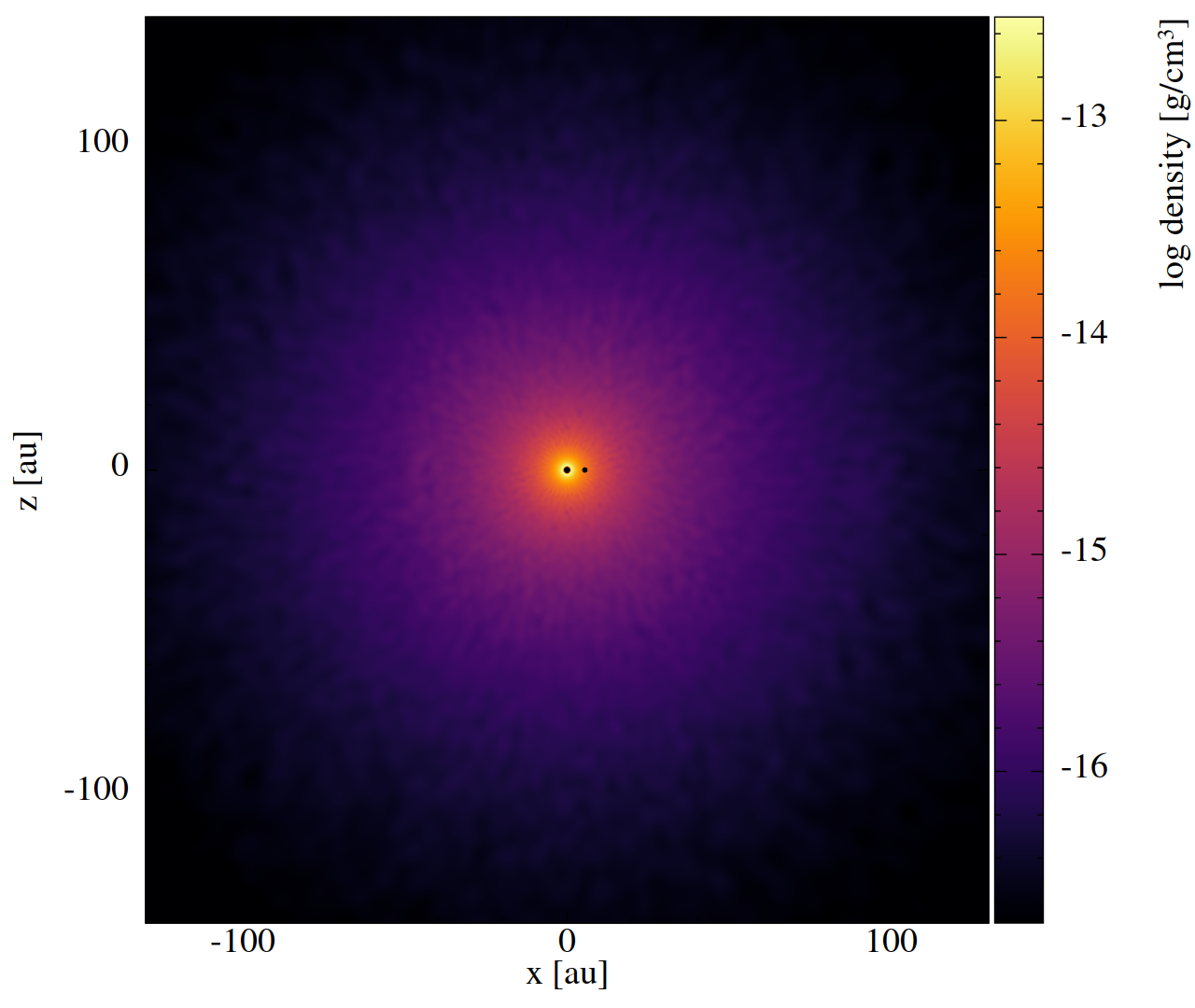}
	\includegraphics[width=0.31\textwidth]{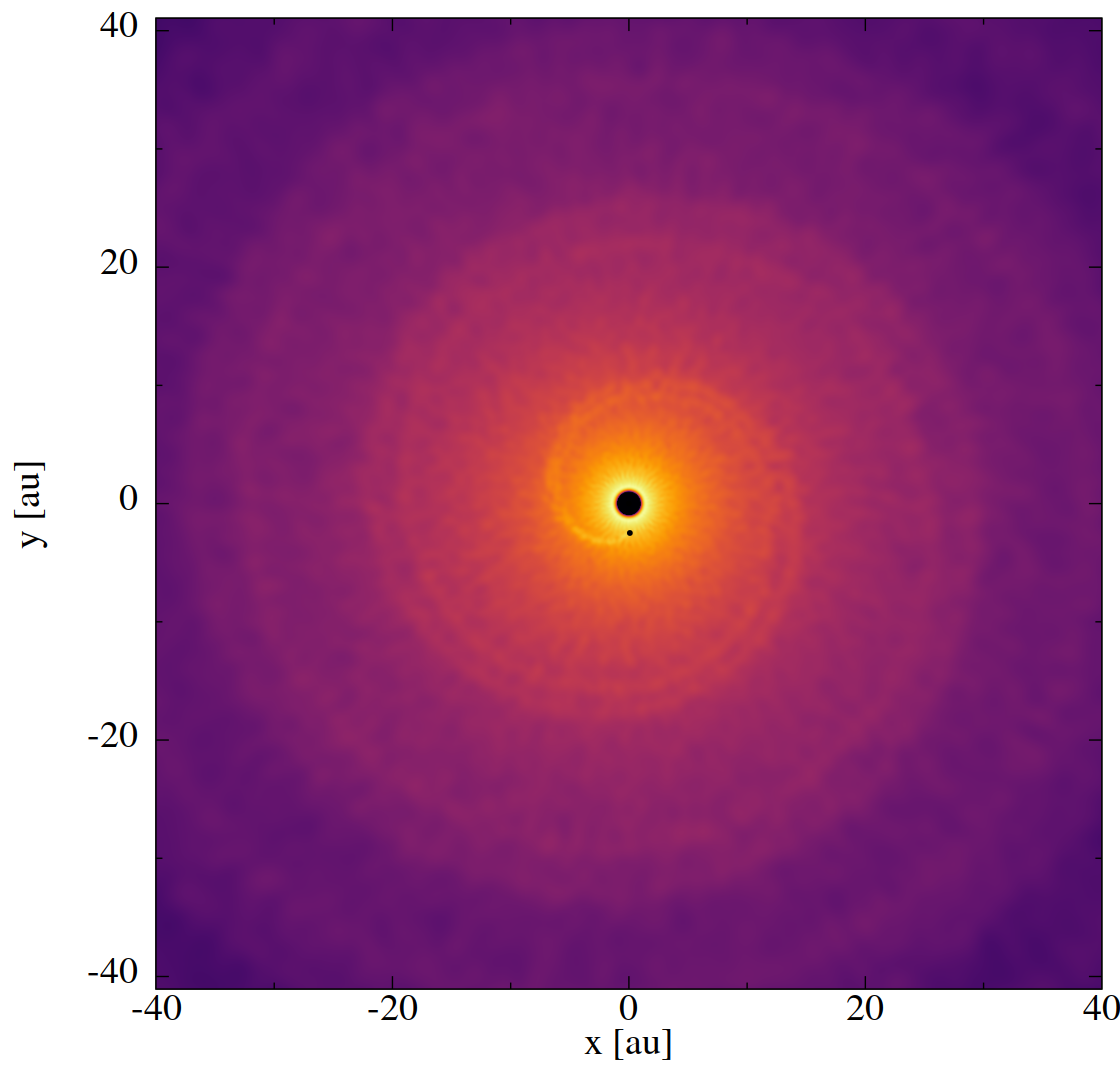}
	\includegraphics[width=0.31\textwidth]{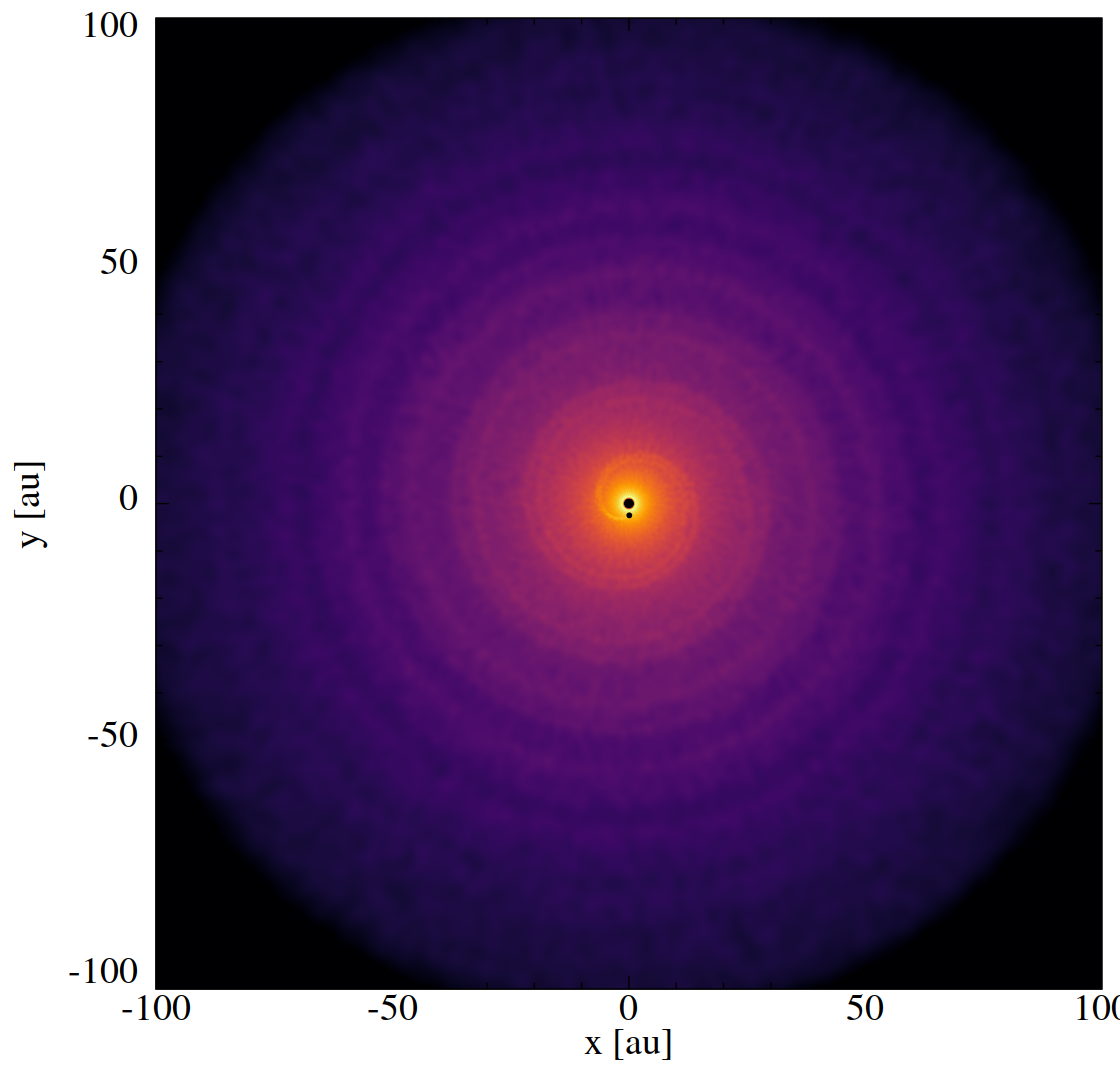}
	\includegraphics[width=0.355\textwidth]{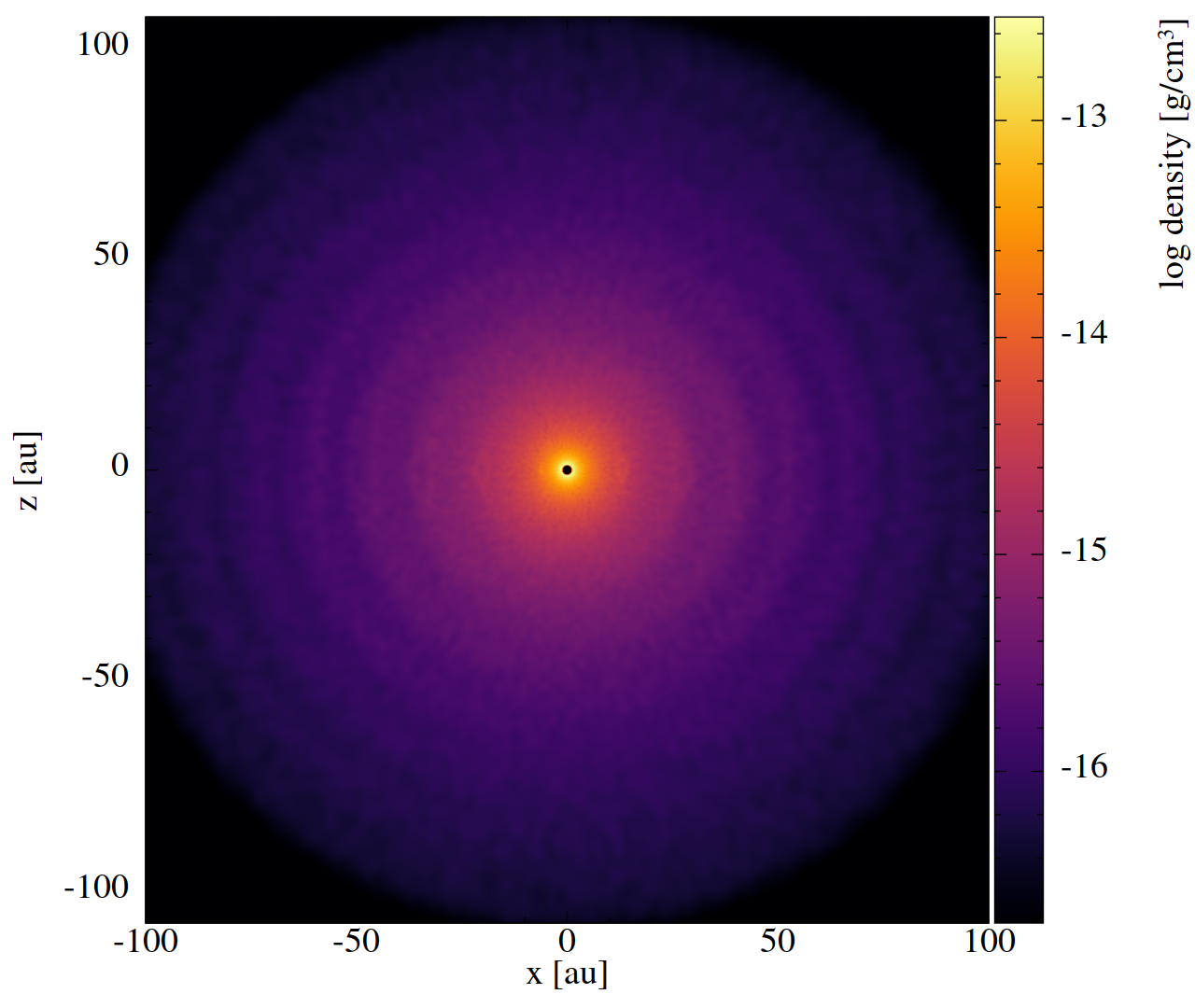}
	\caption{Density distribution of the orbital plane {(\textit{left panels} zoom-in of \textit{middle panels})} and meridional plane (\textit{right panels}). From \textit{top} to \textit{bottom}: models \Cts, \Ots, \Cnp\ and \Ctp. Left and right black dots represent the AGB star and companion, respectively, not to scale. {Snapshot taken after 10 orbits when $a=25\,\AU$ and after 5 when $a=90\,\AU$, at which self-similarity is reached.} }
	\label{fig:gallery1}
\end{figure*}
\begin{figure*}
	\centering
	\includegraphics[width=0.31\textwidth]{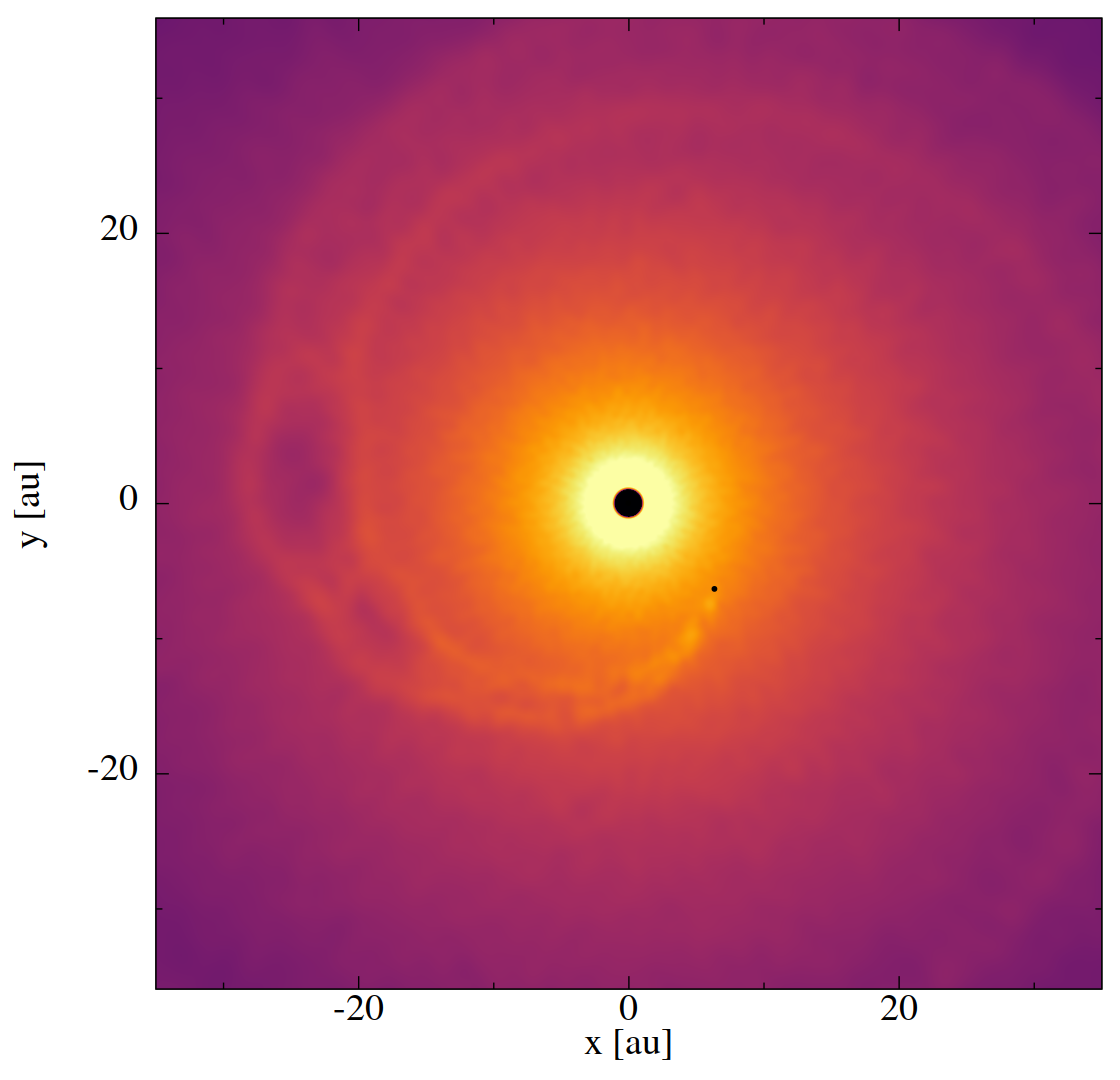}
	\includegraphics[width=0.31\textwidth]{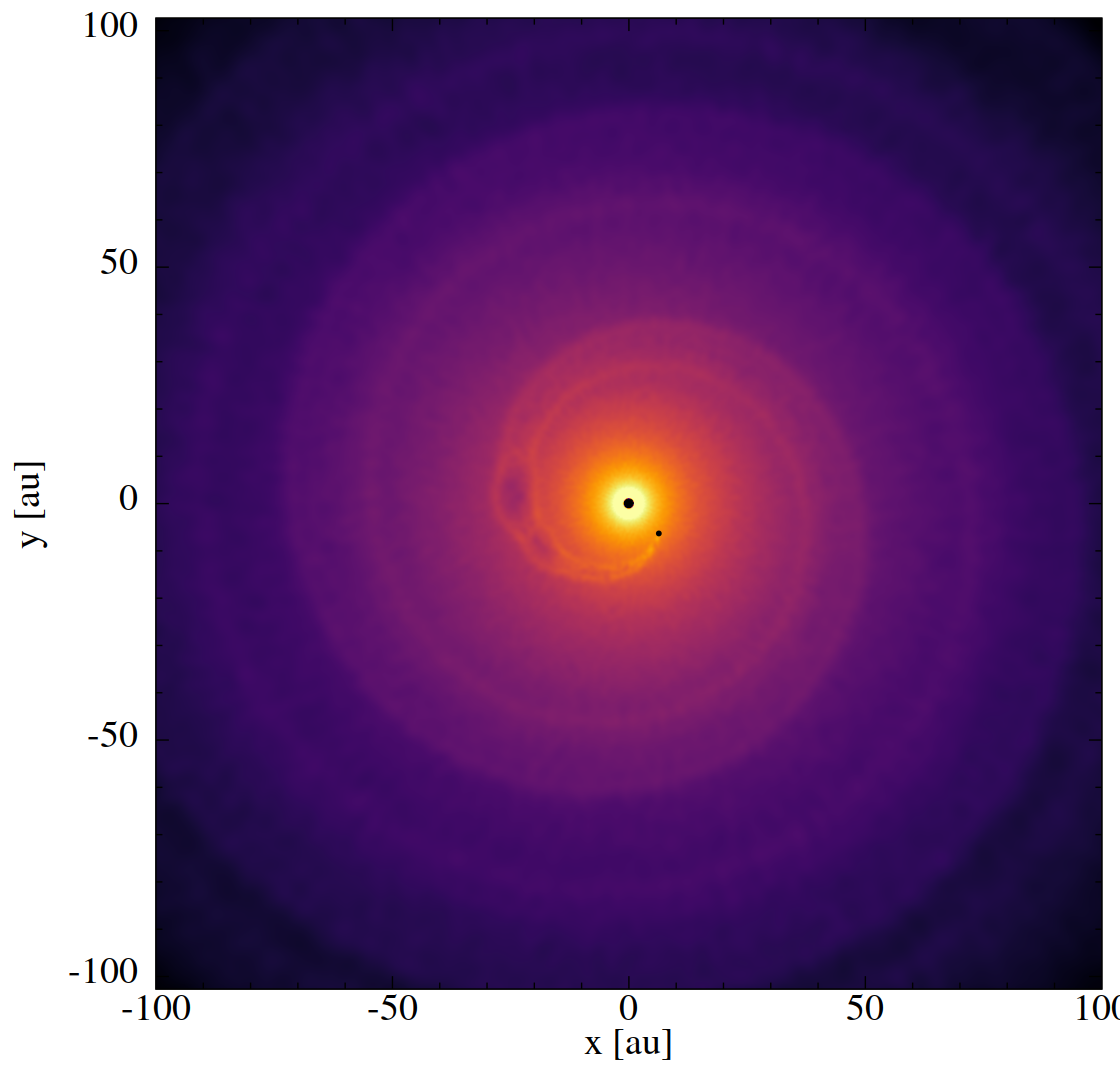}
	\includegraphics[width=0.355\textwidth]{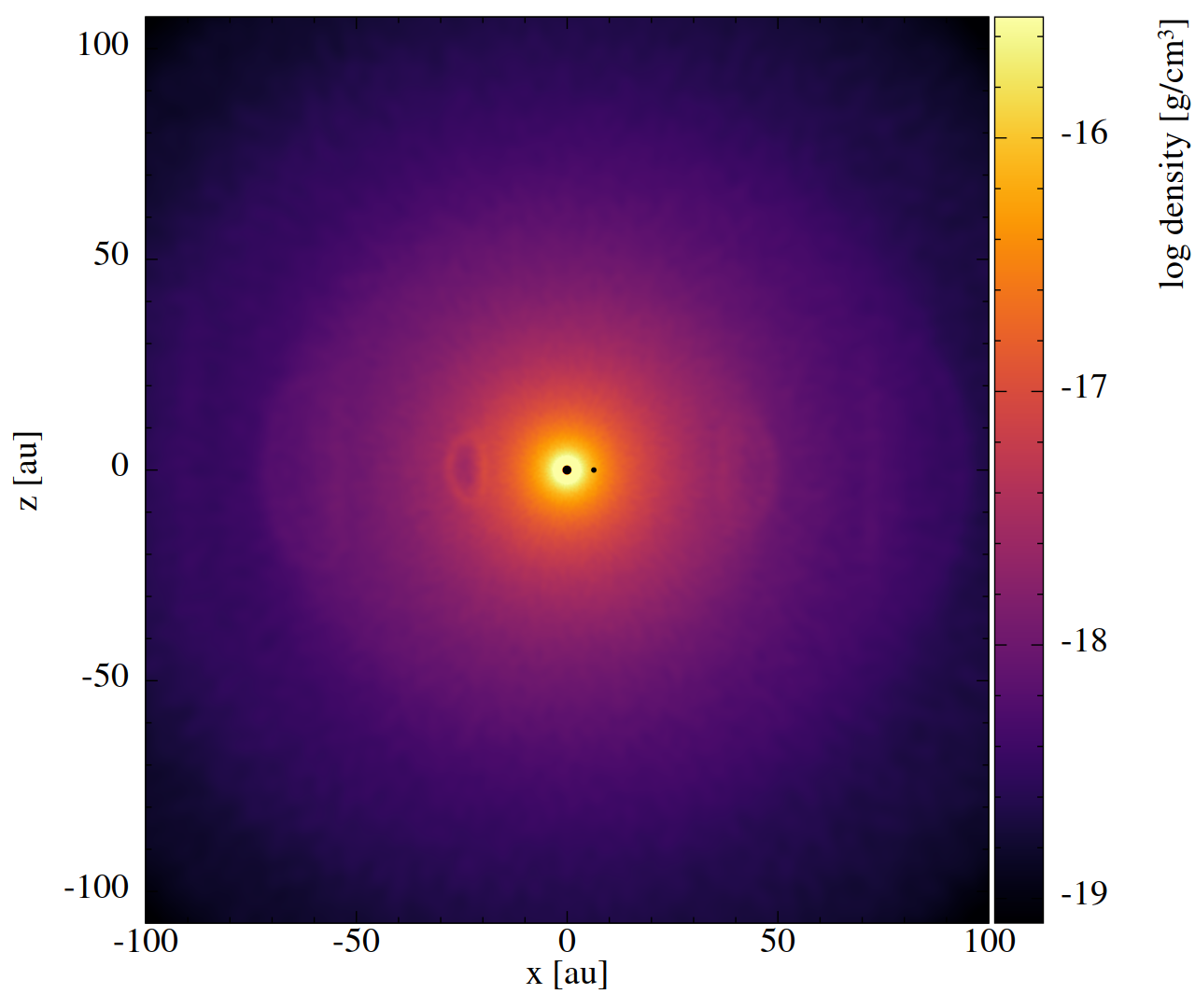}
	\includegraphics[width=0.31\textwidth]{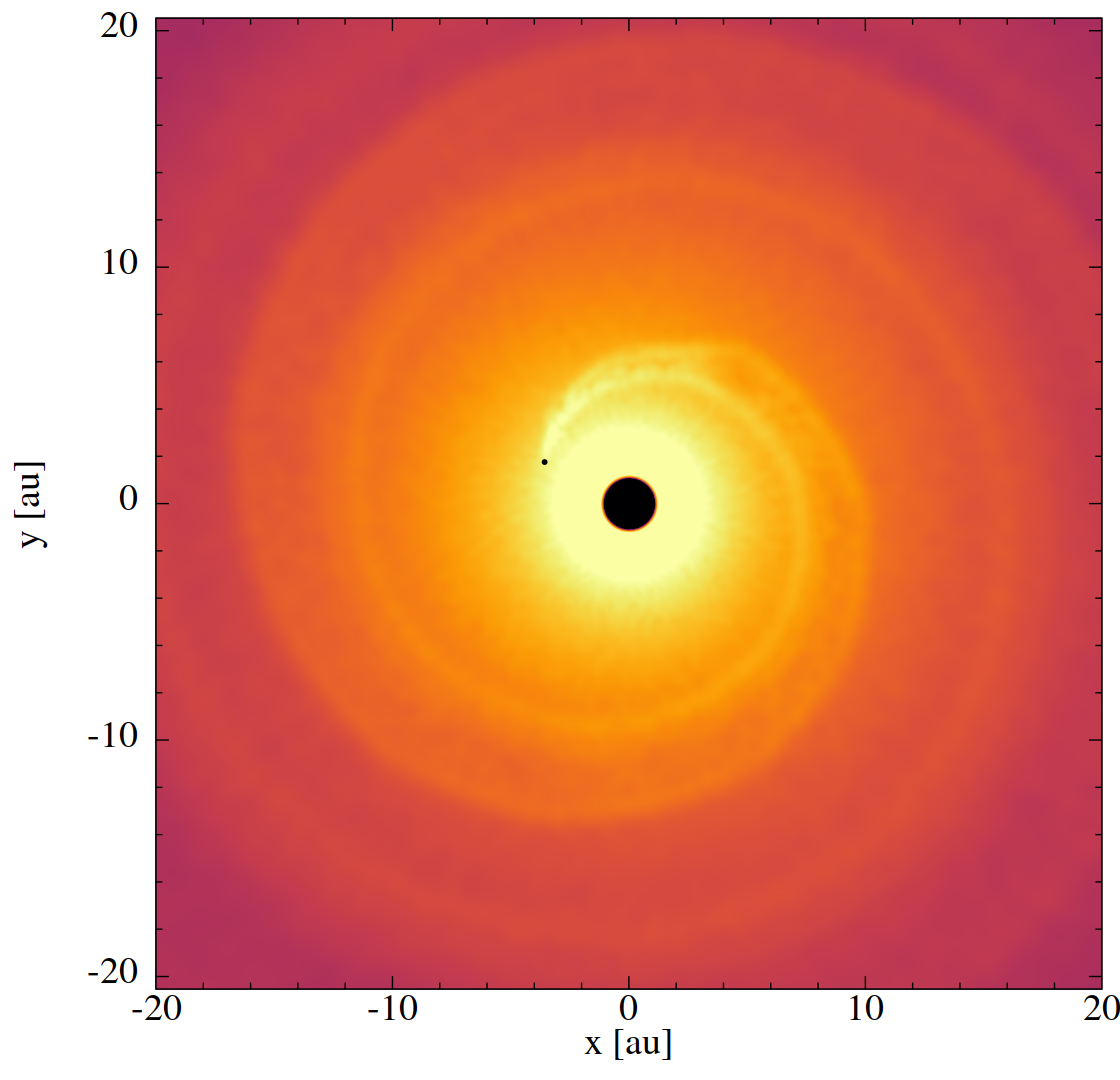}
	\includegraphics[width=0.31\textwidth]{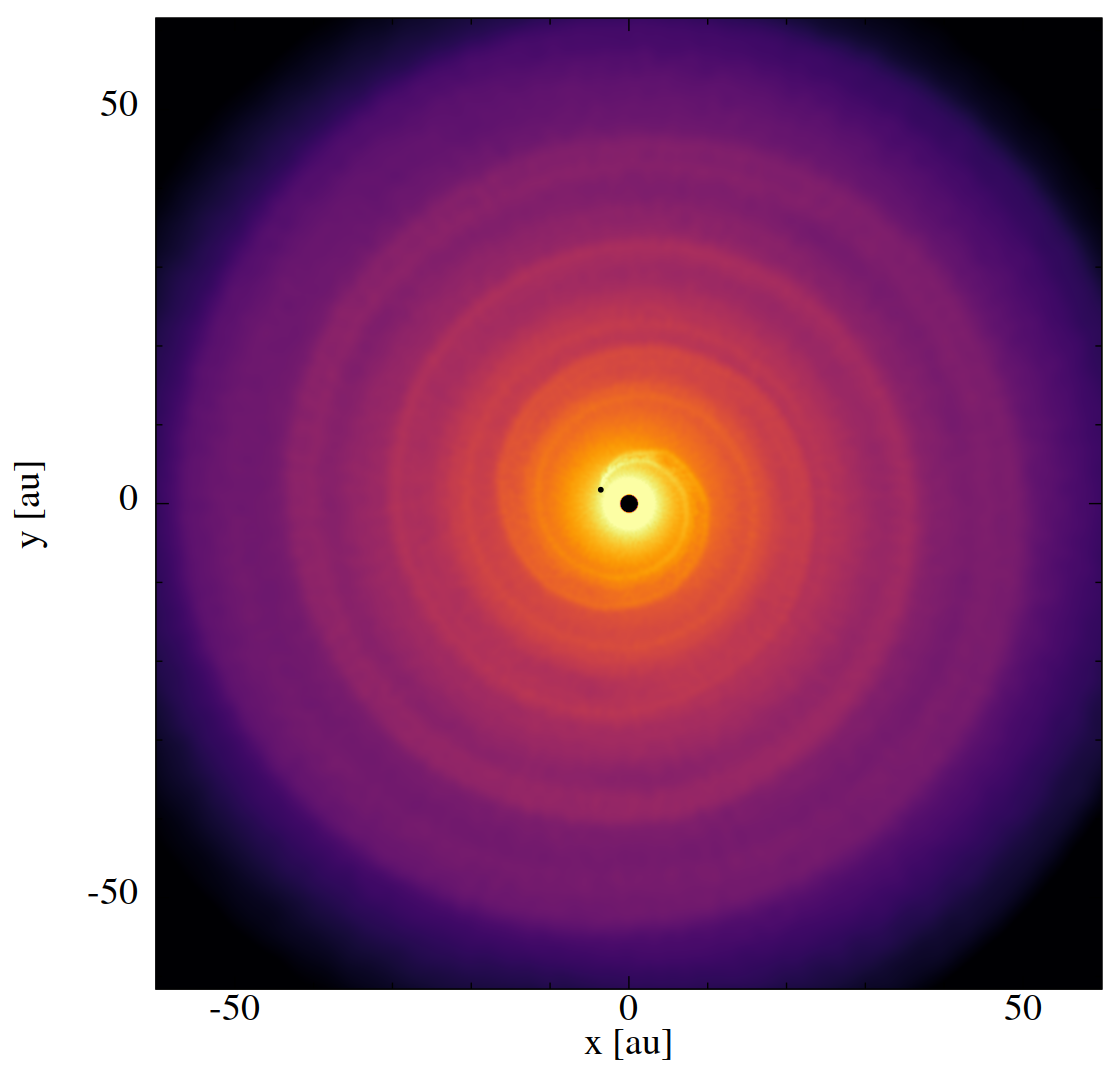}
	\includegraphics[width=0.355\textwidth]{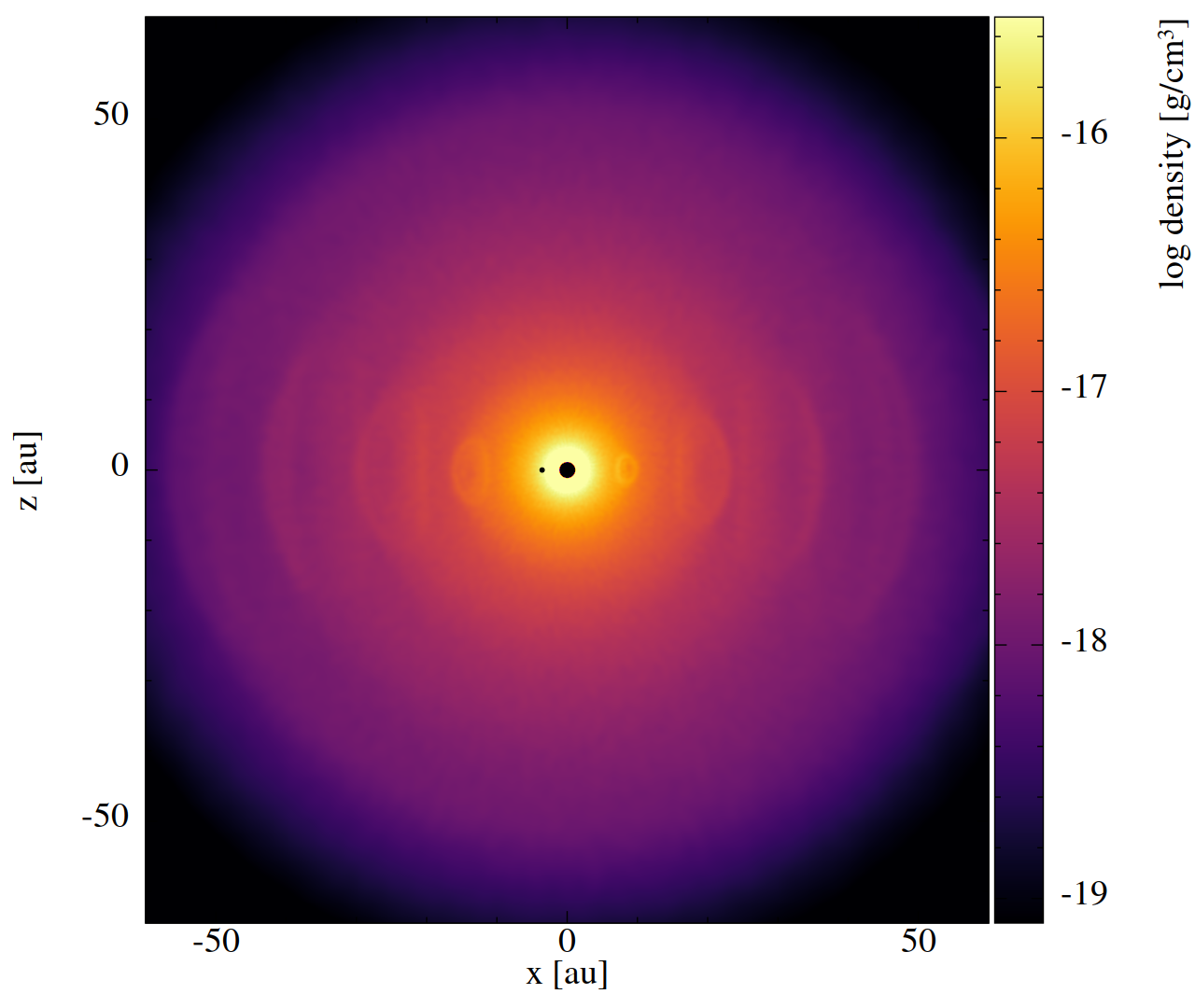}
	\includegraphics[width=0.31\textwidth]{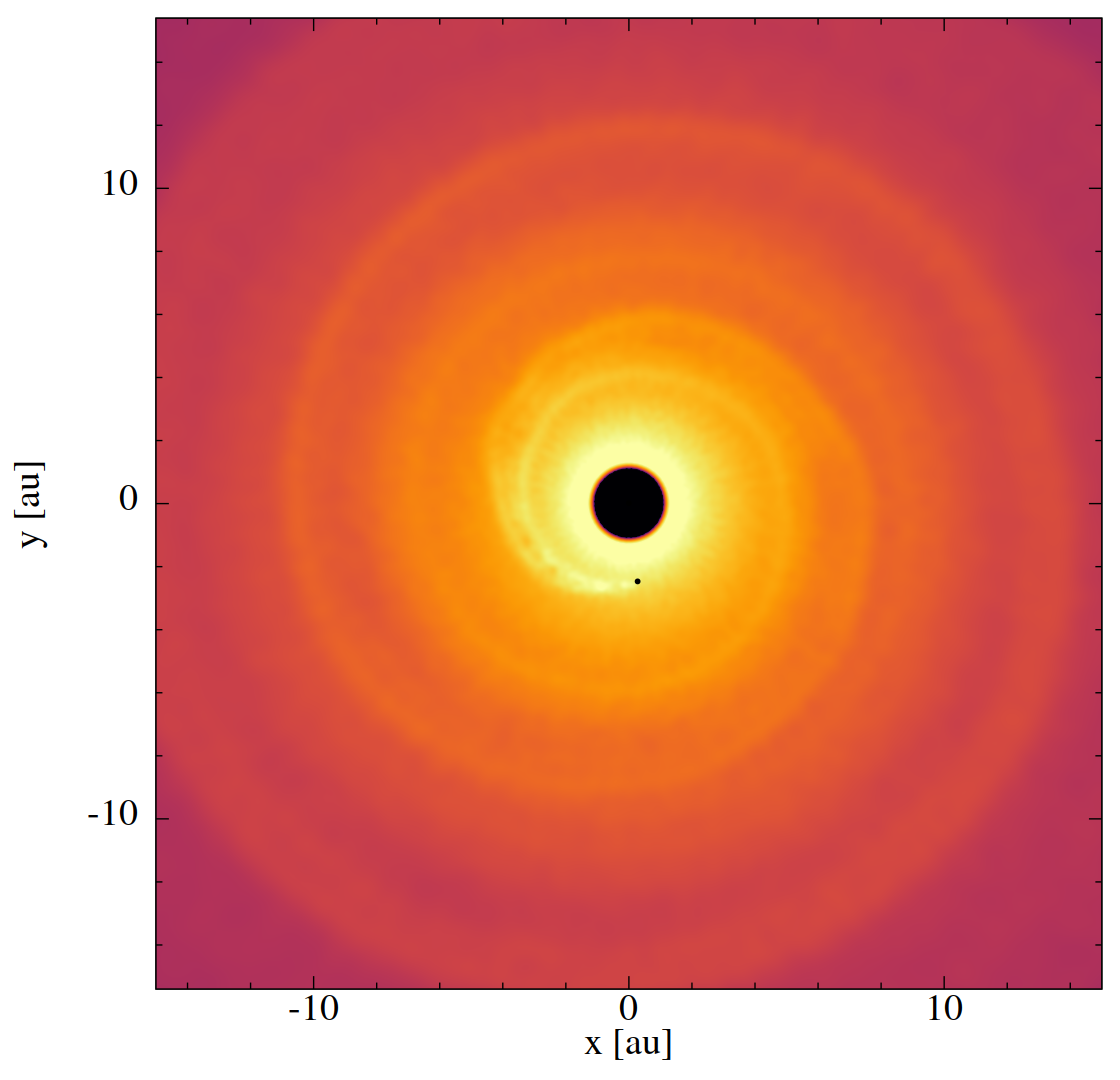}
	\includegraphics[width=0.31\textwidth]{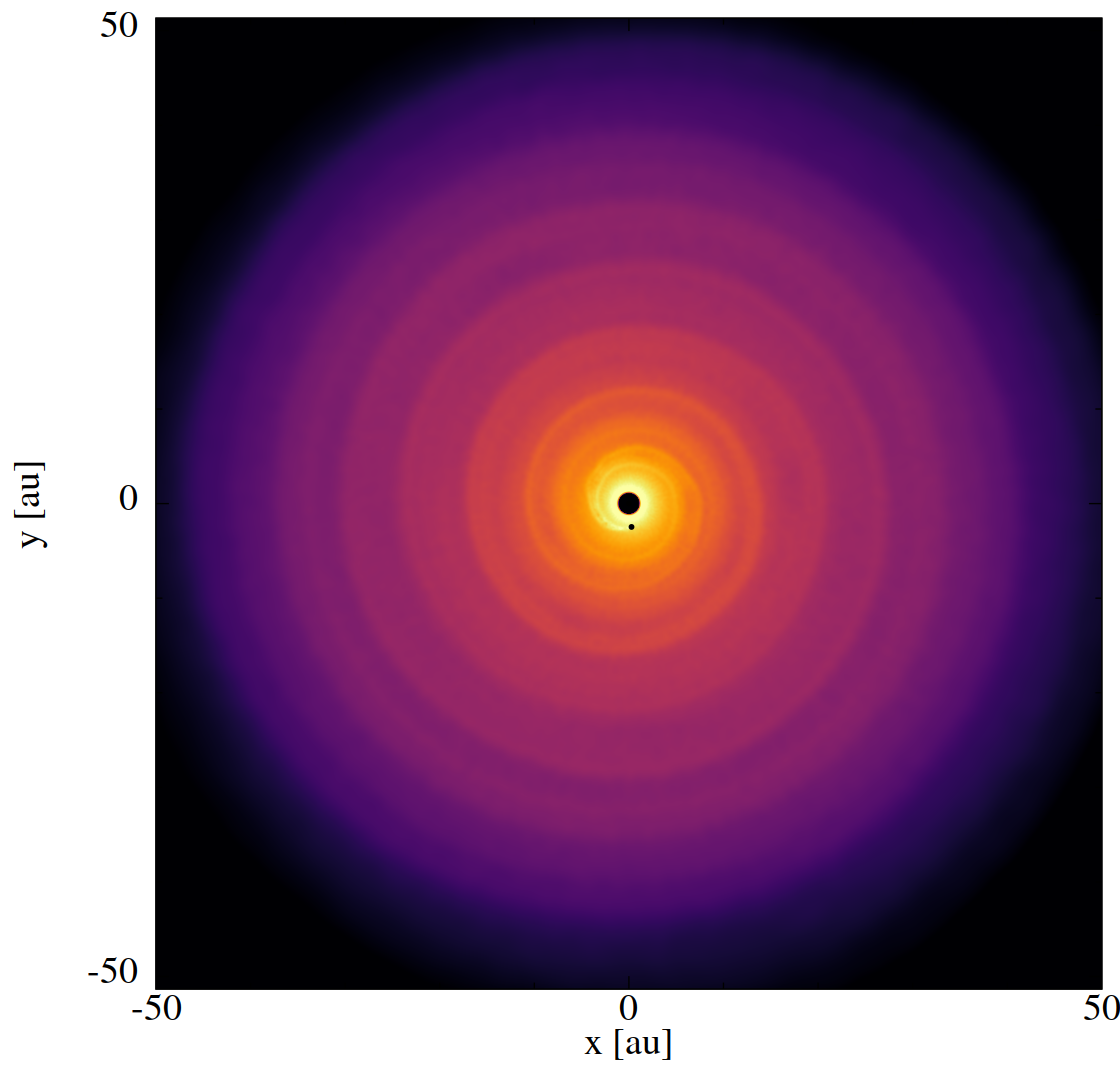}
	\includegraphics[width=0.355\textwidth]{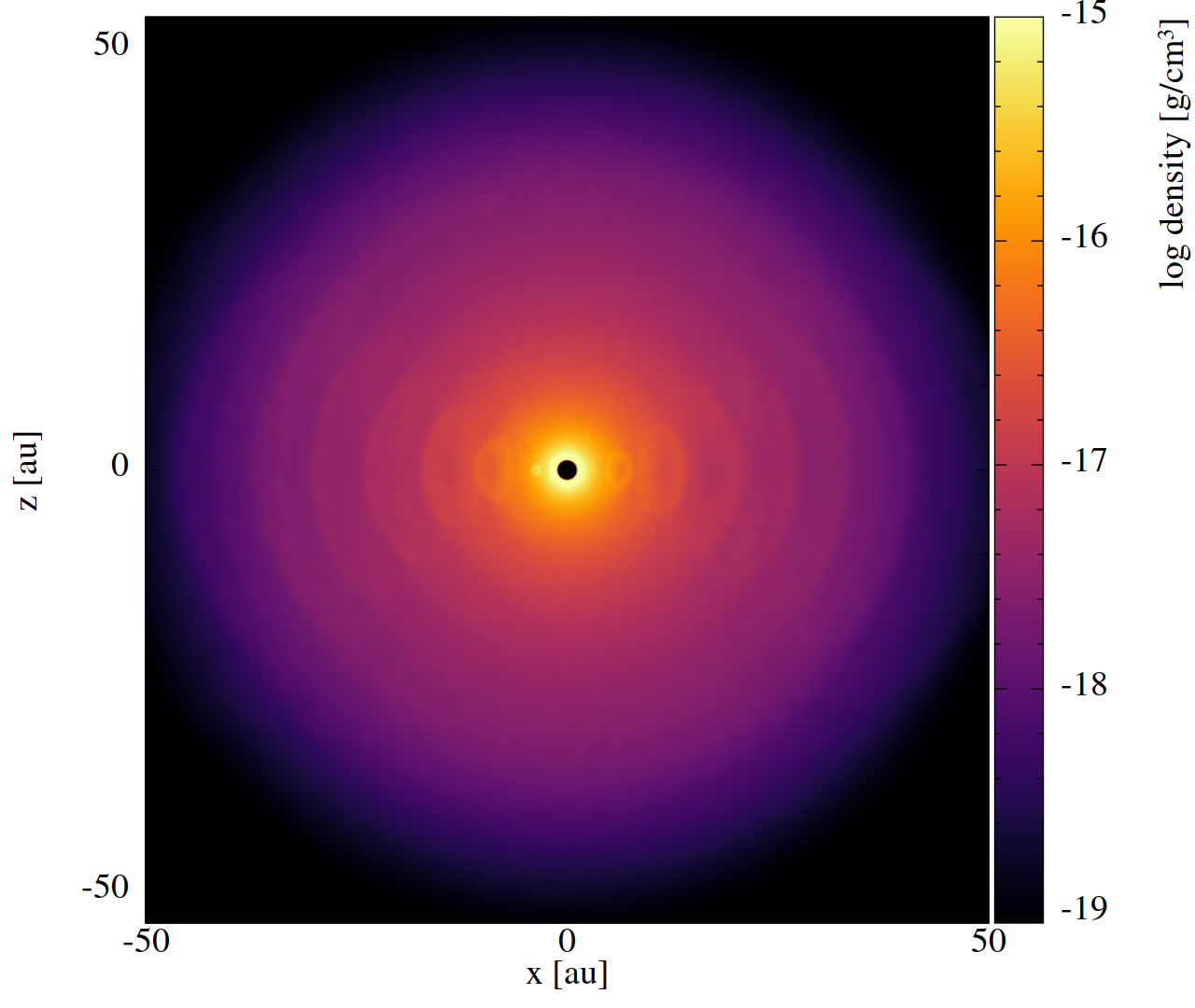}
	\caption{Density distribution of the orbital plane {(\textit{left panels} zoom-in of \textit{middle panels})} and meridional plane (\textit{right panels}). From \textit{top} to \textit{bottom}: models \Onp, \Ofp\ and \Otp. Left and right black dots represent the AGB star and companion, respectively, not to scale. {Snapshot taken after 5, 7 and 10 orbits, respectively, at which self-similarity is reached.} }
	\label{fig:p_slow}
\end{figure*}

\end{document}